\documentclass[aps,superscriptaddress,nofootinbib]{revtex4-1}
\usepackage{tikz,lipsum,lmodern,url,hyperref,xspace,units}

\usepackage{amsmath,amsfonts,amssymb,amsthm}
\allowdisplaybreaks
\usepackage{slashed,extarrows}
\usepackage{commath}
\usepackage{tikz-feynman, contour}
\usepackage{mathrsfs}
\usepackage{graphicx,subcaption} 
\DeclareGraphicsExtensions{.pdf,.png,.jpg}
\usepackage{enumitem}
\usepackage{float,ragged2e,multirow}

\DeclareMathOperator{\Tr}{Tr}
\DeclareMathOperator{\arccot}{arccot}

\newcommand{\ga}{\ensuremath{\gamma\gamma\rightarrow M\bar{M}}\xspace}
\newcommand{\dy}{\ensuremath{q\bar{q}\rightarrow M\bar{M}}\xspace }
\newcommand{\MAD}{\textsc{MadGraph}\xspace}
\newcommand{\Feyn}{\textsc{FeynRules}\xspace}
\newcommand{\Feyncalc}{\textsc{FeynCalc}\xspace}
\newcommand{\Math}{\textsc{Mathematica}\xspace}
\newcommand{\pyt}{\textsc{Python}\xspace}
\newcommand{\fort}{\textsc{Fortran}\xspace}
\newcommand{\Heav}{\textit{Heaviside-Lorentz}\xspace}
\newcommand{\gd}{\ensuremath{g_{\textrm D}}\xspace}
\newcommand{\sqq}{s_{qq}\xspace}
\newcommand{\sgg}{s_{\gamma\gamma}\xspace}
\newcommand{\half}{\nicefrac{1}{2}\xspace}
\newcommand{\dd}{\ensuremath{\mathrm d}}
\newcommand{\pt}{\ensuremath{p_{\mathrm T}}\xspace}

\def\Quad{\hskip1em\relax}
\def\quad{\hskip1em\relax}
\def\qquad{\hskip2em\relax}

\begin{document}
\begin{flushleft} 
KCL-PH-TH/2018-43 \\
IFIC/18-32 
\end{flushleft} 

\title{Monopole production via photon fusion and Drell-Yan processes: M{\small AD}G{\small RAPH} implementation and perturbativity via velocity-dependent coupling and magnetic moment as novel features}

\author{S.~Baines}
\email{stephanie.baines@kcl.ac.uk}

\affiliation{Theoretical Particle Physics and Cosmology Group, Department of Physics, King's College London, Strand, London WC2R~2LS, UK}

\author{N.E.~Mavromatos}
\email{nikolaos.mavromatos@kcl.ac.uk}

\affiliation{Theoretical Particle Physics and Cosmology Group, Department of Physics, King's College London, Strand, London WC2R~2LS, UK}

\author{V.A.~Mitsou}
\email{vasiliki.mitsou@ific.uv.es}

\affiliation{Instituto de F\'isica Corpuscular (IFIC), CSIC -- Universitat de Val\`encia, \\
C/ Catedr\'atico Jos\'e Beltr\'an 2, E-46980 Paterna (Valencia), Spain}

\author{J.L.~Pinfold}
\email{jpinfold@ualberta.ca}

\affiliation{Physics Department, University of Alberta, Edmonton Alberta T6G~2E4, Canada}

\author{A.~Santra}
\email{santra.arka@ific.uv.es}

\affiliation{Instituto de F\'isica Corpuscular (IFIC), CSIC -- Universitat de Val\`encia, \\
C/ Catedr\'atico Jos\'e Beltr\'an 2, E-46980 Paterna (Valencia), Spain}

\vspace{2cm}

\date{\today}

\begin{abstract}

In this work we consider point-like monopole production via photon-fusion and Drell-Yan processes in the framework of an effective $U(1)$ gauge field theory obtained from conventional models describing the interaction of spin~$0,\,\half,\, 1$ magnetically-charged fields with ordinary photons, upon \emph{electric-magnetic dualisation}. We present arguments based on such dualities which support the conjecture of an effective monopole-velocity-dependent magnetic charge. For the cases of spin-\half and spin-1 monopoles, we also include a magnetic-moment term $\kappa$, which is treated as a new phenomenological parameter and, together with the velocity-dependent coupling, allows for a perturbative treatment of the cross-section calculation. We discuss unitarity issues within these effective field theories, in particular we point out that in the spin-1 monopole case only the value $\kappa=1$ may restore unitarity. However from an effective-field-theory point of view, this lack of unitarity should not be viewed as an impediment for the phenomenological studies and experimental searches of generic spin-1 monopoles, given that the potential appearance of new degrees of freedom in the ultraviolet completion of such models might restore it. The second part of the paper deals with an appropriate implementation of photon-fusion and Drell-Yan processes based on the above theoretical scenarios into \MAD UFO models, aimed to serve as a useful tool in interpretations of monopole searches at colliders such as LHC, especially for photon fusion, given that it has not been considered by experimental collaborations so far. Moreover, the experimental implications of such perturbatively reliable monopole searches have been laid out.  
 
\end{abstract}

\maketitle

\section{Introduction}\label{intro}

Eighty seven years since its concrete formulation by Dirac~\cite{dirac} as a quantum mechanical source of magnetic poles, the magnetic monopole remains a hypothetical particle. Although there are concrete field-theoretical models beyond the Standard Model (SM) of particle physics 
which contain concrete monopole solutions~\cite{hpmono,cho,chofinite,you,aruna,you2,sarben,shafi}, these are extended objects with complicated substructure, and their production at collider is either impossible, as their mass range is beyond the capabilities of the latter~\cite{hpmono,shafi}, or extremely suppressed, due to their underlying composite nature~\cite{drukier}. On the other hand, point-like monopoles, originally envisaged by Dirac, are sources of singular magnetic fields for which the
underlying theory, if any, is completely unknown, even though in principle (due to their point-like nature) they could avoid suppression in production. 

In this respect, the spin of the monopole remains a free parameter. One may attempt to obtain an, admittedly heuristic, understanding of their production by 
considering effective field theoretic models for such production mechanisms based on \emph{electric-magnetic duality}. That is deriving the corresponding cross sections from perturbative field-theoretical models describing the interaction of fields of various spins, $S=0,~\half$ and~$1$  with photons, upon the replacement of the electric charge $q_e$ by the magnetic charge $g$, the latter obeying Dirac's quantisation rule
\begin{equation}\label{diracrule}
g q_e = \frac{1}{2} n \, (4\pi \epsilon_0 c)^\xi \, \hbar c , \quad n \in {\mathbb Z},
\end{equation}
where $c$ is the speed of light in vacuo, $\hbar$ is the reduced Planck constant, $\epsilon_0$ is the vacuum permittivity and 
${\mathbb Z}$ is the set of integers (with $n=0$ denoting the absence of magnetic charge). The quantity $\xi$ depends on the system of units used, 
with $\xi=0$ representing the CGS Gaussian system of units, and $\xi=1$ the SI system of units. In natural SI units ($\hbar=c = \epsilon_0=1$), which we adopt here, the fine-structure constant at zero energy scales is given by $\alpha =  \frac{e^2}{4\pi} = \frac{1}{137}$ with $e>0$ the positron charge, from which~\eqref{diracrule} yields
\begin{equation}\label{diracrule2}
g = \frac{1}{2\alpha} n  \, \Big(\frac{e}{q_e} \Big)\, e  = 68.5e \, \Big(\frac{e}{q_e}\Big) \, n \equiv n \,  \frac{e}{q_e}\, \gd, \quad n \in {\mathbb Z},
\end{equation}
with $\gd = 68.5e$ the fundamental Dirac charge. (We note, for completeness, that in composite monopole models~\cite{hpmono,cho}, the magnetic charge can be viewed as a collective coupling of $1/\alpha$ quanta of constituent (non-Abelian $W^{\pm}$-boson and Higgs) fields to a soft photon~\cite{drukier}, which is consistent with the charge quantisation condition~\eqref{diracrule2}.)   

The electric-magnetic duality replacement, which obeys the quantisation rule~\eqref{diracrule}, may be used as a basis for the evaluation of monopole-production cross sections from collisions of SM particles (quarks and leptons). Unfortunately, due to the large value of the magnetic charge~\eqref{diracrule2}, such a replacement renders the corresponding production process non-perturbative, consequently the strong-magnetic-coupling limit \emph{dual} theory is not well defined. Nevertheless, one may attempt to set benchmark scenarios for the cross sections by using tree-level Feynman-like graphs from such dual theories. This is standard practice in all point-like monopole searches at colliders so far~\cite{rpp}. 

\begin{figure}[ht!]
\justify
\begin{subfigure}[b]{0.5\textwidth}
\centering
\begin{tikzpicture}
  \begin{feynman}
  	\vertex (a1);
	\vertex[left=0.3cm of a1] (a11)  {\(q_e\)};
	\vertex[right=2cm of a1] (a2) ;
	\vertex[right=0.3cm of a2] (a22)  {\(q_e\)};
	\vertex[above=1.75cm of a2] (a3);
	\vertex[right=1cm of a3] (a4) {\(l^{-}\)};
	\vertex[below=1.75cm of a2] (a5);
	\vertex[right=1cm of a5] (a6) {\(l^{+}\)};
	
	\vertex[above=1.75cm of a1] (a7);
	\vertex[left=1cm of a7] (a8) {\(q\)};
	\vertex[below=1.75cm of a1] (a9);
	\vertex[left=1cm of a9] (a10) {\(\overline{q}\)};
	
	\diagram* {
	       	(a8) -- [fermion] (a1) -- [fermion] (a10),
		(a1) -- [boson, edge label=\(\gamma\)] (a2),
       		(a6) -- [fermion] (a2) -- [fermion] (a4),
      };
\end{feynman}
\end{tikzpicture}
\caption{Typical DY diagram.\label{fig:dylep}}
\end{subfigure}
~
\begin{subfigure}[b]{0.5\textwidth}
\centering
\begin{tikzpicture}
  \begin{feynman}
  	\vertex (a1);
	\vertex[left=0.3cm of a1] (a11)  {\(q_e\)};
	\vertex[right=2cm of a1] (a2) ;
	\vertex[right=0.3cm of a2] (a22)  {\(g\)};
	\vertex[above=1.75cm of a2] (a3);
	\vertex[right=1cm of a3] (a4) {\(M\)};
	\vertex[below=1.75cm of a2] (a5);
	\vertex[right=1cm of a5] (a6) {\(\overline{M}\)};
	
	\vertex[above=1.75cm of a1] (a7);
	\vertex[left=1cm of a7] (a8) {\(q\)};
	\vertex[below=1.75cm of a1] (a9);
	\vertex[left=1cm of a9] (a10) {\(\overline{q}\)};
	
	\diagram* {
	       	(a8) -- [fermion] (a1) -- [fermion] (a10),
		(a1) -- [boson, edge label=\(\gamma\)] (a2),
       		(a6) -- [fermion] (a2) -- [fermion] (a4),
      };
\end{feynman}
\end{tikzpicture}
\caption{DY monopole production.\label{fig:dy}}
\end{subfigure}

\medskip

\begin{subfigure}[b]{0.5\textwidth}
\centering
\begin{tikzpicture}
  \begin{feynman}
  	\vertex (a1);
	\vertex[above=0.2cm of a1] (a11)  {\(q_e\)};
	\vertex[right=2cm of a1] (a2);
	\vertex[above=0.5cm of a2] (a3)  {\(q\)};
	\vertex[left=2cm of a1] (a4);
	\vertex[above=0.5cm of a4] (a5)  {\(q\)};

	\vertex[below=2.1cm of a1] (a7);	
	\vertex[below=0.2cm of a7] (a22)  {\(q_e\)};
	\vertex[right=2cm of a7] (a8);
	\vertex[below=0.5cm of a8] (a9) {\(\overline{q}\)};
	\vertex[left=2cm of a7] (a81);
	\vertex[below=0.5cm of a81] (a91) {\(\overline{q}\)};
	
	\vertex[below=0.7cm of a1] (a71);	
	\vertex[left=0.1cm of a71] (a711) {\(g\)};	
	\vertex[below=1.4cm of a1] (a72);	
	\vertex[left=0.1cm of a72] (a722) {\(g\)};	
	
	\vertex[right=2cm of a71] (a82) {\(\overline{M}\)};
	\vertex[right=2cm of a72] (a83) {\(M\)};		
	
	\diagram* {
	       	(a5) -- [fermion] (a1) -- [fermion] (a3),
		(a9) -- [fermion] (a7) -- [fermion] (a91),
		(a1) -- [boson, edge label=\(\gamma\)] (a71),
		(a72) -- [boson, edge label=\(\gamma\)] (a7),
		(a82) -- [fermion] (a71) -- [fermion] (a72) -- [fermion] (a83),
      };
\end{feynman}
\end{tikzpicture}
\caption{Three-vertex PF process.\label{fig:pf3}}
\end{subfigure}
~
\begin{subfigure}[b]{0.5\textwidth}
\centering
\begin{tikzpicture}
  \begin{feynman}
  	\vertex (a1);
	\vertex[above=0.2cm of a1] (a11)  {\(q_e\)};
	\vertex[right=2cm of a1] (a2);
	\vertex[above=0.5cm of a2] (a3)  {\(q\)};
	\vertex[left=2cm of a1] (a4);
	\vertex[above=0.5cm of a4] (a5)  {\(q\)};

	\vertex[below=2cm of a1] (a7);	
	\vertex[below=0.2cm of a7] (a22)  {\(q_e\)};
	\vertex[right=2cm of a7] (a8);
	\vertex[below=0.5cm of a8] (a9) {\(\overline{q}\)};
	\vertex[left=2cm of a7] (a81);
	\vertex[below=0.5cm of a81] (a91) {\(\overline{q}\)};
	
	\vertex[below=1cm of a1] (a17);
	\vertex[left=0.1cm of a17] (a171) {\(g^{2}\)};
	
	\vertex[right=2cm of a17] (a82);
	\vertex[above=0.3cm of a82] (a821) {\(\overline{M}\)};
	\vertex[below=0.3cm of a82] (a822) {\(M\)};		
	
	\diagram* {
	       	(a5) -- [fermion] (a1) -- [fermion] (a3),
		(a9) -- [fermion] (a7) -- [fermion] (a91),
		(a1) -- [boson, edge label=\(\gamma\)] (a17) -- [boson, edge label=\(\gamma\)] (a7),
		(a821) -- [fermion] (a17) -- [fermion] (a822),
      };
\end{feynman}
\end{tikzpicture}
\caption{Four-vertex PF process.\label{fig:pf4}}
\end{subfigure}
\caption{Feynman-like tree-level graphs for production processes of monopoles with generic spin $S$. \subref{fig:dylep}: typical SM Drell-Yan process describing charged lepton production from quark-antiquark annihilation; \subref{fig:dy} DY monopole-antimonopole pair production from quark annihilation; \subref{fig:pf3} monopole-antimonopole pair production via photon fusion (for monopole spins 0, \half and 1); \subref{fig:pf4} additional (contact) diagram for monopole-antimonopole pair production via photon-fusion (for spins 0, and 1). The quantities $q_e$ and $g$ denote the electric and magnetic charge, respectively. }
\label{fig:diagrams}
\end{figure}
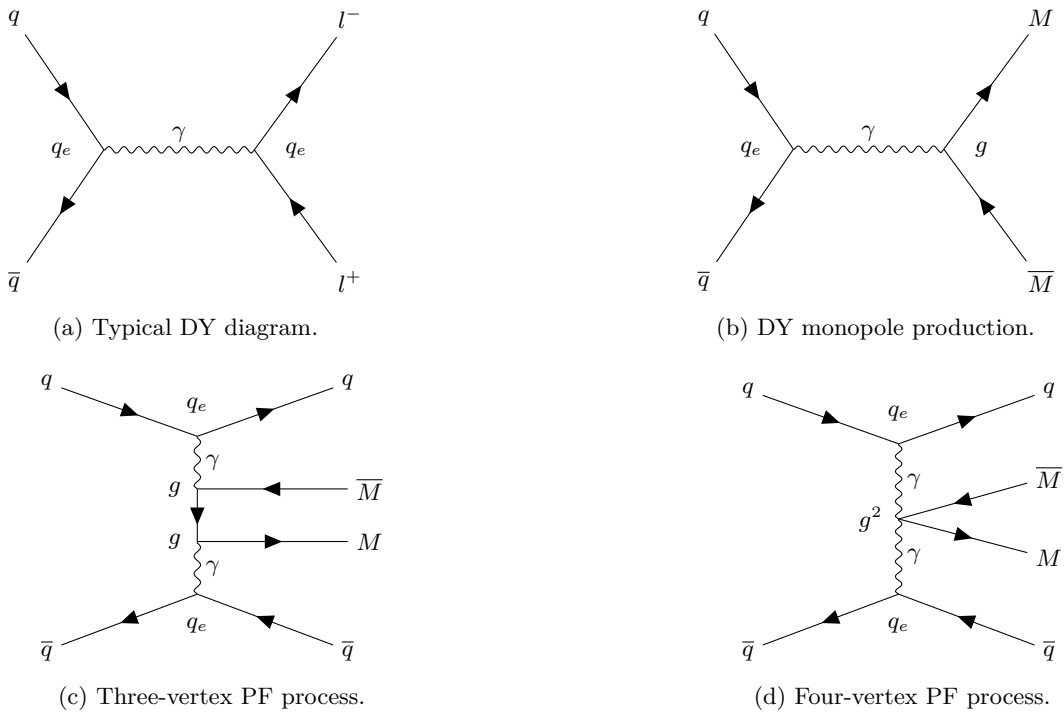

Depending on the spin of the monopole field $M$, typical graphs participating in monopole-antimonopole pair production at LHC from proton-proton ($pp$) collisions are given in fig.~\ref{fig:diagrams}. There are two kinds of such processes: the Drell-Yan (DY) (see fig.~\ref{fig:dy}) and the photon-fusion (PF) induced production (see figs.~\ref{fig:pf3} and~\ref{fig:pf4}).  We also mention, to be complete, that in photon-photon production we have not only elastic but also semi-elastic and inelastic photon-fusion processes. For spin-\half monopoles relevant in this discussion, a comparison between the respective perturbative cross sections has been provided first in ref.~\cite{original} for $p\bar{p}$ collisions at a centre-of-mass energy $\sqrt{s}=1.96$~TeV, and subsequently for $pp$ collisions of $\sqrt{s}$ up to 14~TeV in ref.~\cite{dw}. The conclusion from such analyses was that for $\sqrt{s}=1.96$~TeV the two cross sections are of comparable magnitude~\cite{dw}, whilst for $\sqrt{s}=14$~TeV PF dominates DY by a factor $> 50$, thus stressing the need to utilise the latter in monopole-search interpretations. 

In this work we extend such combined (DY plus PF) studies to also incorporate spin-$0$ and spin-1 monopoles. 
In ref.~\cite{original}, the authors have calculated (in appropriate dualised models) the total cross sections for monopole pair production by photon fusion for three different spin models, spins 0, \half and 1.  Theoretically, the cross sections increase with increasing spin, for a fixed (common) value of the monopole mass. It should be stressed that the expressions for the cross sections are specific to particular definitions of the monopole interactions. 
Specifically, the spin-\half model fixes the particle in a minimally coupled theory, dual to standard QED, thus mirroring the observed behaviour of the electron with a gyromagnetic ratio $g_{Re}=2$. The corresponding magnetic moment $\kappa$ is assumed zero. 
The magnetically charged monopole with spin 1 considered in \cite{original}, on the other hand, is characterised by a non zero magnetic moment term $\kappa=1$, which is the value that characterises the charged $W^{\pm}$ bosons in the SM. In fact the model mirrors the interactions of such bosons with a photon, but in the dual theory, where the electric charge is replaced by the magnetic charge. The value 
$\kappa=1$ is the only one that respects unitarity~\cite{LeeYang}. 
For spin-0 monopoles, the dual theory of which resembles the Scalar Quantum Electrodynamics (SQED), no magnetic moment is allowed. 

In this work, we  generalise the discussion to include an arbitrary value for the magnetic dipole moment for monopoles with spin. The detailed reasoning for this is given in the next section. Hence, we shall treat $\kappa$ as a \emph{new phenomenological parameter}.\footnote{In general one should also add electric dipole moment terms as well. In this work we shall ignore them, assuming them suppressed for brevity, although our analysis can be readily extended to include such terms.}
We note that, setting the issue of unitarity aside,  phenomenological models of charged $W$-bosons interacting with photons with $\kappa \ne 1$ 
have been considered in the past~\cite{Tupper}, where it was demonstrated that the behaviour of the total cross section for the $W$-boson pair production for the $\kappa=1$ case is quite distinct from the $\kappa \ne 1$ cases. In the current work, we shall dualise such models to use them as effective theories for the $\kappa \ne 1$ spin-1 monopole case, generalising the work of \cite{original}. As we shall see, one may allow for some formal 
 large-$\kappa$ limit where, despite the strong magnetic coupling, the associated monopole-pair production cross sections can be made finite. In fact one may give meaning ---under some circumstances to be specified below (specifically, the production of slow monopoles)--- to the perturbative tree-level Feynman-like graphs of the effective theory.
The relevant formalism for arbitrary $\kappa$ and various monopole spins will then be used as a guide for the construction of appropriate \MAD~\cite{MG} algorithms that can be used as tools for data analyses in monopole searches at colliders. We remark for completeness that, for fast relativistic monopoles (characterised by a relative velocity $\beta \simeq 1$) passing through materials, the (large) number of electron-positron pairs produced can be used as a signal for the presence of the monopole~\cite{stodol}. The perturbativity conditions discussed in the present article, which pertain to slowly moving 
monopoles (with $\beta \ll 1$), of relevance to MoEDAL-LHC searches~\cite{moedal-review}, do not apply  to such cases, the study of which requires non-perturbative treatments.

The structure of the article is the following: section~\ref{sectionAmtoKin} is a review of the formal procedure to construct perturbative cross sections from the scattering amplitudes relevant to monopole production. Particular attention is given to a discussion of a rather unresolved current issue, regarding point-like monopole production through scattering of SM particles, namely  the use of an effective magnetic charge coupling that depends on the relative velocity $\beta$ between the monopole and the centre-of-mass of the producing particles (quarks in the case of interest here, see fig.~\ref{fig:diagrams}).  We also motivate the introduction of the magnetic dipole moment for monopoles with spin. In section~\ref{sec:spin}, the differential and total cross sections for monopole-antimonopole pair production via PF and DY processes are derived for various monopole spins. Combined limits of  large magnetic-moment parameter $\kappa$ and small $\beta $ can lead to finite perturbatively valid results, providing some support for the effective formalism. The pertinent Feynman rules are implemented in a dedicated \MAD model, which is described in detail in section~\ref{sec:mad}. In section~\ref{sec:lhc}, the monopole phenomenology at the LHC is discussed, utilising the \MAD UFO models developed in this work, in the context of the above theoretical considerations. Conclusions and outlook are given in section~\ref{sec:concl}, while details on the calculations are provided in appendices~\ref{app:basic} and~\ref{app:xsection}. 

\section{From amplitudes to kinematic distributions}\label{sectionAmtoKin}

We are interested in the electromagnetic interactions of a monopole of spin $S=0,~\half,~1$ with ordinary photons. The corresponding theory is an effective 
$U(1)$ gauge theory which is obtained after appropriate \emph{dualisation} of the pertinent field theories describing the interactions of charged fields of spin-$S$ with photons. 
However, there is a subtle issue here, which we now proceed to discuss, and which will be relevant to our subsequent studies. It concerns a potential dependence of the effective magnetic charge on the relative velocity of the monopole pair and the centre-of-mass of the producing particles, as noted in \cite{sch,milton}. 

To this end, we need first to recall some basic facts of the theory of point-like monopole-matter scattering. Classical (tree-level) scattering of charged particles off massive magnetic monopoles has been studied extensively, using modern quantum field theory (even relativistic) treatments~\cite{shnir,milton,sch,scat}.  As the monopoles of interest are relatively heavy compared to electrons, with masses of order at least TeV, the use of non-relativistic scattering suffices for our purposes in this section. The differential cross section for the classical (non-relativistic) scattering of electrons, representing matter with electric charge $e$ and mass $m$, off a magnetic monopole, with magnetic charge $g$ and mass $M$, reads~\cite{shnir,milton}:
\begin{equation}\label{cross}
\frac{\dd \sigma}{\dd \Omega} = \left(\frac{eg}{c\mu v_0}\right)^2 \sum_{\chi} \frac{1}{4 \sin^4 \! \left(\frac{\chi}{2}\right)} 
\left| \frac{\sin\chi }{\sin\theta } \frac{\dd \chi}{\dd \theta}  \right|~,
\end{equation}
in a frame where the monopole is initially at rest and the electron has an initial velocity $v_0$. In the above formula, $c$ denotes the speed of light in vacuo and $\theta$ is the scattering angle, given in~\cite{milton}: $\cos (\frac{\theta}{2} ) = \cos(\frac{\chi}{2}) \left| \sin \left(\frac{\pi/2}{\cos(\chi/2)}\right)\right|$. 
$\mu = \frac{m\, M}{m + M} $ is the reduced mass of the two body problem at hand, with $\mu \simeq m$ in the cases of monopoles where $M \gg m$ which is of interest here. The angle $\chi = 2 \arccot{(\mu  v_0 b / |\kappa|)}$ defines the (Poincar\'e) cone on which the (classical) trajectory of the electron in the background of the monopole is confined, with $b$ the impact parameter. We note for completeness, that the cross section diverges in two occasions: 
\begin{enumerate}[label=(\roman*)]
\item $\sin \theta =0$ and $\sin \chi \ne 0$, $\theta = \pi$, which occurs for a certain discrete set of cone angles~\cite{sch,milton}, and 
\item when 
$\frac{\dd \theta}{\dd \chi}=0$, which occurs for a certain discrete set of scattering angles $\theta$~\cite{sch,milton}. 
\end{enumerate}
For small scattering angles, $\theta \ll 1$, the differential cross section \eqref{cross} can be approximated by: 
\begin{equation}\label{cross2}
\frac{\dd \sigma}{\dd \Omega} = \left(\frac{e g}{2\mu v_0 c}\right)^2  \frac{1}{(\theta/2)^4}~, \quad \theta \ll 1~.
\end{equation}
The reader should notice that this exhibits a scaling with the inverse square power of the velocity $v_0$, which is different from the standard Rutherford formula of the scattering of electrically-charged particles entailing a $v_0^{-4}$ scaling.

\subsection{Velocity-dependent magnetic charge}

As can be readily seen from~\eqref{cross2}, this expression reduces to the standard Rutherford formula for electron-electron scattering upon the replacement 
\begin{equation}\label{replacement}
  \frac{g}{c}\, \rightarrow \, \frac{e}{v_0} \, .
\end{equation}
This prompted some authors, including Milton, Schwinger and collaborators~\cite{milton,sch}, to conjecture, upon invoking electric-magnetic duality, 
that when discussing the interaction of monopole with matter (electrons or quarks), e.g.\ when discussing propagation of monopoles in matter media used for detection and capture of monopoles, or  considering monopole-antimonopole pair production through DY or PF processes, 
a \emph{monopole-velocity dependent magnetic charge} has to be considered in the corresponding cross section formulae:
\begin{equation}\label{gbeta}
g\,  \rightarrow \,  g \, \frac{v}{c} \equiv g\, \beta .
\end{equation}
We stress again that the above substitution is based on the assumption of electric-magnetic duality, which would lead to the equivalence of the electron-monopole scattering cross-section (\ref{cross2}) with the corresponding Rutherford formula upon applying (\ref{gbeta}). It is not known at present how to derive such an effective magnetic charge in the context of effective field theories (one might think of applying Schwinger-Dyson techniques which could resum the (large) magnetic charge couplings, but such theories are not available, and one cannot simply extend a strongly coupled QED model to the monopole case, upon replacing the electric charge with a magnetic one).  Hence the substitution (\ref{gbeta}) should only be viewed at present as a conjecture, motivated by electric-magnetic duality symmetry.

The replacement \eqref{gbeta} was then used to interpret the  experimental data in collider searches for magnetic monopoles~\cite{milton,kalb,original,dw,vento,atlasmono1,moedalplb}. Due to the lack of a concrete theory for magnetic sources, the results of the pertinent experimental searches can be interpreted in terms of {\it both} a $\beta$-independent and a $\beta$-dependent magnetic charges, and then one may compare the corresponding bounds, as done in the recent searches by the MoEDAL Collaboration~\cite{moedalplb}. The monopole velocity $\beta$ used in this work is given by the \emph{Lorentz invariant expression} in terms of the monopole mass $M$ and the Mandelstam variable $s$ \eqref{Mandelstam}, where $s$ representing the square of the centre-of-mass energy of the fusing incoming particles (photons or (anti)quarks) ($\sqrt{s}=2E_{\gamma/q}$):
\begin{equation}\label{btos}
 \beta = \sqrt{1-\frac{4M^2}{s}}~.
 \end{equation}

 The Lagrangian for each effective model describes the propagation and interactions of a massive monopole field and a photon field. The field theory is chosen according to the spin of the monopole, $S=0,~\half$ or~$1$. As a result of \eqref{gbeta}, the coupling of the monopole to the photon $g(\beta)$ is linearly dependant on the particle boost, $\beta= |\vec{p}| / E_p$, where $|\vec{p}|$ and $E_p$ are the monopole's three-momentum and energy, respectively. 
 
 To incorporate in a unified way both the $\beta$-dependent and $\beta$-independent magnetic couplings of the monopole to photons, we define 
 the magnetic fine structure constant as 
 $\alpha_{g}=g^{2}\beta^{2\delta} / (4\pi)$, where $\delta = 1 (0)$ for 
 $\beta$-(in)dependent couplings \eqref{gbeta}. The monopole Lagrangian of the effective theory can then be recognised as the Lagrangian in an electromagnetic field theory describing the interaction of a spin-$S$ field with photons, with the following substitutions:
\begin{align}
 e \, \mapsto \, g(\beta)~, \qquad & g^2(\beta) = g^2 \beta^{2\delta}, \, \quad \delta=0,1~, \nonumber \\
\alpha_{e}=\frac{e^{2}}{4\pi}  \, \mapsto \,  \,  &\alpha_{g}(\beta) \equiv \frac{g^{2}(\beta)}{4\pi} = \frac{g^{2} \beta^{2\delta}}{4\pi} \equiv \alpha_g^2 \, \beta^{2\delta} ~.
\label{etogb} 
\end{align}
From such a Lagrangian, Feynman rules can be extracted and observables are computed as in standard perturbative treatments. But, as stressed above, this is only a formal procedure, since, given the large value of the magnetic charge, as a consequence of the quantisation rule \eqref{diracrule2}, the interaction coupling is in the non-perturbative regime. Hence, truncated processes, like DY or PF, have no meaning, unless, as we shall discuss below, certain 
formal limits are considered in some special cases.  

An important remark is in order at this point. Since the `velocity' $\beta$ (\ref{btos}) is expressed in terms of Lorentz-invariant Mandelstam variables, the Lorentz invariance of the effective field theory action of the monopole is not affected by the introduction of the effective $\beta$-dependent magnetic charge (\ref{etogb}). However, there is a well known paradox, due to Weinberg~\cite{Weinberg}, who pointed out that the amplitude for a single photon exchange between and electric and a magnetic current (of relevance to DY monopole-antimonopole production processes)  is \emph{neither} Lorentz \emph{nor gauge} invariant due to the r\^ole played by the Dirac string, which contradicts the fact that monopoles appear as consistent soliton solutions in Lorentz and gauge invariant field theories~\cite{hpmono}. It is only recently~\cite{Terning}, that this paradox was arguably resolved, albeit within toy models of monopoles, coupled to photons, with perturbative electric and magnetic couplings. The resolution of Weinberg's paradox in such models is provided by a resummation of soft photons, which was possible due to the pertubatively small magnetic charges involved. Such a resummation resulted in the exponentiation of the Lorentz (and gauge) non-invariant terms pointed out in \cite{Weinberg} to a (Bohm-Aharonov type) phase factor in the respective amplitude. In this sense, the modulus of the amplitude (and hence the associated physical observables, such as cross sections) are Lorentz (and gauge) invariant, with the important result that, upon Dirac quantisation (\ref{diracrule}), the (resummed over soft photons) amplitude itself is Lorentz invariant. It is in such Lorentz (and gauge) invariant frameworks that, as we conjecture, the considerations of the effective, velocity-dependent magnetic charge (\ref{etogb}), may apply, which by the way also implies perturbative magnetic couplings for sufficiently small production velocities of the monopoles in the laboratory frame.

\subsection{The magnetic dipole moment as a novel free parameter for monopoles with spin}

As already mentioned in the introductory section~\ref{intro}, in previous effective-field theory treatments of spin-1 Dirac monopoles~\cite{original}, a magnetic dipole moment with the value $\kappa = 1$ has been introduced mimicking the unitary SM case of $W^{\pm}$ bosons (representing the monopoles), interacting with photons. Lacking an underlying microscopic model for point-like monopoles, the above restrictions in the value of the magnetic moment 
may not necessarily be applied to the monopole field. Indeed, for monopoles with spin, such a parameter may arise, e.g.\ by quantum corrections, in similar spirit to 
the electron case. The difference of course is that in the latter case it is the electric charge of the electron that would play a r\^ole, while in the monopole case it is the magnetic charge which, in view of its large value following the quantisation rule~\eqref{diracrule2}, cannot be treated perturbatively. Nonetheless, a non trivial (possibly large) magnetic moment might be induced in such a case, which might also be responsible for the restoration of unitarity of the effective theory. 
For example, one might hope that the apparent unitarity issues for generic $\kappa \ne 1$ values in the case of spin-1 monopoles can be remedied by embedding the corresponding theory in microscopic ultraviolet complete models beyond the SM, in much the same way as unitarity is restored in the case of the $W^\pm$ gauge bosons interacting with photons in the SM case.  

There is an additional reason as to why a non-trivial value for the parameter $\kappa$ for the monopoles might be feasible. 
Although the full microscopic r\^ole of $\kappa$ still needs to be determined, something which is obscured at present due to the lack of a microcsopic theory of point-like Dirac monopoles, nonetheless its presence is arguably consistent with the charge quantisation rule~\eqref{diracrule}. This is due to the fact that, as we shall discuss, a magnetic dipole moment does not contribute to the singular part of the magnetic field of the monopole, which is responsible for the charge quantisation~\cite{shnir}. This can be argued as follows: as it is well known~\cite{shnir}, Dirac's quantisation of charge in the presence of a monopole can be discussed by considering a particle of electric charge $q_e$ (e.g.\ electron) moving along a loop far away from the monopole centre, whose area is pierced by the Dirac string. Placing a magnetic dipole moment vector $\vec \mu_{\rm D}$ 
at the origin of a coordinate system, where the monopole is assumed at rest, will induce, according to standard (classical) electromagnetism, a magnetic field 
\begin{equation}\label{dipolemf}
\vec B_{\rm D} = \frac{\mu_0}{4\pi \, r^3} \, |\vec \mu_{\rm D} |\, \Big(2 \, \cos \theta \, \widehat{\mathbf{r}} + \sin \theta \, \widehat{\mathbf{\theta}} \Big), 
\end{equation}
where $\mu_0$ is the magnetic permittivity of the vacuum (`free space'), the symbol $\widehat{\ }$ denotes a unit vector, and $(r, \theta, \phi)$ are the usual spherical polar coordinates. The formula is valid for large distances compared to the dipole longitudinal dimension. Writing $\vec B_{\rm D} = \vec \nabla \times \vec A_{\rm D}$, we can determine the corresponding vector potential as $\vec A_{{\rm D}} = \frac{\mu_0}{4\pi} \, \frac{ \vec \mu_D \times \vec r}{r^3}$, at large distances $r$.  One can readily confirm that $\vec \nabla \cdot \vec{B}_{\rm D} = 0$. 

On the other hand, due to the monopole's magnetic charge, there is a magnetic field contribution, which however, due to the (singular) Dirac string, requires proper regularisation~\cite{shnir}. Upon doing so, one obtains for the regularised monopole magnetic field 
\begin{equation}\label{magmono}
\vec B^{{\rm reg}}_{\rm monopole} = \vec B_{\rm monopole} + \vec B_{\rm sing} = \frac{g}{r^2} \, \widehat r - 4\pi \, g \, \widehat n \, \theta(z) \, \delta(x) \delta (y) , 
\end{equation}
for a Dirac string along the $z$-axis, in which case the unit vector $\widehat n =(0, 0, 1)$ also lies along that axis. 
The regularised form 
of the monopole's magnetic field intensity yields the correct formula $\vec \nabla \cdot \vec B^{\rm reg}_{\rm monopole} = 4\pi \, g \, \delta^{(3)}(\mathbf{r})$, implying that the magnetic monopole is the source of a field. 

If one considers a charged particle looping the Dirac string far away from the position of the monopole, one would then observe that it is the singular part of the magnetic field~\eqref{magmono}, $\vec B_{\rm sing}$, which contributes to the phase of the electron wavefunction~\cite{shnir}, $q_e \oint_L \dd\vec x \cdot \vec A = \int_{\Sigma (L)} \dd\sigma \cdot \vec B_{\rm sing} = 4\pi q_e  g $. The magnetic dipole moment does \emph{not} contribute to the singular part of the magnetic field, and thus the charge quantisation~\eqref{diracrule} is not affected. We note that, as a result of the $r^3$ suppression, the contributions of~\eqref{dipolemf} would be subdominant, at large distances $r$ from the monopole centre, as compared to  those  of~\eqref{magmono}.

It is worth remarking at this stage that one can also view~\cite{Thober} the quantisation rule~\eqref{diracrule} itself as a consequence of representing the (non-physical) singular string solenoid, assumed in the original Dirac's construction~\cite{dirac}, as a collection of (small) \emph{fictitious} current loops (with an area perpendicular to the solenoid's axis). Each one of these loops will induce a magnetic moment $I A$, with $I$ the current and $A$ the area of the loop (assumed vanishing in this case). Assuming a uniform magnetic moment per unit length ${\mathcal M}$ for the Dirac string, then, and taking into account that the solenoid may be viewed as the limiting case of a magnetic dipole of infinite length, one may apply the aforementioned formula~\eqref{dipolemf} in this case to derive the singular magnetic field of the monopole itself, in which case the magnetic charge 
is obtained as $g \propto {\mathcal M}$~\cite{Thober}. However, we stress, that, the contribution of the induced magnetic moment of the quantum effective theory of a monopole with spin, whose strength depends on the parameter $\kappa$, which we shall discuss in this work, is independent of that due to the magnetic charge $g$, as explained above. Lacking though an underlying fundamental theory for the point-like monopole the determination of $\kappa$ is at present not possible.  

Before closing this section, we would like to present an equivalent, yet less elaborate, way\footnote{We thank V.~Vento for a discussion on this point.} to see the irrelevance of the magnetic dipole moment for the quantisation rule~\eqref{diracrule}, which avoids the use of Dirac strings. To this end, one covers the three-space surrounding the monopole by two hemispheres, with appropriate gauge potentials defined in each of them, whose curl yields the corresponding magnetic field strengths. 
For the magnetic monopole gauge potential one has the expressions~\cite{shnir}: 
\begin{equation}
\vec A_S = g\, (1 -  \cos\theta)\, \vec \nabla \phi = \frac{g\, (1 -  \cos\theta)}{r\, \sin\theta}\, \widehat \phi, \qquad \theta \in [0, \frac{\pi}{2} + \delta ), \quad \delta \to 0^+,
\end{equation}
for the south hemisphere, which is singular at the south pole $\theta = \pi$, 
and 
\begin{equation}
\vec A_N =   -g\, (1 +  \cos\theta)\, \vec \nabla \phi   = \frac{-g\, (1 +  \cos\theta)}{r\, \sin\theta} \, \widehat  \phi, \qquad \theta \in (\frac{\pi}{2}-\delta, \pi ], \quad \delta \to 0^+, 
\end{equation}
for the north hemisphere, which is singular at the north pole $\theta=0$. These two patches overlap $\frac{\pi}{2}-\delta \, < \, \theta \, < \frac{\pi}{2} + \delta, \, \delta \to 0^+,$ and, as is well known, the 
difference of 
\begin{equation}\label{pdiff}
\vec A_S - \vec A_N = \vec \nabla f = \frac{2g}{r\, \sin\theta} \widehat \phi, 
\end{equation}
yields a singular gauge transformation at $\theta =0, \pi$, which contributes to the phase  $q_e \oint_L \dd\vec x \cdot \vec A$ of the charged particle wavefunction,  the requirement of single-valuedness of which yields the rule~\eqref{diracrule}. 

On the other hand, as already mentioned, the vector potential corresponding to the magnetic moment, for large distances $r$ from the centre of the sphere where the monopole is located, is of the form 
\begin{equation}
\vec A_{\rm D} = \frac{\mu_0}{4\pi} \, \frac{\vec \mu_{\rm D} \times \vec r}{r^3} = \frac{\mu_0}{4\pi} \, \frac{|\vec \mu_{\rm D}| \sin\theta}{r^2} \, \widehat \eta, 
\end{equation}
with $\widehat \eta$ the unit vector perpendicular to the plane of $\vec r $ and $\vec \mu_{\rm D}$ (assumed parallel to the $z$-axis); this is not singular at the poles $\theta=0,\pi$ (in fact it vanishes there). The total potential in each hemisphere is then given by the corresponding sum $\vec A_i + \vec A_{\rm D}, i=S,N$. Hence, the magnetic moment does not contribute to the difference, and thus it does not affect the wavefunction phase, which is associated only with the monopole part~\eqref{pdiff}. 

\section{Cross sections for spin-$S$ monopole production \label{sec:spin}}

In this section we derive the pertinent Feynman rules and then proceed to give  expressions for the associated differential and total production cross sections for monopole fields of various spins. The pertinent expressions are evaluated using the package \Feyncalc~\cite{FeynCalc} in \Math. We commence 
the discussion with the well-studied cases of scalar (spin-0) and fermion (spin-\half) monopole cases, but extend the fermion-monopole case to include an arbitrary magnetic moment term  $\kappa \ne 0$. Then we proceed to discuss the less studied case of a spin-$1$ monopole including an arbitrary magnetic moment term $\kappa$. We consider both $\beta$-dependent and $\beta$-independent magnetic couplings. In an attempt to make some sense of the perturbative estimates, we discuss, where appropriate, various formal limits of weak $\beta$ and large $\kappa$ for which the pertinent cross sections remain finite. In each spin case we present both PF and DY cross sections, which will help us present a comparison at the end of the section.

\subsection{Scalar monopole}

The first model studied in this work, is the one for massive spin-0 monopole interacting with a massless $U(1)$ gauge field representing the photon. Searches at the LHC~\cite{atlasmono1,atlasmono2,moedal, moedalplb} and other colliders have set upper cross-section limits is this scenario assuming Drell-Yan production. The Lagrangian describing the electromagnetic interactions of the monopole is given simply by a dualisation of the SQED Lagrangian
\begin{equation}\label{spinzero}
\mathcal{L}=-\frac{1}{4}F^{\mu\nu}F_{\mu\nu}+(D^{\mu}\phi)^{\dagger}(D_{\mu}\phi)-M^2\phi^{\dagger}\phi ,
\end{equation}
where $D_{\mu}=\partial_{\mu}-ig(\beta)\mathcal{A}_{\mu}$, $\mathcal{A}_{\mu}$ is the photon field,  whose field strength (Maxwell) tensor is $F_{\mu\nu}=\partial_{\mu}\mathcal{A}_{\nu}-\partial_{\nu}\mathcal{A}_{\mu}$ and $\phi$ is the scalar monopole field. There are two interaction vertices associated with theory. The three- and four-point vertices are illustrated in fig.~\ref{FeynmanRulesSpinZero} in appendix~\ref{app:xsection}, where the cross-section calculations are detailed. These interactions are the only couplings generated between the spin-0 monopole and the $U(1)$ gauge field. 

\subsubsection{Pair production of spin-0 monopoles via photon fusion}\label{SpinZeroSection}

There are three possible graphs contributing to scalar monopole production by PF, a $t$-channel, $u$-channel and seagull graph shown in fig.~\ref{Spinzerographs}. Their respective matrix amplitudes are given by eqs.~\eqref{MatAmplSpinZero} in appendix~\ref{app:xsection}. $M$ is the spin-0 boson mass, $\varepsilon_{\lambda}(q_1)$ and $\varepsilon_{\lambda^{'}}(q_2)$ are the photon polarisations, $p_{1}$ and $p_{2}$ are the monopoles four-momenta such that $p_{i_{\mu}}^{2}=M^{2}$, and $q_{1}$ and $q_{2}$ are the photons four-momenta such that $q_{i_{\mu}}^{2}=0$, as defined in fig.~\ref{Spinzerographs}.

\begin{figure}[ht!]
\justify
\begin{subfigure}[b]{0.5\textwidth}
\centering
\begin{tikzpicture}
  \begin{feynman}
  	\vertex (a1) {\(\mathcal{A}_{\rho}, \epsilon_{\lambda'}\)};
	\vertex [right=2.25cm of a1] (a2);
	\vertex [right=2cm of a2] (a3){\(\overline{M}\)};
	\vertex [above=2cm of a2] (a5);
	\vertex [right=2cm of a5] (a6){\(M\)};
	\vertex [left=2cm of a5] (a4){\(\mathcal{A}_{\sigma}, \epsilon_{\lambda}\)};
	\diagram* {
	(a3) -- [scalar, edge label=\(p_{2_{\mu}}\)] (a2) -- [scalar, edge label=\(k_{\pi}\)] (a5) -- [scalar, edge label=\(p_{1_{\nu}}\)] (a6),
	(a4) -- [boson, edge label=\(q_{1_{\sigma}}\)] (a5),
	(a1) -- [boson, edge label=\(q_{2_{\rho}}\)] (a2)
         };
\end{feynman}
\end{tikzpicture}
\caption{$t$-channel.\label{fig:tchan}}
\end{subfigure}
~
\begin{subfigure}[b]{0.5\textwidth}
\centering
\begin{tikzpicture}
  \begin{feynman}
  	\vertex (a1) {\(\mathcal{A}_{\rho} ,q_{2_{\rho}},  \epsilon_{\lambda'}\)};
	\vertex [right=2.5cm of a1] (a2);
	\vertex [right=2cm of a2] (a3){\(\overline{M}\)};
	\vertex [above=2cm of a2] (a5);
	\vertex [right=2cm of a5] (a6){\(M\)};
	\vertex [left=2cm of a5] (a4){\(\mathcal{A}_{\sigma} ,q_{1_{\sigma}},\epsilon_{\lambda}\)};
	\diagram* {
	(a3) -- [scalar, edge label=\(p_{2_{\mu}}\)] (a2) -- [scalar, edge label=\(\widetilde{k_{\pi}}\)] (a5) -- [scalar, edge label=\(p_{1_{\nu}}\)] (a6),
	(a4) -- [boson] (a2),
	(a1) -- [boson] (a5)
      }; 
\end{feynman}
\end{tikzpicture}
\caption{$u$-channel.\label{fig:uchan}}
\end{subfigure}
~
\begin{subfigure}[b]{1\textwidth}
\centering
\begin{tikzpicture}
\begin{feynman}
  	\vertex (a2);
	\vertex[left=1.75cm of a2] (a3);
	\vertex[below=1cm of a3] (a4) {\(\mathcal{A}_{\rho} , q_{2_{\rho}}, \epsilon_{\lambda'}\)};
	\vertex[right=1.75cm of a2] (a5);
	\vertex[below=1cm of a5] (a6) {\(\overline{M}\)};
	\vertex[above=1cm of a3] (a8) {\(\mathcal{A}_{\sigma} ,q_{1_{\sigma}},  \epsilon_{\lambda}\)};
	\vertex[above=1cm of a5] (a7) {\(M\)};
	\diagram* {
       		(a6) -- [scalar, edge label=\(p_{2_{\mu}}\)] (a2) -- [scalar, edge label=\(p_{1_{\nu}}\)] (a7),
		(a8) -- [boson] (a2) -- [boson] (a4)
      };
\end{feynman}
\end{tikzpicture}
\caption{Four-vertex diagram.\label{fig:seagull}}
\end{subfigure}
\caption{Feynman-like graphs for: \subref{fig:tchan} $t$-channel;  \subref{fig:uchan} $u$-channel; and \subref{fig:seagull} seagull processes encompass all the contributions to the matrix amplitude of scalar particle production by PF. The variable definitions are given in the text.}
\label{Spinzerographs}
\end{figure}
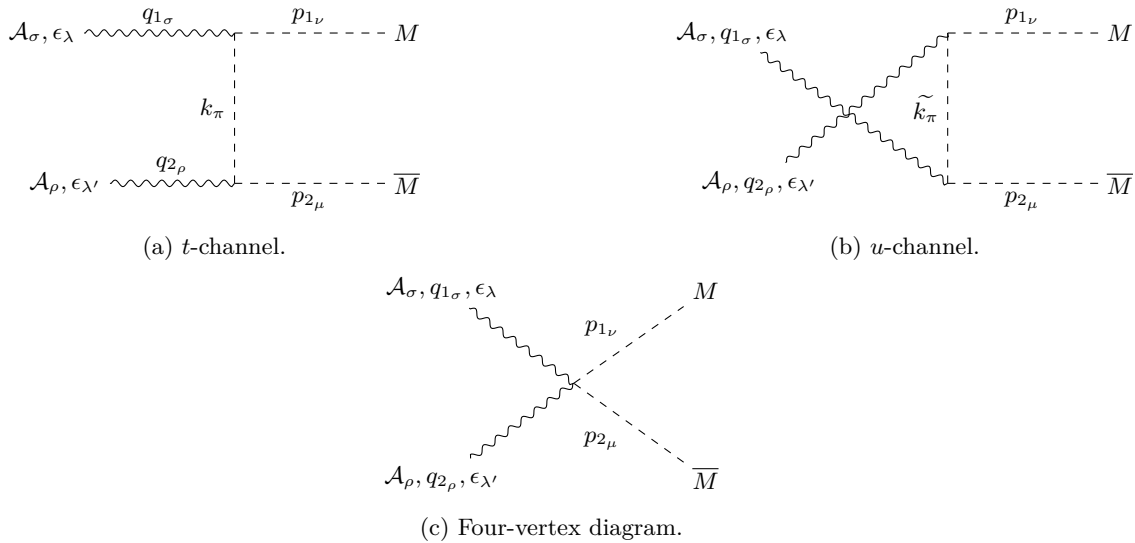

\begin{figure}[ht!]\centering
\includegraphics[width=0.95\textwidth]{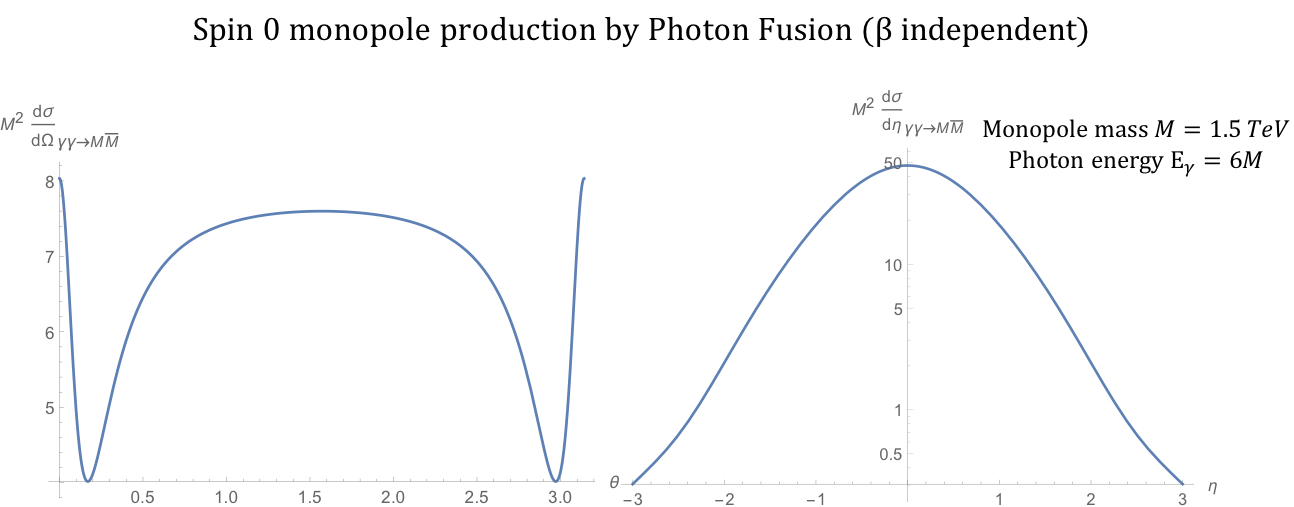}
\caption{Differential cross section distributions for the production of spin-0 monopoles of mass $M=1.5$~TeV via PF at $\sqrt{\sgg}=2E_{\gamma}$, where $E_{\gamma}=6M$, as a function of the scattering angle $\theta$ (left) and the pseudorapidity $\eta$ (right). }\label{PFdsigmaSpin0}
\end{figure}

After some calculation, detailed in appendix~\ref{app:xsection}, the differential cross section for the spin-0 monopole-antimonopole production is reduced to:
\begin{equation}\label{dsigmadomegazero}
	\frac{\dd\sigma_{\gamma\gamma\rightarrow M\overline{M}}^{S=0}}{\dd\Omega} = \frac{\alpha_{g}^{2}(\beta)\beta}{2\sgg} \left[ 1+\left(1-\frac{2(1-\beta^{2})}{ 1-\beta^{2}\cos^{2}\theta}\right)^{2}\right],
\end{equation}
which in terms of the pseudorapidity $\eta$, defined in appendix~\ref{app:basic}, becomes:
\begin{align}\label{dsigmadomegazerorap}
	\frac{\dd\sigma_{\gamma\gamma\rightarrow M\overline{M}}^{S=0}}{\dd\eta}  = \frac{\pi\alpha_{g}^{2}(\beta)\beta}{\sgg \cosh^{2}\eta} \left[ 1+\left(1-\frac{2(1-\beta^{2})}{1-\beta^{2}\tanh^{2}\eta}\right)^{2} \right].
\end{align}
The differential distributions as a function of the scattering angle and the pseudorapidity are shown in fig.~\ref{PFdsigmaSpin0} for a hypothetical monopole of mass $M=1.5$ TeV at $\sqrt{\sgg}=2E_{\gamma}$, where $E_{\gamma}=6M$. The pair production is mostly central as is the case for various beyond-SM scenarios. Such distributions are used for the validation of the simulation package discussed in section~\ref{sec:mad}.

\begin{figure}[ht!]\centering
\includegraphics[width=0.55\textwidth]{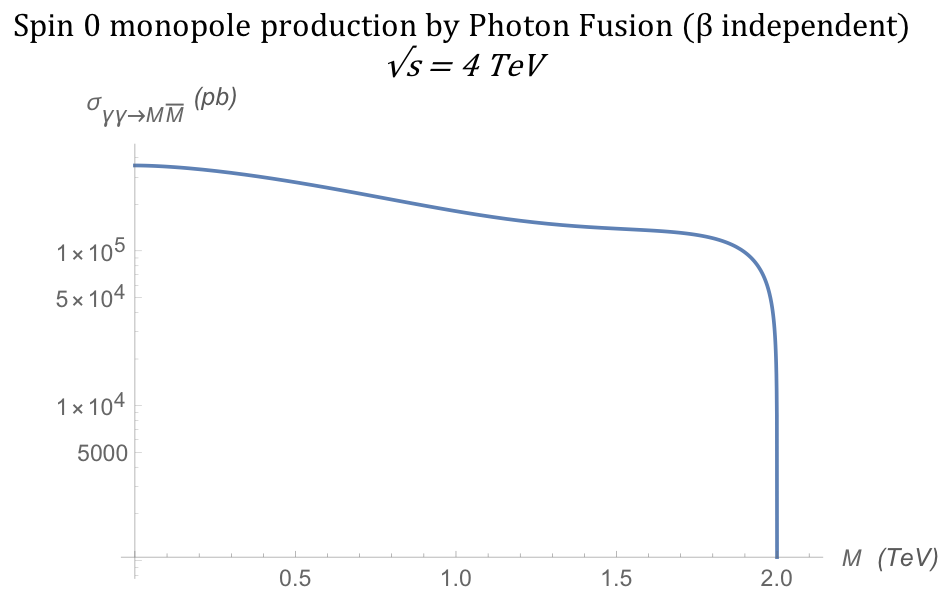}
\caption{Total cross section for the production of spin-0 monopoles via the PF process as a function of the monopole mass $M$ at $\sqrt{\sgg}=4$ TeV.}\label{PFsigmaSpin0}
\end{figure}

After integrating over the solid angle, as discussed in section \ref{sectionAmtoKin}, the total cross section becomes~\cite{original} 
\begin{equation} \label{tcrosssection}
	\sigma^{S=0}_{\gamma\gamma\rightarrow M\overline{M}} =\frac{4\pi\alpha_{g}^{2}(\beta)\beta}{\sgg}\left[2-\beta^{2}-\frac{1}{2\beta}(1-\beta^{4})\ln{\left(\frac{1+\beta}{1-\beta}\right)}\right],
\end{equation}
and is shown in fig.~\ref{PFsigmaSpin0} for the selected energy of $\sqrt{\sgg}=4~{\rm TeV}$. The cross section drops rapidly with increasing mass and disappears sharply at the kinematically forbidden limit of $M > \sqrt{\sgg}/2$.  Again this result is compared against the \MAD implementation prediction in section~\ref{sec:mad}.

\subsubsection{Pair production of spin-0 monopoles via Drell-Yan}

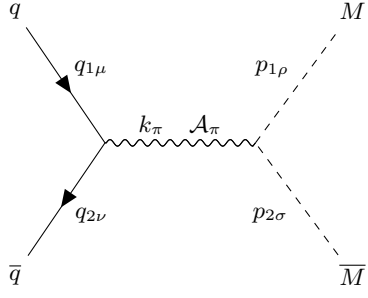
\begin{figure}[ht]
\begin{tikzpicture}
  \begin{feynman}
  	\vertex (a1);
	\vertex[right=2cm of a1] (a2);
	\vertex[above=1.75cm of a2] (a3);
	\vertex[right=1cm of a3] (a4) {\(M\)};
	\vertex[below=1.75cm of a2] (a5);
	\vertex[right=1cm of a5] (a6) {\(\overline{M}\)};
	
	\vertex[above=1.75cm of a1] (a7);
	\vertex[left=1cm of a7] (a8) {\(q\)};
	\vertex[below=1.75cm of a1] (a9);
	\vertex[left=1cm of a9] (a10) {\(\overline{q}\)};
	
	\diagram* {
	       	(a8) -- [fermion, edge label=\(q_{1\mu}\)] (a1) -- [fermion, edge label=\(q_{2\nu}\)] (a10),
		(a1) -- [boson, edge label=\(k_{\pi} \quad \mathcal{A}_{\pi}\)] (a2),
       		(a6) -- [scalar, edge label=\(p_{2\sigma}\)] (a2) -- [scalar, edge label=\(p_{1\rho}\)] (a4),
		(a1) -- [boson] (a2),
      };
\end{feynman}
\end{tikzpicture}
\caption{Feynman-like diagram representing the DY process in a scalar monopole theory. The variable definitions are given in the text. }\label{FeynScalar}
\end{figure}

The Feynman-like diagram in fig.~\ref{FeynScalar} shows the DY process in the case of a scalar monopole. The quarks $q\overline{q}$ annihilate to a photon $\mathcal{A}_{\pi}$, which decays to a $M\overline{M}$ pair in the $s$-channel. The quark lines are supplemented by momentum 4-vectors $q_{1\mu}$ and $q_{2\mu}$, where $q_{1,2}^2=m^2$ and the scalar monopole lines have momentum 4-vectors $p_{1\mu}$ and $p_{2\mu}$, where $p_{1,2}^2=M^2$ on shell. The centre-of-mass energy of the colliding quarks is $k_{\pi}k^{\pi}=\sqq$. The three-point vertex in this model is illustrated again in fig.~\ref{vetexScalara}(a), and the vertex for the $q\overline{q}\mathcal{A}_{\mu}$ coupling in fig.~\ref{vetexScalara}(b). 

\begin{figure}[ht]\centering
\includegraphics[width=0.95\textwidth]{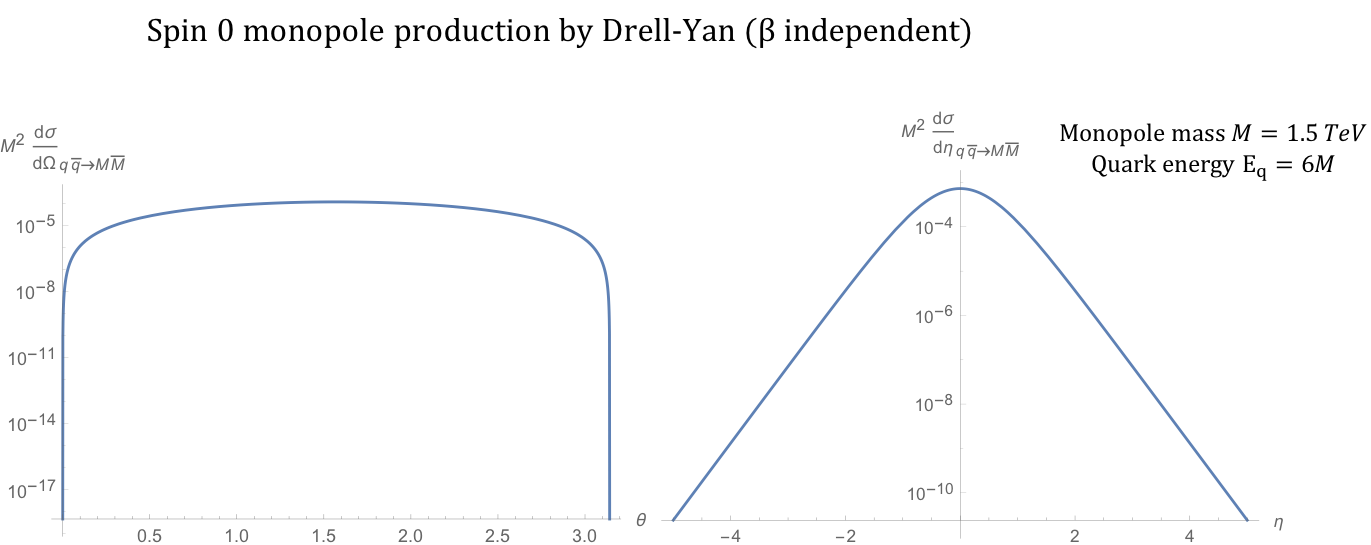}\\
\caption{Differential cross section distributions for the production of spin-0 monopoles of mass $M=1.5$~TeV via DY from massless quarks with $\beta_q=1$ at $\sqrt{\sqq}=2E_q$, where $E_q=6M$, as a function of the scattering angle $\theta$ (left) and the pseudorapidity $\eta$ (right). }\label{DYdsigmadetaSpin0}
\end{figure}

 After some calculations detailed in appendix~\ref{app:xsection} and in particular in eqs.~\eqref{DrellYanMatScalar}-\eqref{Xsecdeff0}, the differential cross section for DY scalar monopole production yields
 \begin{align}\label{xsecdeff}
	\frac{\dd\sigma_{q\overline{q}\rightarrow M\overline{M}}^{S=0}}{\dd\Omega} = \frac{5\alpha_{g}(\beta)\alpha_{e}}{72 \sqq}\beta^3(1-\cos^2\theta),
\end{align}
where $\beta$ is defined in eq.~\eqref{btos}.
The distribution is shown in fig.~\ref{DYdsigmadetaSpin0} for a monopole with a mass $M=1.5$~TeV at a centre-of-mass energy $\sqrt{\sqq}=2E_{q}$ where $E_{q}=6M$.

In terms of the pseudorapidity $\eta$, the differential cross section reads:
\begin{equation}\label{dsigdrap}
	\begin{split}
	\frac{\dd\sigma_{q\overline{q}\rightarrow M\overline{M}}^{S=0}}{\dd\eta} = \frac{5\pi\alpha_{g}(\beta)\alpha_{e}}{36 \sqq \cosh^2\eta}\beta^3(1-\tanh^2\eta),
	\end{split}
\end{equation}
and is also plotted in fig.~\ref{DYdsigmadetaSpin0}. As expected for the production of scalar particles, the distribution is almost flat with respect to the scattering angle $\theta$. When compared to the corresponding kinematic distributions from the PF case in fig.~\ref{PFdsigmaSpin0}, the DY case exhibits a more central distribution.

\begin{figure}[ht]\centering
\includegraphics[width=0.55\textwidth]{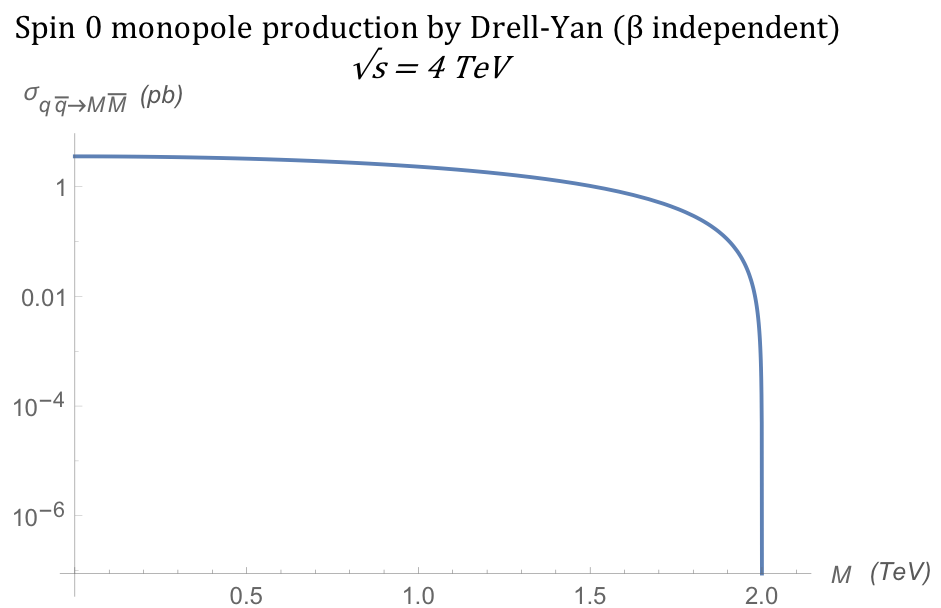}
\caption{Total cross section for the pair production of spin-0 monopoles via the DY process, in which $\beta_q=1$, as a function of the monopole mass $M$ at $\sqrt{\sqq}=4$ TeV.}\label{DYsigmaSpin0}
\end{figure}

Finally, the total cross section is evaluated by integrating eq.~\eqref{xsecdeff} over the solid angle $\dd \Omega=\dd\phi \, \dd\!\cos\theta$:
 \begin{equation}\label{xsecspin0}
 \sigma_{q\overline{q}\rightarrow M\overline{M}}^{S=0} = \frac{5\pi \alpha_{g}(\beta)\, \alpha_{e}}{27 \sqq}\beta^3,
\end{equation}
and is shown graphically at a centre-of-mass energy $\sqrt{\sqq}=4$ TeV in fig.~\ref{DYsigmaSpin0}.
A note of caution here: in an experimental setup for high energy collisions, the result~\eqref{xsecspin0} is valid for opposite-sign hadrons, e.g.\ $p\bar{p}$ at the Tevatron; it will be doubled in the case of same-hadron colliding bunches, such as in proton-proton collisions at the LHC. A comparison of the DY versus the PF cases, as far as the differential/total cross sections is concerned, is reserved for section~\ref{sec:comparison}. 

\subsection{Spin-\half point-like monopole with arbitrary magnetic moment term}

The phenomenology of a monopole with spin~\half, examined in this section, is the most thoroughly studied case~\cite{kalb,original,dw,vento}, however so far only the Drell-Yan production process has been explored in collider searches~\cite{atlasmono1,atlasmono2,moedal,moedalplb}. This type of monopole resembles a magnetic dual to the electron. The electromagnetic interactions of the monopole with photons, are described by model $U(1)$ gauge theory for a spinor field $\psi$ representing the monopole interacting with the massless $U(1)$ gauge field $A^{\mu}$, representing the photon.  In the cases discussed in the existing literature,
the effective Lagrangian describing the interactions of the spinor monopole with photons is taken from standard QED upon imposing electric-magnetic duality, in which there is no bare magnetic-moment term (at tree level). However for our analysis here, we shall insert in the Lagrangian a magnetic moment generating term, 
\begin{equation}\label{magmomferm}
{\mathcal L}_{\text{mag.~moment}} = - i\frac{1}{4}\, g(\beta)\kappa F_{\mu\nu}\overline{\psi}[\gamma^{\mu},\gamma^{\nu}]\psi,
\end{equation} 
to keep the treatment general. The origin of the magnetic moment of the monopole is not known, so it would be na\"ive to assume it is generated only through anomalous quantum-level spin interactions as for the electron in QED.\footnote{For instance, it is known that such terms have a geometrical (gravitational) origin in 4-dimensional effective field theories obtained from (Kaluza-Klein) compactification of higher-dimensional theories, such as brane/string universes~\cite{ichinose}. Moreover, as mentioned in the introduction, one could also include in the Lagrangian a CP-violating electric dipole moment (EDM) term for the spinor monopole, parametrised by a parameter $\eta$, 
${\mathcal L}_{\rm EDM} =   \frac{1}{4}\, g(\beta)\eta F_{\mu\nu}\overline{\psi}\, \gamma^5\, [\gamma^{\mu},\gamma^{\nu}]\psi $.  In this work, we assume such EDM terms suppressed, compared to the magnetic-moment-$\kappa$ terms, which can be arranged by assuming appropriate limits of parameters in the underlying microscopic theory, e.g.~\cite{ichinose}. Nonetheless, our analysis can be extended appropriately to include both $\kappa$ and $\eta$ parameters.} In the event $\kappa=0$, the Dirac Lagrangian is recovered. Thus, the effective Lagrangian for the spinor-monopole-photon interactions takes the form
\begin{equation}\label{12lag}
\mathcal{L}=-\frac{1}{4}F_{\mu\nu}F^{\mu\nu}+\overline{\psi}(i\slashed{D}-m)\psi-i\frac{1}{4}\, g(\beta)\, \kappa \, F_{\mu\nu}\overline{\psi}[\gamma^{\mu},\gamma^{\nu}]\psi,
\end{equation}
where $F_{\mu\nu}$ is the electromagnetic field strength tensor, the total derivative is $\slashed{D}=\gamma^{\mu}(\partial_{\mu}-ig(\beta)\mathcal{A}_{\mu})$ and $[\gamma^{\mu},\gamma^{\nu}]$ is a commutator of $\gamma$ matrices. 
The magnetic coupling $g(\beta)$ is  given in \eqref{etogb}, and is at most (depending on the case considered) 
linearly dependent on the monopole boost, $\beta=|\vec{p}| / E_p$, where $|\vec{p}|$ and $E_p$ are the monopole momentum and energy, respectively. 
The effect of the magnetic-moment term is observable through its influence on the magnetic moment at tree level which is
\begin{equation}
\mu_{M}=\frac{g(\beta)}{2M}2(1+2\tilde{\kappa})\hat{S}, \quad \hat{S}=\frac{1}{2},
\end{equation}
where $M$ is the spinor-monopole mass, $\hat{S}$ is the spin expectation value and the corresponding ``gyromagnetic ratio'' $g_R = 2( 1 + 2\tilde{\kappa})$. The dimensionless constant $\tilde{\kappa}$ is defined such that
\begin{equation}\label{ktilde}
\kappa=\frac{\tilde{\kappa}}{M}. 
\end{equation}
Noticeably, the parameter $\kappa$ in \eqref{12lag}  is not dimensionless, but has units of inverse mass which breaks the renormalisability of the theory. This may not be a serious obstacle for considering the case $\kappa\neq0$, if the pertinent model is considered in the context of an effective field theory embedded in some yet unknown microscopic theory, in which the renormalisation can be recovered.~\footnote{In case one adds an EDM $\eta$-term for the spinor monopole, the corresponding dimensionless parameter $\tilde \eta$ can also be defined in analogy with (\ref{ktilde}), i.e. $\eta=\frac{\tilde \eta}{M}$.}

\subsubsection{Spinor pair production via photon fusion}

The monopole couples to the photon at a three-particle vertex through a Feynman rule that is shown in fig.~\ref{FeynRuleS12} in appendix~\ref{app:xsection} together with details of the amplitudes that lead to the cross-section calculation. The parameter $\kappa$ influences the amplitude in the second term of \eqref{FeynRuleS12eq}, ensuring the $\kappa$-dependence of the observables. PF occurs through $t$-channel and $u$-channel processes as shown in fig.~\ref{spinhalfgraphs}, with the $\kappa$-dependent matrix amplitudes stated in eqs.~\eqref{matrixel} of appendix~\ref{app:xsection}. The photon momenta are $q_1$, $q_2$, the monopole momenta are $p_1$, $p_2$, whilst $k$, $\tilde{k}$ are the $t$- and $u$-channel exchange momenta, respectively.

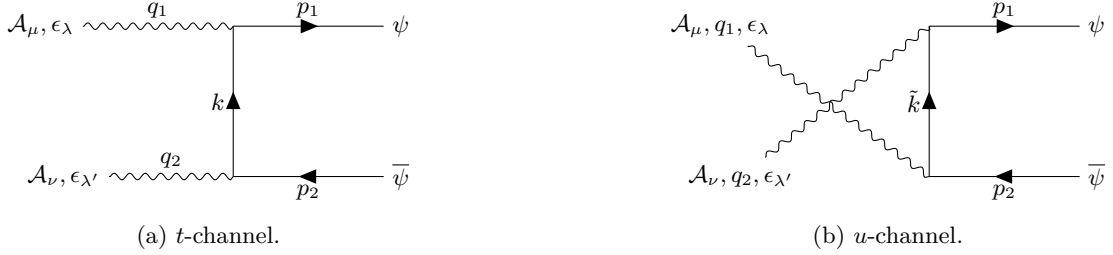
\begin{figure}[ht!]
\justify
\begin{subfigure}[b]{0.5\textwidth}
\centering
\begin{tikzpicture}
  \begin{feynman}
  	\vertex (a1) {\(\mathcal{A}_{\nu}, \epsilon_{\lambda'}\)};
	\vertex [right=2.25cm of a1] (a2);
	\vertex [right=2cm of a2] (a3){\(\overline{\psi}\)};
	\vertex [above=2cm of a2] (a5);
	\vertex [right=2cm of a5] (a6){\(\psi\)};
	\vertex [left=2cm of a5] (a4){\(\mathcal{A}_{\mu}, \epsilon_{\lambda}\)};
	\diagram* {
	(a3) -- [fermion, edge label=\(p_{2}\)] (a2) -- [fermion, edge label=\(k\)] (a5) -- [fermion, edge label=\(p_{1}\)] (a6),
	(a4) -- [boson, edge label=\(q_{1}\)] (a5),
	(a1) -- [boson, edge label=\(q_{2}\)] (a2)
         };
\end{feynman}
\end{tikzpicture}
\caption{$t$-channel.\label{fig:tchan-pf-half}}
\end{subfigure}
\begin{subfigure}[b]{0.5\textwidth}
\centering
\begin{tikzpicture}
  \begin{feynman}
  	\vertex (a1) {\(\mathcal{A}_{\nu},  q_{2},  \epsilon_{\lambda'}\)};
	\vertex [right=2.5cm of a1] (a2);
	\vertex [right=2cm of a2] (a3){\(\overline{\psi}\)};
	\vertex [above=2cm of a2] (a5);
	\vertex [right=2cm of a5] (a6){\(\psi\)};
	\vertex [left=2cm of a5] (a4){\(\mathcal{A}_{\mu} ,  q_{1},  \epsilon_{\lambda}\)};
	\diagram* {
	(a3) -- [fermion, edge label=\(p_{2}\)] (a2) -- [fermion, edge label=\(\tilde{k}\)] (a5) -- [fermion, edge label=\(p_{1}\)] (a6),
	(a4) -- [boson] (a2),
	(a1) -- [boson] (a5)
      }; 
\end{feynman}
\end{tikzpicture}
\caption{$u$-channel.\label{fig:uchan-pf-half}}
\end{subfigure}
\caption{Feynman-like graphs for the $t$-channel~\subref{fig:tchan-pf-half} and $u$-channel~\subref{fig:uchan-pf-half} show the contributions to the matrix amplitude for pair production of spin-\half monopoles by PF. The variable definitions are given in the text. }
\label{spinhalfgraphs}
\end{figure}

\begin{figure}[ht!]\centering
\includegraphics[width=0.95\textwidth]{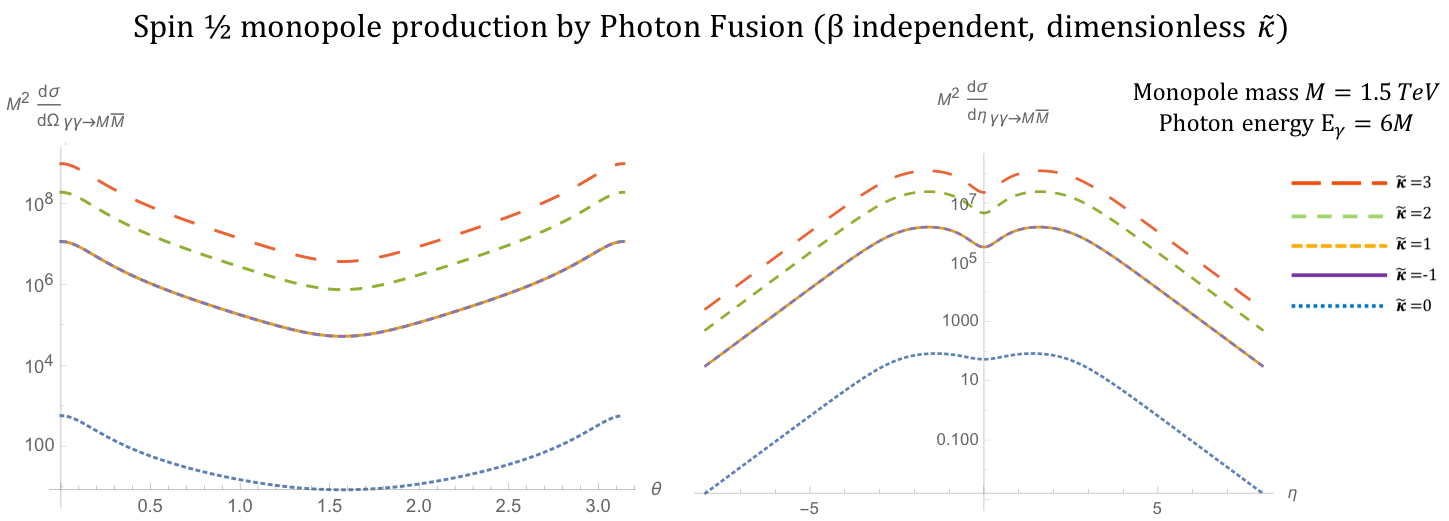}
\caption{Angular distributions for monopole-antimonopole pair production via PF for the case of a fermionic monopole with spin~\half and mass $M=1.5$~TeV as a function of the scattering angle $\theta$ (left) and the pseudorapidity $\eta$ (right) at $\sqrt{\sgg}=2E_{\gamma}$, where $E_{\gamma}=6M$, and for various values of the parameter $\tilde{\kappa}$.  The case $\tilde{\kappa}=0$ is analogous to the SM expectation and is clearly distinctive as the only unitary and renormalisable case.}\label{LeeYangDsigDstuff12label}
\end{figure}

The differential cross section distributions are computed as described in section~\ref{sectionAmtoKin}, with more detail given in appendix~\ref{app:xsection}. The $\kappa$-dependent forms are given below:
\begin{equation}\label{dxsec12}
\begin{split}
 \frac{\dd\sigma^{S=\frac{1}{2}}_{\gamma\gamma\rightarrow M\overline{M}}}{\dd\Omega} &=\frac{  \alpha_g^2(\beta) \beta}{4 \sgg (1-\beta^2 \cos ^2\theta)^2}(-\beta^6 \kappa ^4 \sgg^2 \cos ^6\theta-2 \beta^4 (\kappa ^4 \sgg^2+4)\\
 &+\beta^2 (48 \kappa  \sqrt{\sgg-\beta^2 \sgg}+2 \kappa ^4 \sgg^2+32 \kappa ^2 \sgg+8)-\beta^4 \cos ^4\theta  ((2 \beta^2+3) \kappa ^4 \sgg^2\\
 &+8 \kappa ^2 \sgg+4)+\beta^2 \cos ^2\theta (2 \beta^4 \kappa ^4 \sgg^2+8 \beta^2 (5 \kappa ^2 \sgg+1)-48 \kappa  \sqrt{\sgg-\beta^2 \sgg}\\
 &+3 \kappa ^4 \sgg^2-60 \kappa ^2 \sgg-8)+(\kappa ^2 \sgg-2)^2),
\end{split}
\end{equation}
with the standard (dual QED) case $\kappa=0$ being given in refs.~\cite{original,dw}:
\begin{equation}\label{standardqed12}
 \frac{\dd\sigma^{S=\frac{1}{2}}_{\gamma\gamma\rightarrow M\overline{M}}}{\dd\Omega} =\frac{  \alpha_g^2(\beta) \beta}{4 \sgg (1-\beta^2 \cos ^2 \theta)^2}(-8 \beta^4+8\beta^2 -4\beta^4 \cos ^4\theta +8 \beta^4 \cos ^2\theta  -8\beta^2 \cos ^2\theta  +4), \qquad  \kappa=0.
\end{equation}
The angular distributions for monopole-antimonopole pair production by PF are shown in fig.~\ref{LeeYangDsigDstuff12label} for spin~\half and for various values of the parameter $\tilde{\kappa}$. The monopole mass is set to $M=1.5$ TeV and the photon energy is $E_{\gamma}=6M$. This is the expectation if the monopole is truly dual to the electron in that it is in every way just the magnetic counterpart of well-known electric sources. The distribution shape is quite distinct from that of the scalar-monopole case (see fig.~\ref{PFdsigmaSpin0}) exhibiting a depression at $\eta\simeq0$ for all $\tilde{\kappa}$ values. The case $\tilde{\kappa}=0$ represents the SM expectation for electron-positron pair production if the coupling is substituted for the electric charge $e$, i.e.\ restoring the Lagrangian to simple Dirac QED. This case is clearly distinctive as the only unitary and renormalisable case. It is observed that the vertical heights of the curves for the differential cross section change with $\kappa$. Furthermore, the $\kappa=1$ and $\kappa=-1$ cases are totally equivalent, so a degeneration between positive and negative $\kappa$ values is evident.

\begin{figure}[ht!]\centering 
\includegraphics[width=0.55\textwidth]{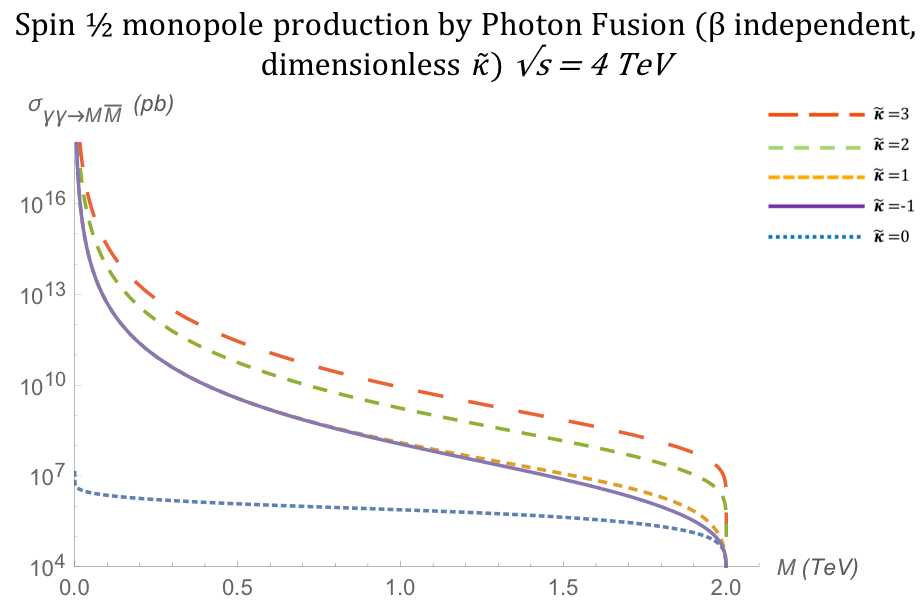}
\caption{Total cross section for the pair production of spin-\half monopoles via the PF process, as a function of the monopole mass $M$ for different values of $\tilde{\kappa}$ at $\sqrt{\sgg}=4$~TeV. }\label{LeeYangsig12label}
\end{figure}

The total cross section for arbitrary $\kappa$, also shown graphically in fig.~\ref{LeeYangsig12label}, is given by:
\begin{equation}\label{totxsec12}
\begin{split}
 \sigma^{S=\frac{1}{2}}_{\gamma\gamma\rightarrow M\overline{M}}&=\frac{\pi  \alpha_g^2(\beta)}{3\sgg} \Bigg[ 3 \beta^4 \kappa ^4 \sgg^2 \ln\Big(\frac{1-\beta}{1+\beta}\Big)+6 \beta^4 \ln \Big(\frac{1-\beta}{1+\beta}\Big)-7 \beta^3 \kappa ^4 \sgg^2+12 \beta^3-6 \beta^2 \kappa ^4 \sgg^2 \ln\Big (\frac{1-\beta}{1+\beta}\Big)\\
 &+6 \beta^2 \kappa ^2 \sgg \ln \Big(\frac{1-\beta}{1+\beta}\Big)-72 \beta \kappa  \sqrt{(1-\beta^2) \sgg}-36 \beta^2 \kappa  \sqrt{(1-\beta^2) \sgg} \ln\Big(\frac{1-\beta}{1+\beta}\Big)\\
 &-36 \kappa  \sqrt{(1-\beta^2) \sgg} \ln \Big(\frac{1-\beta}{1+\beta}\Big)-15 \beta \kappa ^4 \sgg^2-9 \kappa ^4 \sgg^2 \ln\Big(\frac{1-\beta}{1+\beta}\Big)\\
 &-132 \beta \kappa ^2 \sgg-60 \kappa ^2 \sgg \ln \Big(\frac{1-\beta}{1+\beta}\Big)-24 \beta-18 \ln \Big(\frac{1-\beta}{1+\beta}\Big) \Bigg].
\end{split}
\end{equation}
Setting $\kappa=0$, the expression~\eqref{totxsec12} reduces to that given in refs.~\cite{original,dw,vento}, used in standard monopole searches at colliders for data interpretation~\cite{atlasmono1,atlasmono2,moedal,moedalplb}. 

The high-energy limit for \eqref{dxsec12},expressed by $\sgg\rightarrow \infty$, leads to the approximations $\beta^n\rightarrow 1$ for $n>2$ and  $\beta\simeq 1-\frac{2M^2}{\sgg}$. In this limit, we observe that the differential cross section \eqref{dxsec12} diverges, except, unsurprisingly, when recovering the Dirac model with $\kappa=0$: 
\begin{align}
\begin{split}
	\frac{\dd\sigma^{S=\frac{1}{2}}_{\gamma\gamma\rightarrow M\overline{M}}}{\dd\Omega} \, \xrightarrow{\sgg\rightarrow \infty} & \, \frac{  \alpha_g^2(\beta) (1-\frac{2M^2}{\sgg}) }{4  (1-\cos^2 \theta )}\\&\quad\quad\quad\left(\kappa ^4 \sgg  \cos ^4 \theta +6 \kappa ^4 \sgg \cos ^2 \theta +\kappa ^4 \sgg+8 \kappa ^2  \cos ^2 \theta +28 \kappa ^2 + \frac{1}{\sgg} 4 \big(\cos ^2 \theta + 1\big) \right)\nonumber
\end{split}\\
	 \xrightarrow{\sgg\rightarrow \infty} & \, \sgg, \quad\quad \kappa \neq 0. \label{diffcross12sinf} \nonumber \\
 	\frac{\dd\sigma^{S=\frac{1}{2}}_{\gamma\gamma\rightarrow M\overline{M}}}{\dd\Omega} \quad \xrightarrow{\sgg\rightarrow \infty} &\quad  \frac{\alpha_g^2(\beta)}{\sgg} \frac{1 + \cos ^2 \theta}{1-\cos^2 \theta}, \quad\quad \kappa=0.
 \end{align}
 The total cross section \eqref{totxsec12}  carries the same high energy behaviour. It is finite for $\kappa=0$, and diverges for all other values of $\kappa$:
\begin{equation}\label{txs}
	\sigma^{S=\frac{1}{2}}_{\gamma\gamma\rightarrow M\overline{M}} \quad \xrightarrow{\sgg\rightarrow \infty} \quad  
	\begin{cases}
		\sgg &, \quad\quad \kappa\neq 0, \nonumber \\
		\dfrac{4 \pi  \alpha_g^2(\beta) }{\sgg} \Big[ \ln \Big(\dfrac{\sgg}{M^2}\Big) -1 \Big] &, \quad\quad \kappa=0.  
	\end{cases}
\end{equation}
For a constant value of $\sgg$, as assumed in fig.~\ref{LeeYangsig12label}, the high-energy limit may be approximated by $M \rightarrow 0$, where the cross section becomes finite only for the value $\kappa=0$ and diverges in other $\kappa$ values, as expected from \eqref{txs}. 
  
Hence, a unitarity requirement may isolate the $\kappa=0$ model from the others as the only viable theory for the spin-\half monopole, unless the model is viewed as an effective field theory. In that case, the value of $\kappa$ can be used as a window to extrapolate some characteristics of the extended model in which unitarity is restored. Also, as already mentioned, $\kappa$ is not dimensionless, hence, a non-zero $\kappa$ clearly makes this (effective) theory non-renormalisable.
 
\subsubsection{Pair production of spin-\half monopoles via Drell-Yan}

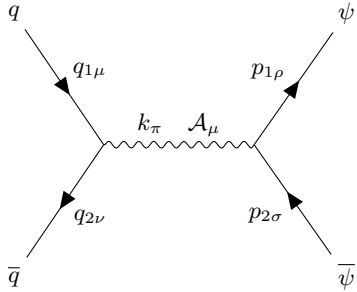
\begin{figure}[ht!]
\centering
\begin{tikzpicture}
  \begin{feynman}
  	\vertex (a1);
	\vertex[right=2cm of a1] (a2);
	\vertex[above=1.75cm of a2] (a3);
	\vertex[right=1cm of a3] (a4) {\(\psi\)};
	\vertex[below=1.75cm of a2] (a5);
	\vertex[right=1cm of a5] (a6) {\(\overline{\psi}\)};
	
	\vertex[above=1.75cm of a1] (a7);
	\vertex[left=1cm of a7] (a8) {\(q\)};
	\vertex[below=1.75cm of a1] (a9);
	\vertex[left=1cm of a9] (a10) {\(\overline{q}\)};
	
	\diagram* {
	       	(a8) -- [fermion, edge label=\(q_{1\mu}\)] (a1) -- [fermion, edge label=\(q_{2\nu}\)] (a10),
		(a1) -- [boson, edge label=\(k_{\pi} \quad \mathcal{A}_{\mu}\)] (a2),
       		(a6) -- [fermion, edge label=\(p_{2\sigma}\)] (a2) -- [fermion, edge label=\(p_{1\rho}\)] (a4),
      };
\end{feynman}
\end{tikzpicture}
\caption{Feynman-like diagram for the DY process where $q\overline{q}\rightarrow \psi\overline{\psi}$ via a virtual photon in the $s$-channel. The variable definitions are given in the text. }\label{FeynSpinor}
 \end{figure}

Monopole pair production through the $s$-channel is also possible for fermionic monopoles through the annihilation of quarks into a photon, which decays to the monopole-antimonopole pair. The relevant Feynman rules are displayed in fig.~\ref{vetexSpinora} with the $\kappa$-dependent matrix amplitude given in eq.~\eqref{Matampspinor}, both in appendix~\ref{app:xsection}. The complete DY process is shown in fig.~\ref{FeynSpinor}. The exchange energy in the centre-of-mass frame is $k_{\pi}k^{\pi}=\sqq$.

The kinematic distributions are computed with details given in appendix~\ref{app:xsection} and are shown in fig.~\ref{LeeYangDsigDstuff12labeldy}. The monopole is treated as the magnetic dual to the electron. Analytically, the differential cross section becomes:
\begin{align}\label{Xsecdeffspin12}
	\frac{\dd\sigma^{S=\frac{1}{2}}_{q\overline{q}\rightarrow M\overline{M}}}{\dd\Omega} &=\frac{5 \alpha_e \alpha_g(\beta) }{36 \sqq}\Big[\beta^3 (\cos ^2\theta -\kappa ^2 \sqq \cos ^2\theta -\kappa ^2 \sqq-1)+\beta(4 \kappa  \sqrt{\sqq-\beta^2 \sqq}+2 \kappa^2 \sqq+2)\Big],
\end{align}
where $\beta_q=1$ on the right as the quarks are assumed to be massless (compared to the heavy monopoles), and the index has been dropped on $\beta_p\rightarrow\beta$. Still, the magnetic coupling depends on $\beta$ as $\alpha_g(\beta)\propto \beta^2$ in the velocity-dependent magnetic charge case. 

\begin{figure}[ht!]\centering
\includegraphics[width=0.95\textwidth]{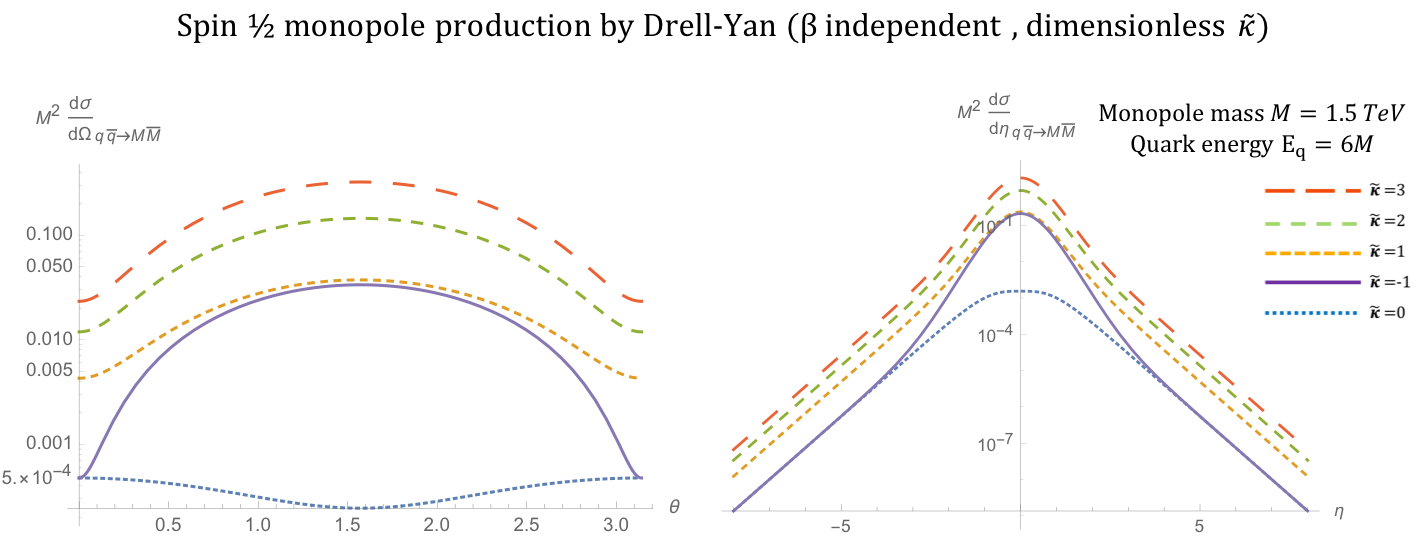}
\caption{Angular distributions for monopole-antimonopole pair production via DY for the case of a fermionic monopole with spin~\half and mass $M=1.5$~TeV as a function of the scattering angle $\theta$ (left) and the pseudorapidity $\eta$ (right) at $\sqrt{\sqq}=2E_q$, where $E_q=6M$, and for various values of the parameter $\tilde{\kappa}$. The case $\tilde{\kappa}=0$ represents the SM expectation for electron-positron pair production if the coupling is substituted for the electric charge $e$. }\label{LeeYangDsigDstuff12labeldy}
\end{figure}

As observed in the previous section, the value of $\kappa$ affects the unitarity and renormalisability of the model. In the spin-\half model, this parameter has dimensions, destroying the renormalisability except when $\kappa=0$, in which case the magnetic moment is only a by-product of anomalous spin interactions within a minimally coupling field theory. 
But unitarity \emph{for this process} is maintained for all values of $\kappa$. This becomes apparent in the high energy limit $\sqq\rightarrow \infty$. Taking the expression \eqref{Xsecdeffspin12} to first order in $\beta$ in this limit, we observe that it is finite for all $\kappa$,
\begin{equation}
	\frac{\dd\sigma^{S=\frac{1}{2}}_{q\overline{q}\rightarrow M\overline{M}}}{\dd\Omega} \quad \xrightarrow{\sqq\rightarrow \infty}  \quad 
	\begin{cases}
\dfrac{5 \beta \alpha_e \alpha_g(\beta)}{36 }\kappa ^2 \, (1-\cos ^2\theta ),  & \quad \kappa\neq0, \\
\dfrac{5 \beta \alpha_e \alpha_g(\beta)}{36 \sqq}(1+\cos ^2 \theta ),  &  \quad \kappa=0.
\end{cases}
\end{equation}

\begin{figure}[ht!]\centering
\includegraphics[width=0.55\textwidth]{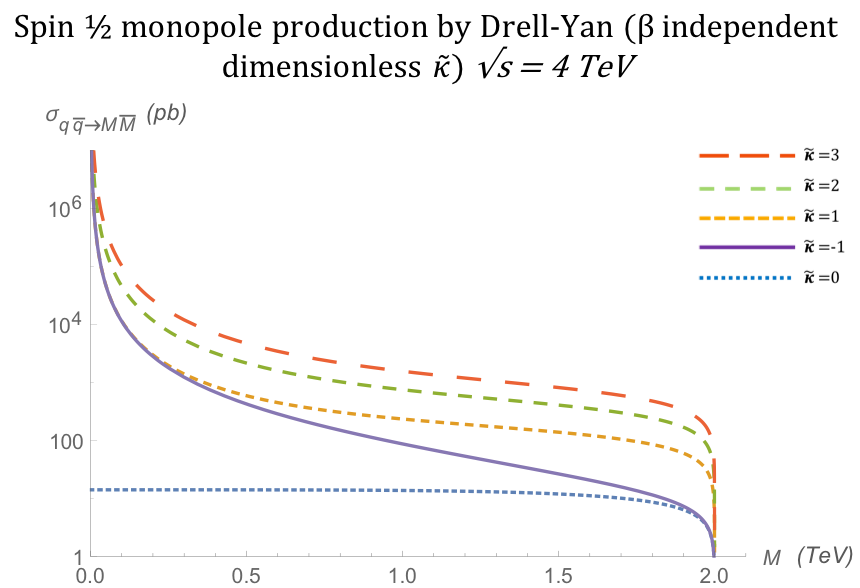}
\caption{Total cross section for the pair production of spin-\half monopoles via the DY process, as a function of the monopole mass $M$ for different values of $\tilde{\kappa}$ at $\sqrt{\sqq}=4$~TeV. }\label{LeeYangsig12labeldy}
\end{figure}

The total cross section follows the same trend. Its full form is
\begin{equation}\label{thisone}
\sigma^{S=\frac{1}{2}}_{q\overline{q}\rightarrow M\overline{M}}=\frac{10 \pi  \beta \alpha _e \alpha _g(\beta) }{27 \sqq}\left(3-\beta^2-(2 \beta ^2-3) \kappa ^2 \sqq+6 \kappa  \sqrt{\sqq-\beta ^2 \sqq}\right),
\end{equation}
which is displayed graphically in fig.~\ref{LeeYangsig12labeldy}. In the high-energy limit, it reduces to
\begin{equation}\label{thatone}
\sigma^{S=\frac{1}{2}}_{q\overline{q}\rightarrow M\overline{M}} \xrightarrow{\sqq\rightarrow \infty}
\frac{10 \pi  \beta \alpha _e \alpha _g(\beta) }{27\sqq}(\kappa ^2 \sqq+2-12\kappa M),
\end{equation}
which is finite for all values of $\kappa$. In fig.~\ref{LeeYangsig12labeldy}, where $\sgg$ is constant, the high-energy limit is approximated at $M \rightarrow 0$, where the cross section becomes finite for the value $\kappa=0$ and diverges as $\sim M^{-2}$ in other $\kappa$ values, bearing in mind that $\kappa$ is related to the monopole mass as in \eqref{ktilde}. It should be noted however that DY alone is not the only mechanism for producing $\psi\overline{\psi}$ pairs and on its own cannot define the unitarity of the theory as a whole. In the previous section, it was shown how this model violates unitarity when considering pair production by PF.

It is worth noting that monopole production in a high energy collider sees twice the production cross section for collisions with same-sign incoming beams, such as proton-proton collisions at the LHC, in contrast to opposite-sign hadron colliders, such as the Tevatron, which maintain the exact cross section as given in \eqref{thisone}. 

\subsection{Vector monopole with arbitrary magnetic moment term \label{sec:spin1}}

Monopoles of spin-1 have been addressed for the first time in colliders recently by the MoEDAL experiment for the Drell-Yan production~\cite{moedalplb}. A monopole with a spin $S=1$ is postulated as a \emph{massive} vector meson $W_{\mu}$ interacting only with a massless gauge field $\mathcal{A}_{\mu}$ in the context of a gauge invariant Proca field theory. As mentioned previously, lacking a fundamental theory for point-like magnetic poles,  we keep the treatment general by including a magnetic moment term in the effective Lagrangian, proportional to $\kappa$, which is a free phenomenological parameter. Unlike the spin-\half monopole case, however, for the vector monopole the magnetic moment parameter $\kappa$ is dimensionless. The case $\kappa=0$ corresponds to a pure Proca Lagrangian, and $\kappa=1$ to that of the SM $W_{\mu}$ boson in a Yang-Mills theory with spontaneous symmetry breaking. In this respect, our approach resembles early phenomenological studies of charged $W^{\pm}$-boson production in the SM through PF, where the magnetic moment of the $W$-boson was kept free~\cite{Tupper}, different from the value $\kappa=1$ dictated by unitarity. The aim of such analyses was to determine measurable (physical) quantities in purely electromagnetic SM processes, that were sensitive to the value of $\kappa$, and more or less independent of the Higgs field and the neutral gauge boson $Z^0$. These quantities were the angular distributions at sufficiently high energies, whose behaviour for the unitarity-imposed value $\kappa=1$ was found to be quite distinct from the case $\kappa \ne 1$. As we shall see in our case, for certain formal limits of large $\kappa$ and slowly moving monopoles, one may also attempt to make sense of the perturbative DY or PF processes of monopole-antimonopole pair production, when velocity-dependent magnetic charges are employed. 

 The pertinent effective Lagrangian, obtained by imposing electric-magnetic duality on the respective Lagrangian terms for the interaction of $W^{\pm}$ gauge bosons with photons in the generalised SM framework, as described above~\cite{Tupper}, takes the form:
	\begin{align}\label{GaugeInvLag}
	& \mathcal{L}=-\xi(\partial_{\mu}W^{\dagger\mu})(\partial_{\nu}W^{\nu})-\frac{1}{2}(\partial_{\mu}\mathcal{A}_{\nu})(\partial^{\nu}\mathcal{A}_{\mu})-\frac{1}{2}G_{\mu\nu}^{\dagger}G^{\mu\nu}-M^{2}W_{\mu}^{\dagger}W^{\mu}-ig(\beta)\kappa F^{\mu\nu}W^{\dagger}_{\mu}W_{\nu}~,
	\end{align}
where the symbol $\dagger$ denotes the hermitian conjugate, and $G^{\mu\nu}=(D^{\mu}W^{\nu}-D^{\nu}W^{\mu})$, with $D_{\mu}=\partial_{\mu}-ig(\beta)\, \mathcal{A}_{\mu}$ the $U(1)$ covariant derivative, which provides 
the coupling of the (magnetically charged) vector field $W_{\mu}$ to the gauge field $\mathcal{A}_{\mu}$, playing the role of the ordinary photon.  
The parameter $\xi$ is a gauge-fixing parameter. 
The magnetic coupling is considered in the general form of \eqref{etogb}, so as to cover both the $\beta$-dependent and $\beta$-independent cases in a unified formalism. The tensor $F^{\mu\nu}$ represents the Abelian electromagnetic field strength (Maxwell). 

The magnetic and quadrupole moments are given respectively by
\begin{subequations}
\begin{align}
\mu_M=\frac{g({\beta})}{2M}&(1+\kappa)\hat{S}, \quad \hat{S}=1, \label{magmom}\\
Q_E&=-\frac{g(\beta)\kappa}{M^2}\label{quadmom}~,
\end{align}
\end{subequations}
where $\hat{S}$ is the monopole spin expectation value and the corresponding ``gyromagnetic ratio'' $g_R = 1 + \kappa$. The phenomenological moment term $-ig(\beta)\kappa F^{\mu\nu}W^{\dagger}_{\mu}W_{\nu}$ is highly divergent and contributes correction terms to the magnetic and quadruple moments in eqs.~\eqref{magmom} and \eqref{quadmom}, respectively. As mentioned already, following ref.~\cite{Tupper}, we treat $\kappa$ as a free phenomenological parameter of the theory.\footnote{Corrections to the magnetic moment of the monopole could also arise through anomalous spin interactions at a quantum level. For $\beta$-independent magnetic charges, these are uncontrollable, as perturbation theory fails. However, if one accepts velocity-dependent couplings \eqref{replacement}, then for slowly-moving monopoles such loop corrections can be made subleading. Moreover, as for the case of spinor monopoles, one could also add an EDM term for the vector monopole~\cite{silenko}, ${\mathcal L}_{\rm EDM} = i g(\beta)\eta \widetilde F^{\mu\nu}W^{\dagger}_{\mu}W_{\nu}$, where $\widetilde F^{\mu\nu} = \frac{1}{2}\epsilon^{\mu\nu\rho\sigma} F_{\rho\sigma}$, is the dual Maxwell tensor, with $\epsilon^{\mu\nu\rho\sigma}$ the totally antisymmetric Levi-Civita tensor in four space-time dimensions. Such terms are assumed suppressed in our analysis, although the latter can be straightforwardly extended to include them.} For $\kappa \ne 1$, the theory is known to be non unitary, and is plagued by 
ultraviolet divergences in the self-energy loop graphs in this model, making the quadrupole moment infinite in a non-renormalisable way. To tackle such divergences, in the pre-SM era, Lee and Yang~\cite{LeeYang} proposed the effective Lagrangian \eqref{GaugeInvLag}, and demonstrated that such divergences are removed through the inclusion of the gauge fixing term with the gauge fixing parameter $\xi \ne 0$, but at a cost
of introducing a negative metric (and thus ghosts, reflecting the unitarity issue for $\kappa \ne 1$). In this ``$\xi$-limiting formalism'', as it is called, the 
observables are evaluated from the $\xi$-dependent Lagrangian before taking the limit $\xi \rightarrow 0$. The quadruple moment becomes finite at one-loop level~\cite{Tupper}. Unitarity in this formalism is held only at energy scales $E^2\leq M^2 / \xi$, the rest mass energy of a single ghost state. For our purposes, this could be an acceptable assumption for an effective field theory considered valid up to a cut-off scale $\Lambda^2 = M^2 / \xi$. 
In general, lacking a concrete fundamental theory on magnetic poles, we shall ignore the unitarity issue when we consider the incorporation of an arbitrary magnetic moment $\kappa$ in our construction. 

At this point, we should also mention that in general, there is another unitarity issue that arises within the context of the PF or DY cross sections in the dual effective theories we consider here. Even in the unitary $\kappa=1$ case inspired by SM physics, the associated cross section for large values of $\beta$ violates the unitarity bound for a cross section dominated by a single partial wave of angular momentum $J$:
\begin{equation}\label{unibound}
\sigma_J  \le \frac{4\pi}{s} (2J + 1)~.
\end{equation}
For sufficiently small values of $\beta$ on the other hand, such unitarity bounds are respected. 
In fact this is a feature that characterises the cross sections of all three cases of monopole spin $S=0,\,\half,\,1$ and not only the vector case~\cite{original}. To tackle such an issue, within a phenomenological effective model for the monopoles, one may assume the existence of appropriate form factors that depend on the energy of the photon-monopole interaction~\cite{original}. We shall not pursue such issues further in this section, but we 
mention that such form factors will be included in  the \MAD generator simulating monopole production at colliders, as we shall discuss in section \ref{sec:lhc}. 

After this parenthesis, we come back to the spin-1 monopole-production process via PF within the context of the model \eqref{GaugeInvLag}. Restricting our attention to tree level, the gauge fixing parameter $\xi$ is redundant and the $\xi$-independent interaction vertices are given in fig.~\ref{FeynRulesFig}. These are used to evaluate the $\kappa$-dependent PF Born amplitudes, $\mathcal{A}_{\mu}\mathcal{A}_{\nu}\rightarrow W_{\mu}W^{\dagger}_{\nu}$ and DY $\psi\overline{\psi}\rightarrow W_{\mu}W^{\dagger}_{\nu}$. As already mentioned, we introduce a velocity-dependent magnetic 
coupling $g=g(\beta)$ corresponding to a magnetic charge linearly dependent on the monopole boost $\beta=|\vec{p}| / E_p$ where $\vec{p}$ and $E_p$ are the monopole momentum and energy, respectively.  

\subsubsection{Pair production of spin-1 monopoles via photon fusion}

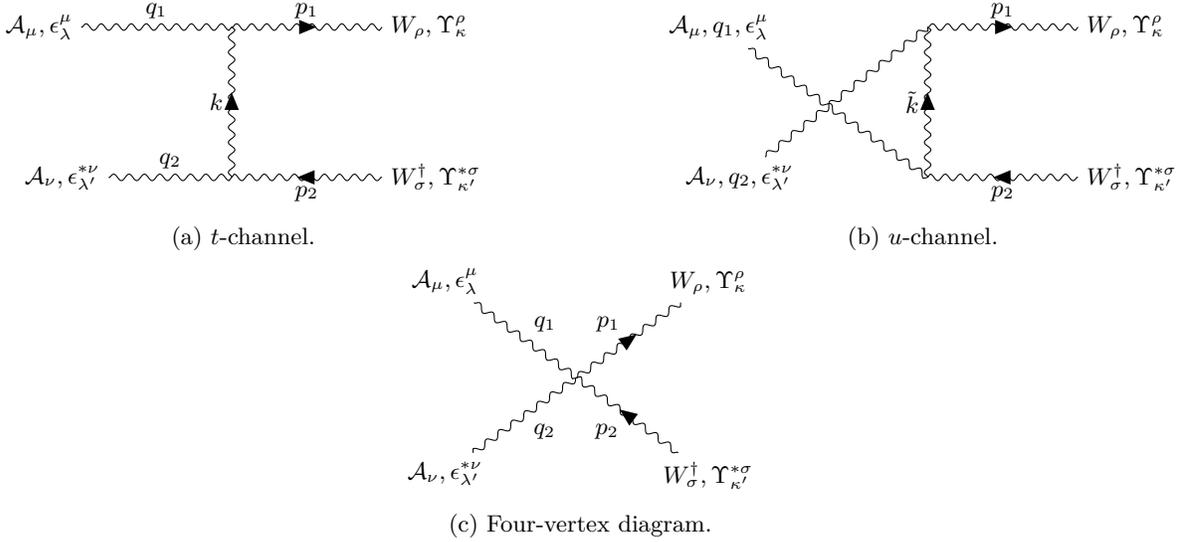
\begin{figure}[ht!]
\justify
\begin{subfigure}[b]{0.5\textwidth}
\centering
\begin{tikzpicture}
  \begin{feynman}
  	\vertex (a1) {\(\mathcal{A}_{\nu}, \epsilon^{*\nu}_{\lambda'}\)};
	\vertex [right=2.25cm of a1] (a2);
	\vertex [right=2cm of a2] (a3){\(W_{\sigma}^{\dagger},  \Upsilon^{*\sigma}_{\kappa'}\)};
	\vertex [above=2cm of a2] (a5);
	\vertex [right=2cm of a5] (a6){\(W_{\rho},  \Upsilon^{\rho}_{\kappa}\)};
	\vertex [left=2cm of a5] (a4){\(\mathcal{A}_{\mu},  \epsilon^{\mu}_{\lambda}\)};
	\diagram* {
	(a3) -- [charged boson, edge label=\(p_{2}\)] (a2) -- [charged boson, edge label=\(k\)] (a5) -- [charged boson, edge label=\(p_{1}\)] (a6),
	(a4) -- [boson, edge label=\(q_{1}\)] (a5),
	(a1) -- [boson, edge label=\(q_{2}\)] (a2)
         };
\end{feynman}
\end{tikzpicture}
\caption{$t$-channel.\label{fig:tchan-pf-one}}
\end{subfigure}
\begin{subfigure}[b]{0.5\textwidth}
\centering
\begin{tikzpicture}
  \begin{feynman}
  	\vertex (a1) {\(\mathcal{A}_{\nu} ,  q_{2},   \epsilon^{*\nu}_{\lambda'}\)};
	\vertex [right=2.5cm of a1] (a2);
	\vertex [right=2cm of a2] (a3){\(W_{\sigma}^{\dagger}, \Upsilon^{*\sigma}_{\kappa'}\)};
	\vertex [above=2cm of a2] (a5);
	\vertex [right=2cm of a5] (a6){\(W_{\rho}, \Upsilon^{\rho}_{\kappa}\)};
	\vertex [left=2cm of a5] (a4){\(\mathcal{A}_{\mu} ,  q_{1},   \epsilon^{\mu}_{\lambda}\)};
	\diagram* {
	(a3) -- [charged boson, edge label=\(p_{2}\)] (a2) -- [charged boson, edge label=\(\tilde{k}\)] (a5) -- [charged boson, edge label=\(p_{1}\)] (a6),
	(a4) -- [boson] (a2),
	(a1) -- [boson] (a5)
      }; 
\end{feynman}
\end{tikzpicture}
\caption{$u$-channel.\label{fig:uchan-pf-one}}
\end{subfigure}
\begin{subfigure}[b]{1\textwidth}
\centering
\begin{tikzpicture}
\begin{feynman}
  	\vertex (a2);
	\vertex[left=1.75cm of a2] (a3);
	\vertex[below=1cm of a3] (a4) {\(\mathcal{A}_{\nu},  \epsilon^{*\nu}_{\lambda'}\)};
	\vertex[right=1.75cm of a2] (a5);
	\vertex[below=1cm of a5] (a6) {\(W_{\sigma}^{\dagger}, \Upsilon^{*\sigma}_{\kappa'}\)};
	\vertex[above=1cm of a3] (a8) {\(\mathcal{A}_{\mu}, \epsilon^{\mu}_{\lambda}\)};
	\vertex[above=1cm of a5] (a7) {\(W_{\rho}, \Upsilon^{\rho}_{\kappa}\)};
	\diagram* {
       		(a6) -- [charged boson, edge label=\(p_{2}\)] (a2) -- [charged boson, edge label=\(p_{1}\)] (a7),
		(a8) -- [photon, edge label=\(q_{1}\)] (a2) -- [photon, edge label=\(q_{2}\)] (a4)
      };
\end{feynman}
\end{tikzpicture}
\caption{Four-vertex diagram.\label{fig:seagull-pf-one}}
\end{subfigure}
\caption{Feynman-like diagrams for the $t$-channel~\subref{fig:tchan-pf-one}, $u$-channel~\subref{fig:uchan-pf-one} and seagull diagram~\subref{fig:seagull-pf-one}  that contribute to the production of a vector-monopole pair, $WW^{\dagger}$, from the fusion of two gauge bosons. The variable definitions are given in the text. }
\label{FeynGraphsFig}
\end{figure}

Monopole-antimonopole pairs are generated at tree level by photons fusing in the $t$-channel, $u$-channel and at a 4-point vertex depicted by the Feynman-like graphs in fig.~\ref{FeynGraphsFig}. The matrix amplitude for each process is given in eqs.~\eqref{MatrixAmp} in fig.~\ref{FeynRulesFig} in appendix~\ref{app:xsection}. $k$ and $\tilde{k}$ are the exchange momenta of the $t$- and $u$-channel processes, respectively. The monopoles have polarisation vectors $\Upsilon(p_1)_{\kappa}, \Upsilon(p_2)_{\kappa'}$ and momentum 4-vectors $p_{1\mu}$ and $p_{2\mu}$, where (on mass-shell): $p_{1,2}^2=M^2$. The photons polarisation vectors are $\epsilon(q_1)_{\lambda},\epsilon(q_2)_{\lambda'}$ and their momentum 4-vectors are $q_{1\mu}$ and $q_{2\mu}$, $q_{1,2}^2=0$. Details of the calculations of the analytic expressions for the kinematic distributions and cross section are given in appendix~\ref{app:xsection}. 

\begin{figure}[ht!]\centering
\includegraphics[width=0.95\textwidth]{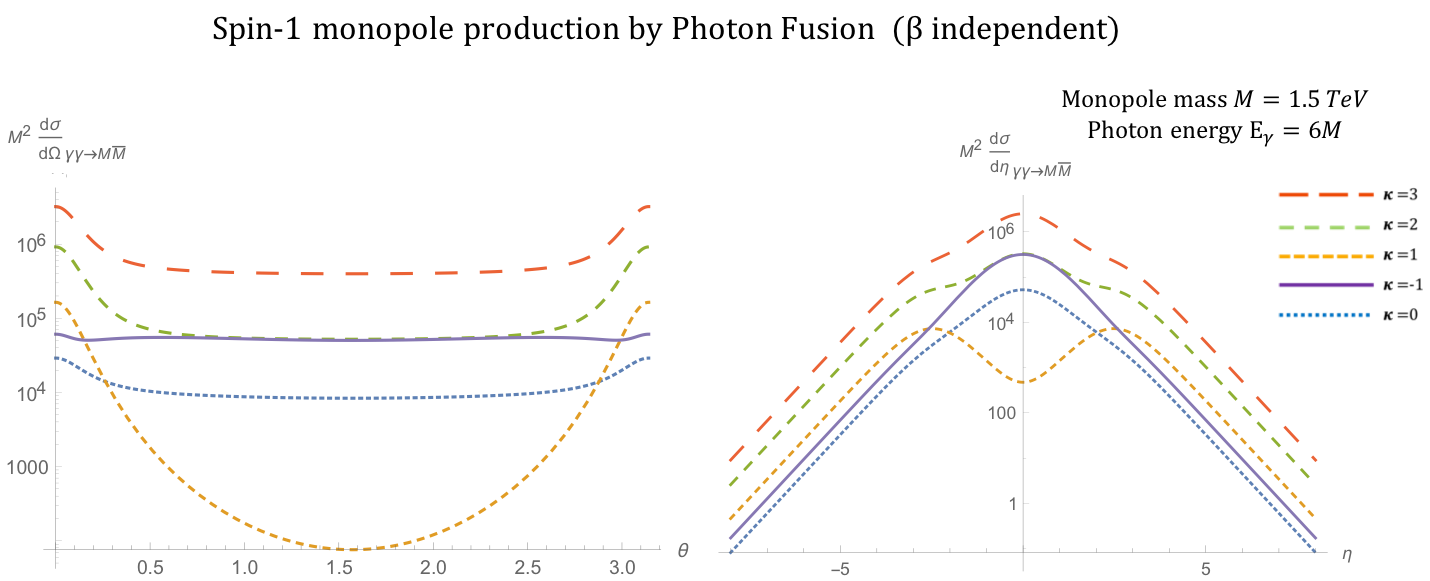}
\caption{\label{LeeYangDsigDomega1pflabel} Angular distributions for monopole-antimonopole pair production via PF for the case of a monopole with spin~1 and mass $M=1.5$~TeV as a function of the scattering angle $\theta$ (left) and the pseudorapidity $\eta$ (right) at $\sqrt{\sgg}=2E_{\gamma}$, where $E_{\gamma}=6M$, and for various values of the phenomenological parameter $\kappa$. The case $\kappa=1$ represents the SM expectation for the pair production of spin-1 $W^{\pm}$ gauge bosons and is distinctive as the only unitary and renormalisable case.}
\end{figure}

The phenomenological parameter $\kappa$ enters in the expressions of the differential cross sections given in \eqref{dxsec1k}. These kinematic distributions are plotted as functions of the kinematic variables $\theta$ and $\eta$ in fig.~\ref{LeeYangDsigDomega1pflabel} assuming a monopole mass $M=1.5$~TeV and centre-of-mass energy $\sqrt{\sgg}=2E_{\gamma}$ with $E_{\gamma}=6M$. The monopole boost $\beta$ is defined in \eqref{btos}, where the centre-of-mass energy is understood to be the energy of the fusing photons, i.e.\  $\sqrt{s}=\sqrt{\sgg}=2E_{\gamma}$.

\begin{equation}\label{dxsec1k}
\begin{split}
\frac{\dd\sigma^{S=1}_{\gamma\gamma\rightarrow M\overline{M}}}{\dd\Omega}=&\frac{\alpha_g^2(\beta) \beta}{16 \left(1-\beta^2\right)^2 \sgg \left(1-\beta^2 \cos ^2\theta \right)^2} \Bigg\{48 \beta^8+\beta^6 (\kappa-1)^4 \cos ^6 \theta \\
&-144 \beta^6+2 \beta^4 \left(3 \kappa^4+28 \kappa^3+42 \kappa^2-4 \kappa+79\right)-2 \beta^2 \left(11 \kappa^4+60 \kappa^3+58 \kappa^2+12 \kappa+35\right)\\
&+\beta^4 \left[24 \beta^4+2 \beta^2 \left(\kappa^4+12 \kappa^3-10\kappa^2-20 \kappa-7\right)+9 \kappa^4-36 \kappa^3+22 \kappa^2+28 \kappa+1\right] \cos ^4 \theta \\
&-\beta^2 \bigg[48 \beta^6+2 \beta^4 \left(\kappa^4+4 \kappa^3-34 \kappa^2-28 \kappa-55\right)\\
&-4 \beta^2 \left(3 \kappa^4-42 \kappa^2-8 \kappa-29\right)+35 \kappa^4-44 \kappa^3-78 \kappa^2-12 \kappa-29\bigg] \cos ^2 \theta \\
&+29 \kappa^4+44 \kappa^3+46 \kappa^2+12 \kappa+21\Bigg\} 
\end{split}
\end{equation}

The kinematic distributions change as the parameter $\kappa$ is varied. The reader can readily see from fig.~\ref{LeeYangDsigDomega1pflabel} the distinct behaviour of the kinematic distribution in the unitarity-respecting case $\kappa=1$, that shows a depression around $\eta=0$, as compared to the peaks in the cases where $\kappa \ne 1$. 
This is in agreement with the situation characterising $W^+W^-$ production in the SM case~\cite{Tupper}. However as we shall see in section~\ref{kappaMad}, when the PDF of the photon in the proton is taken into account, these shape differences are smoothed out in $pp$ collisions including the (discernible here) $\kappa=1$ case. For $\kappa=1$ the expression \eqref{dxsec1k} becomes:
\begin{equation}\label{dxsec1k1}
\frac{\dd\sigma^{S=1}_{\gamma\gamma\rightarrow M\overline{M}}}{\dd\Omega}=\frac{\alpha_g(\beta)^2 \beta}{2 \sgg(1-\beta^2 \cos ^2 \theta )^2} \Big(3\beta^4 (\cos ^4 \theta -2\cos ^2 \theta +2)+\beta^2(16 \cos^2 \theta -6)+19\Big), \qquad \kappa=1,
\end{equation}
with $\beta$ given in \eqref{btos} and $\alpha_g(\beta)$ in \eqref{etogb}. 

As mentioned previously, the parameter $\kappa$ influences the unitarity and renormalisability of the effective theory. 
It is important to notice that 
the differential cross section in the $\kappa=1$ case \eqref{dxsec1k1} is the \emph{only finite} solution in the ultraviolet limit $\sgg \to \infty$.
Indeed, in the high energy limit, $\sgg\rightarrow \infty$, one may approximate $\beta^4 \simeq 1$ and $\beta \simeq 1-\frac{2M^2}{\sgg}$, implying that the angular distribution \eqref{dxsec1k1} falls off as $\sgg^{-1}$: 
\begin{align}
\frac{\dd\sigma^{S=1}_{\gamma\gamma\rightarrow M\overline{M}}}{\dd\Omega} \quad \xrightarrow{\sgg\rightarrow \infty} \quad  \frac{\alpha_g^2(\beta)}{2\sgg( 1-\cos^2 \theta )^2}\left(3\cos^4 \theta +10\cos^2 \theta +19\right), \quad\ \kappa=1.
\end{align}
For all other $\kappa$ values one obtains a differential cross section proportional to $\sgg$, which diverges linearly with $\sgg \to \infty$:
\begin{align}
\frac{\dd\sigma^{S=1}_{\gamma\gamma\rightarrow M\overline{M}}}{\dd\Omega} \quad  \xrightarrow{\sgg\rightarrow \infty} \quad  \sgg, \quad\quad \kappa\neq1.
\end{align}

\begin{figure}[ht!]\centering
\includegraphics[width=0.55\textwidth]{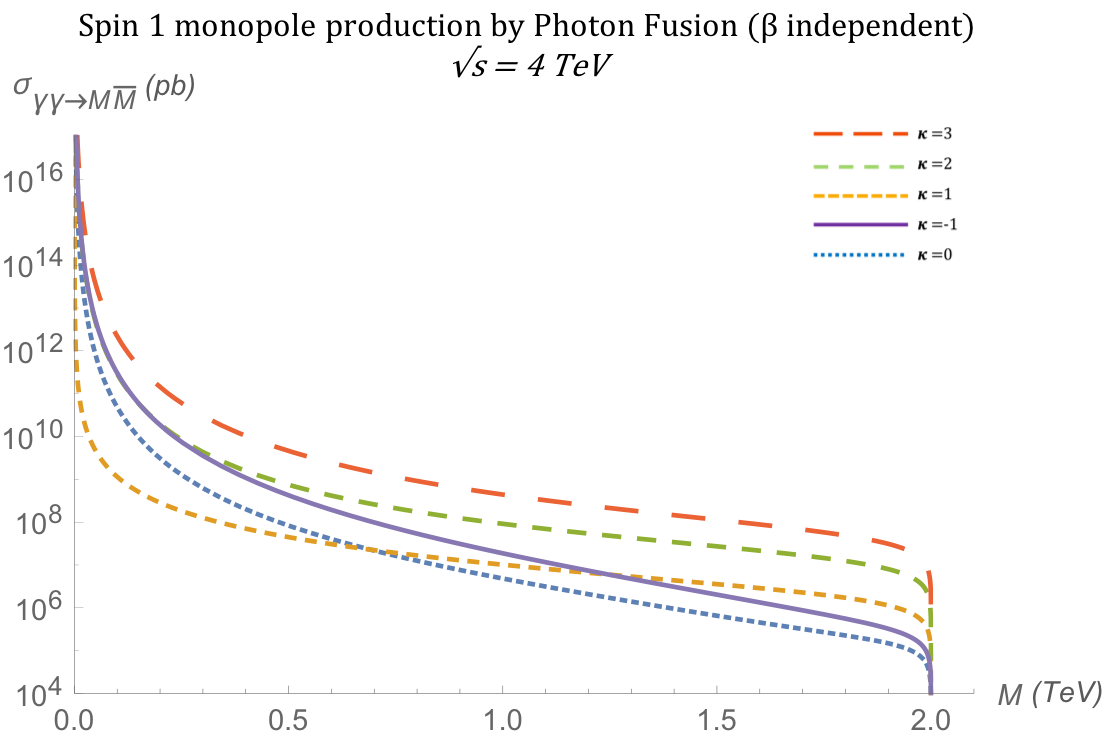}
\caption{\label{LeeYangsig1pflabel} Total cross section for the pair production of spin-1 monopoles via the PF process, as a function of the monopole mass $M$ for different values of $\kappa$ at $\sqrt{\sgg}=4$~TeV. }
\end{figure}

The total cross section for general $\kappa$ is given by
\begin{equation}\label{totxsec1k}
\begin{split}
\sigma_{\gamma\gamma\rightarrow M\overline{M}}^{S=1}&=\frac{\pi  \alpha^2_g(\beta)}{12 (1-\beta^2)^2 \sgg} \Bigg[ -72 \beta^7 + 288 \beta^5 -\beta^3 (-\kappa ^4+4 \kappa ^3+282 \kappa ^2+196 \kappa +263)\\
&-6 (1-\beta^2) (6 \beta^6-6 \beta^4+\beta^2 (\kappa ^4+8 \kappa ^3+2 \kappa ^2-8 \kappa -9)-4 \kappa ^4-16 \kappa ^3+16 \kappa ^2+8 \kappa +2) \ln \left(\frac{1+\beta}{1-\beta}\right)\\
& +3\beta (13 \kappa ^4-20 \kappa ^3+110 \kappa ^2+44 \kappa +29)\Bigg],
\end{split}
\end{equation}
which for the $\kappa=1$ case reduces to~\cite{original}
\begin{equation}\label{reproRusakovich1}
\begin{split}
\sigma_{\gamma\gamma\rightarrow M\overline{M}}^{S=1}=&\frac{\pi \alpha_g^2(\beta)\beta}{\sgg}\left(2\frac{3\beta^4-9\beta^2+22}{1-\beta^2}-\frac{3(1-\beta^4)}{\beta}\ln\left(\frac{1+\beta}{1-\beta}\right)\right).
\end{split}
\end{equation}
The total cross section is shown graphically in fig.~\ref{LeeYangsig1pflabel} for various $\kappa$ values. In the high energy limit $\sgg\rightarrow \infty$,  \emph{only} the total cross section for $\kappa=1$ is \emph{finite}, in similar spirit to the differential cross section behaviour:
\begin{equation}
\sigma^{S=1}_{\gamma\gamma\rightarrow M\overline{M}} \quad \xrightarrow{\sgg\rightarrow \infty} \quad  
\begin{cases}
\dfrac{8\pi\alpha_g^2(\beta)}{M^2} &, \quad \kappa=1, \nonumber \\ 
 \sgg &, \quad  \kappa \neq 1.
\end{cases}
\end{equation}
We stress once again that the $\kappa=1$ case is the SM result for vector boson scattering with photons in which the coupling term $-ig F^{\mu\nu}W^{\dagger}_{\mu}W_{\nu}$ naturally arises in $SU(2)\times U(1)$ gauge theory with spontaneous symmetry breaking. This value for $\kappa$ also restores renormalisability and unitarity in the absence of the negative metric and the $\xi$-gauge fixing. The meson adopts a (tree-level) gyromagnetic ratio $g_R=2$, which is the value associated with the $W^{\pm}$ gauge boson of the SM.

\subsubsection{Pair production of spin-1 monopoles via Drell-Yan}

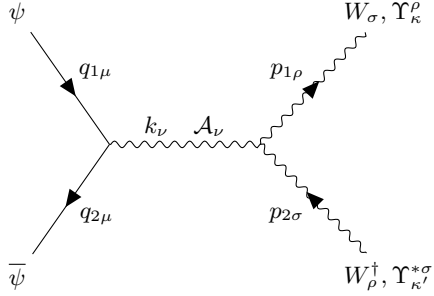
\begin{figure}[ht!]
\centering
\begin{tikzpicture}
  \begin{feynman}
  	\vertex (a1);
	\vertex[right=2cm of a1] (a2);
	\vertex[above=1.75cm of a2] (a3);
	\vertex[right=1cm of a3] (a4) {\(W_{\sigma}, \Upsilon^{\rho}_{\kappa}\)};
	\vertex[below=1.75cm of a2] (a5);
	\vertex[right=1cm of a5] (a6) {\(W_{\rho}^{\dagger},  \Upsilon^{*\sigma}_{\kappa'}\)};
	
	\vertex[above=1.75cm of a1] (a7);
	\vertex[left=1cm of a7] (a8) {\(\psi\)};
	\vertex[below=1.75cm of a1] (a9);
	\vertex[left=1cm of a9] (a10) {\(\overline{\psi}\)};
	
	\diagram* {
	       	(a8) -- [fermion, edge label=\(q_{1\mu}\)] (a1) -- [fermion, edge label=\(q_{2\mu}\)] (a10),
		(a1) -- [boson, edge label=\(k_{\nu} \Quad \mathcal{A}_{\nu}\)] (a2),
       		(a6) -- [charged boson, edge label=\(p_{2\sigma}\)] (a2) -- [charged boson, edge label=\(p_{1\rho}\)] (a4)
      };
\end{feynman}
\end{tikzpicture}
\caption{Feynman-like diagram representing the DY process of the spin-1 monopole. The quarks annihilate to a photon through an electromagnetic process, which subsequently decays to a monopole-antimonopole pair. The variable definitions are given in the text. }\label{S1DrellYan}
\end{figure}

As discussed previously, another mechanism contributing to the production of monopole-antimonopole pairs is quark-antiquark annihilation through the $s$-channel, also known as Drell-Yan, drawn in fig.~\ref{S1DrellYan}, for which the relevant Feynman rules are given in fig.~\ref{FeyRule1DY}. The quarks each have a mass $m$ considered small compared to the monopole mass, $M$, and are characterised by momentum 4-vectors $q_{1\mu}$ and $q_{2\mu}$, where on mass shell one has $q_{1,2}^2=m^2$. Similarly, the mesons have mass $M$ each and are characterised by momentum 4-vectors $p_{1\mu}$ and $p_{2\mu}$, where on mass-shell  one has $p_{1,2}^2=M^2$. The centre-of-mass energy of the quark-antiquark pair is $k_{\nu}k^{\nu}=\sqq$.

The differential cross section distributions are computed as defined in section \ref{sectionAmtoKin}. The expression for $\beta$ \eqref{btos} is formally valid here as well, with the understanding that $s = \sqq$ now represents the Mandelstam variable for the initial quark-antiquark pair (see fig.~\ref{S1DrellYan}). In the approximation of negligible quark masses when compared to the monopole mass $M$, the differential cross section reads:
\begin{align}\label{Xsecdeffspin1}
	\frac{\dd\sigma^{S=1}_{q\overline{q}\rightarrow M\overline{M}}}{\dd\Omega}&=\frac{5 \beta ^3 \alpha _e \alpha _g(\beta)}{288 (1-\beta ^2) M^2}\Bigg\{3\beta^4 (1-\cos^2 \theta)-\beta^2 \big[ 2\kappa^2(\cos^2 \theta+1)+8\kappa-4\cos^2 \theta+8 \big] \nonumber \\
				& +2\kappa^2(3-\cos^2 \theta)+8\kappa-\cos^2 \theta+5\Bigg\}.
\end{align}
Neglecting quark masses set $\beta_q=1$ so the index on the monopole boost $\beta_p$ has been dropped. Also, the $\beta$ dependance of the magnetic coupling is made apparent.

\begin{figure}[ht!]\centering
\includegraphics[width=0.95\textwidth]{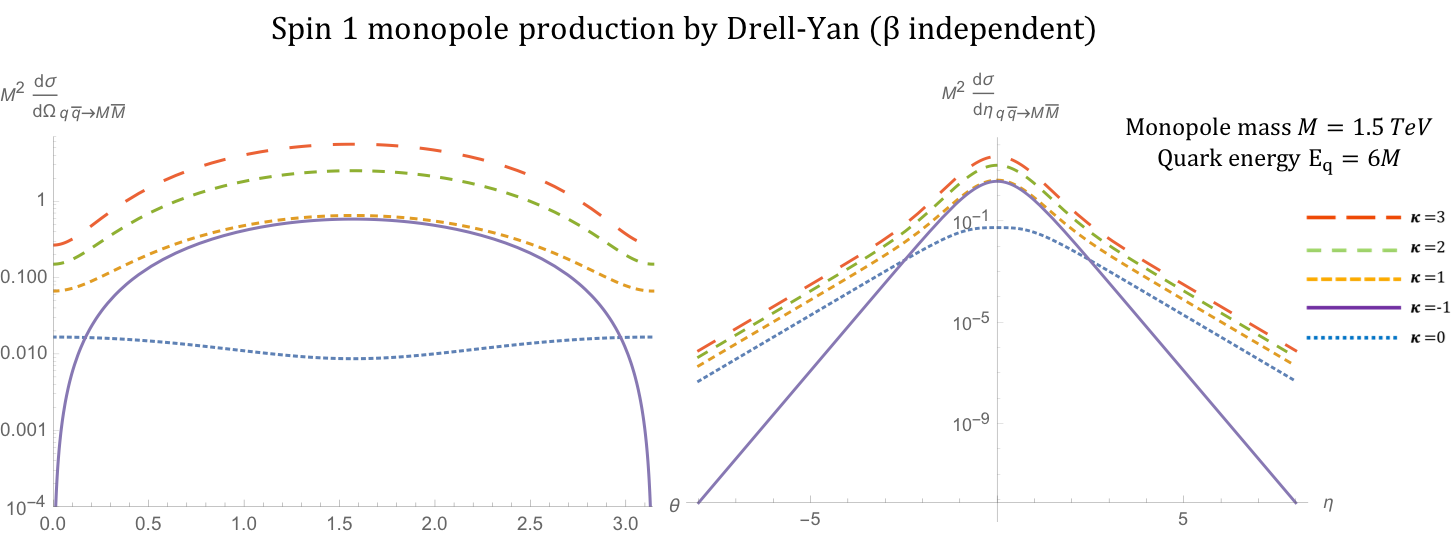}
\caption{Angular distributions for monopole-antimonopole pair production via DY for the case of a monopole with spin~\half and mass $M=1.5$~TeV as a function of the scattering angle $\theta$ (left) and the pseudorapidity $\eta$ (right) at $\sqrt{\sqq}=2E_q$, where $E_q=6M$, and for various values of the parameter $\kappa$. The case $\kappa=1$ represents the SM expectation for the DY pair production of spin-1 $W^{\pm}$ bosons.}\label{LeeYangDsigDomega1label}
\end{figure}

The kinematic distributions are plotted in fig.~\ref{LeeYangDsigDomega1label} for various values of the parameter $\kappa$ with $\beta_q=1$ as expected for massless quarks. As in previous spin models, a monopole mass of $M=1.5$~TeV and $\sqrt{\sqq}=2E_q$ with $E_q=6M$ is assumed. This parameter influences the shape of the distributions and the convergence of the cross section as $\sqq \rightarrow \infty$. Indeed, 
\begin{equation}\label{div1}
\frac{\dd\sigma^{S=1}_{q\overline{q}\rightarrow M\overline{M}}}{\dd\Omega} \quad \xrightarrow{\sqq\rightarrow \infty} \quad 
\begin{cases} 
\sqq &, \quad \kappa\neq 0, \nonumber \\
\dfrac{5 \alpha _e \alpha _g(\beta)}{288 M^2}(2 \cos^2 \theta +2) &,  \quad \kappa= 0,
\end{cases} 
\end{equation}
where the limit was taken such that $\beta^4\simeq 1$, $\beta^3\simeq 1$ and keeping $\beta^2= 1-\frac{4M^2}{\sqq}$. 

\begin{figure}[ht!]\centering
\includegraphics[width=0.55\textwidth]{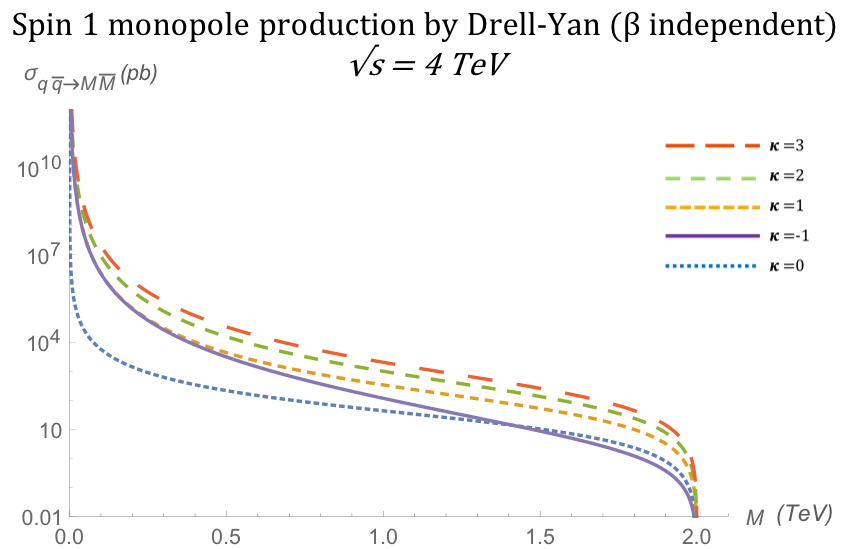}
\caption{Total cross section for the pair production of spin-1 monopoles via the DY process, as a function of the monopole mass $M$ for different values of $\kappa$ at $\sqrt{\sqq}=4$~TeV. }\label{LeeYangsig1label}
\end{figure}

The total cross section for arbitrary values of $\kappa$ is given by:
\begin{equation}\label{totxsecdyk}
 \sigma^{S=1}_{q\overline{q}\rightarrow M\overline{M}}=\frac{5 \pi \sqq \alpha _e \alpha _g(\beta)}{432 M^4}\Big(1-\frac{4M^2}{\sqq} \Big)^{\frac{3}{2}}\left[ 8\kappa^2-(4\kappa^2+12\kappa+10)  \Big(1-\frac{4M^2}{\sqq} \Big)+12\kappa+3\Big(1-\frac{4M^2}{\sqq} \Big)^2+7 \right],
\end{equation}
and it is plotted as a function of the monopole mass in fig.~\ref{LeeYangsig1label}. In the high energy limit, it becomes 
\begin{equation}\label{tcsdyuv}
\sigma^{S=1}_{q\overline{q}\rightarrow M\overline{M}} \quad \xrightarrow{\sqq\rightarrow \infty} \quad  \frac{5\pi\alpha_e\alpha_g(\beta)}{108M^4}\sqq\left(\kappa^2+(4\kappa^2+12\kappa+4)\frac{M^2}{\sqq}\right),
\end{equation}
which converges only for $\kappa=0$. In fig.~\ref{LeeYangsig12labeldy}, we observe the $\kappa$-dependent behaviour at $M \rightarrow 0$ as expected from \eqref{tcsdyuv}: the cross section diverges as $\sim M^{-2}$ for $\kappa=0$ and more rapidly as $\sim M^{-4}$ for other $\kappa$ values. 

The $\kappa=1$ case represents the scattering of monopoles within a totally renormalisable and unitary field theory without gauge fixing, as uniquely found in the SM. For this case, in harmony with the previous subsection, the explicit expressions for the kinematic distribution and the total cross section are given respectively in the following equations:
\begin{align}\label{dy1dsec}	\frac{\dd\sigma^{S=1}_{q\overline{q}\rightarrow M\overline{M}} }{\dd\Omega} &=\frac{5 \beta ^3 \alpha _e \alpha _g(\beta)}{288 (1-\beta ^2) M^2}\Bigg[3\beta^4\big(1-\cos^2 \theta\big)-\beta^2\big(18-2\cos^2 \theta\big)-3\cos^2 \theta+19\Bigg], \qquad \kappa=1, 
\end{align}
and 
\begin{equation}\label{dy1tcsec}
\sigma^{S=1}_{q\overline{q}\rightarrow M\overline{M}}=\frac{5 \pi \sqq \alpha _e \alpha _g(\beta)}{432 M^4}\beta^3\Big(3 \beta^4-26\beta^2+27\Big), \qquad \kappa=1.
\end{equation}
It should be noted that the unitarity of the $\kappa=1$ case becomes apparent only if other SM processes are included in the total amplitude, hence the divergence seen for $\kappa=1$ in the high energy limit (cf.\ \eqref{div1}, \eqref{tcsdyuv}) is not surprising. We also remark that the DY cross section increases by a factor of two when considering processes generated by high energy collisions of identical hadrons, e.g.\ protons to protons.

\subsection{Comparison of various spin models and production processes}\label{sec:comparison} 

\begin{figure}[ht!]
\includegraphics[width=\textwidth]{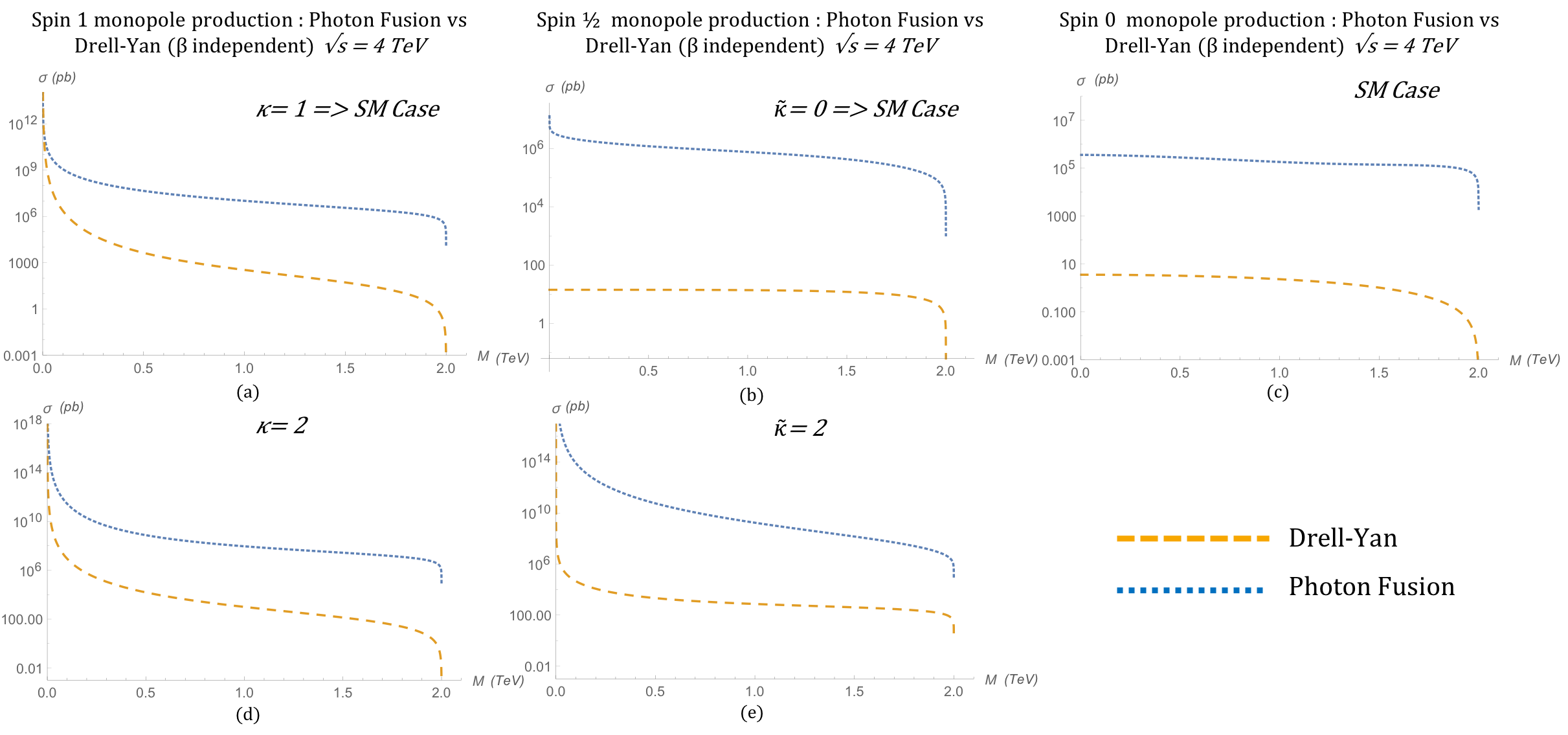}\centering
\caption{Comparison between the cross sections of monopole pair production for PF and DY processes varying with monopole mass $M$ at $\sqrt{s_{qq/\gamma\gamma}}=4$ TeV for $\beta$-independent coupling. (a) the spin-1 monopole in the SM-like case where $\kappa=1$; (b) the spin-\half monopole SM-like case for which $\tilde{\kappa}=0$; (c) the spin-0 monopole case, which does not have magnetic moment; (d) the spin-1 monopole cross section with $\kappa=2$; (e) the spin-\half monopole cross section with $\tilde{\kappa}=2$. }\label{PFvsDY}
\end{figure}

This section is brought to a close by briefly comparing the cross section distributions discussed in this work. Firstly, it is important to point out that the dominant production process is PF by a large margin at $\sqrt{s_{qq/\gamma\gamma}}=4$ TeV. This is seen in fig.~\ref{PFvsDY} for all spin models. In particular, graphs~\ref{PFvsDY}(a) and~\ref{PFvsDY}(b) show the SM-like cases for which $\kappa=1$ and $\tilde{\kappa}=0$ represent the $S=1$ and $S=\half$ SM-like cases, respectively. Graph~\ref{PFvsDY}(c) shows the spin-0 monopole case, the only one for which there is no magnetic moment. This corroborates the assertion in~\cite{dw}. Graphs~\ref{PFvsDY}(d) and~\ref{PFvsDY}(e) demonstrate that this behaviour is maintained for all values of $\kappa$ and $\tilde{\kappa}$ by choosing the non-distinctive value equal to two.

\begin{figure}[ht!]\centering
\includegraphics[scale=0.6]{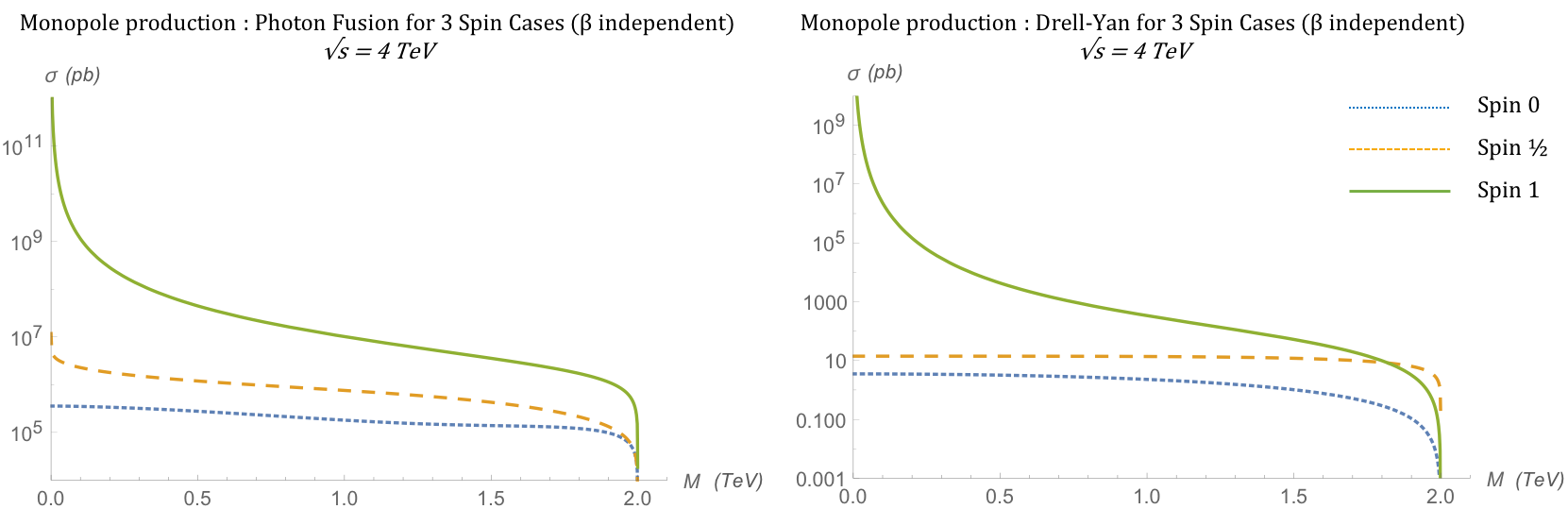}
\caption{Comparison of the cross section between all three spin models at $\sqrt{s_{qq/\gamma\gamma}}=4~{\rm TeV}$ varying with monopole mass $M$ for PF (left) and DY (right) and for $\beta$-independent coupling. In the $S=\half$ and $S=1$ cases, the SM values $\tilde{\kappa}=0$ and $\kappa=1$, respectively, are drawn, while there is no magnetic moment in the spin-0 case. }\label{Comparing3Spins}
\end{figure}

Secondly, it is apparent that the cross section for monopole production increases with the spin of the monopole most of the mass range, as observed in fig.~\ref{Comparing3Spins}, if the SM-like cases for the magnetic-moment parameters are chosen. This observation supports the findings of ref.~\cite{dw}. As shown in figs.~\ref{LeeYangsig12label} and~\ref{LeeYangsig12labeldy} for a fermionic monopole, the trend is maintained for $\tilde{\kappa}>0$ for all masses. For a vector monopole, on the other hand, the cross-section ordering is not consistent across the monopole mass for varying $\kappa$ values, as evident from figs.~\ref{LeeYangsig1pflabel}) and~\ref{LeeYangsig1label}. More discussion on the phenomenological implications of the magnetic-moment parameter will follow in section~\ref{kappaMad}, this time in the context of proton-proton collisions.

\subsection{Perturbatively consistent  limiting case of large $\kappa$ and small $\beta$} 
\label{kappaThe}

As discussed in section~\ref{intro}, the non-perturbative nature of the large magnetic Dirac charge of the monopole invalidate any perturbative treatment based on Drell-Yan calculations of the pertinent cross sections and hence any result based on the latter is only indicative, due to the lack of any other concrete theoretical treatment. This situation may be resolved if thermal production in heavy-ion collisions ---that does not rely on perturbation theory--- is considered~\cite{affleck,arttu}. Another approach is discussed here involving  a specific limit of the parameters $\kappa$ and $\beta$ of the effective models of vector and spinor monopoles, used above, in the case of a \emph{velocity-dependent magnetic charge} \eqref{etogb}. In this limit, the perturbative truncation of the monopole pair production processes, described by the Feynman-like graphs of figs.~\ref{spinhalfgraphs},~\ref{FeynSpinor},~\ref{FeynGraphsFig} and~\ref{S1DrellYan}, becomes meaningful provided the monopoles are slowly moving, that is $\beta \ll 1$.  In terms of the centre-of-mass energy $\sqrt{s_{\gamma\gamma/qq}}$, such a condition on $\beta$ implies, on account of eq.~\eqref{btos}, that the monopole mass is around $2M \simeq \sqrt{s_{\gamma\gamma/qq}} + \mathcal O (\beta^2)$. It should be noted at this point that, in collider production of monopole-antimonopole pairs considered in this work, $s_{\gamma\gamma/qq}$ is not definite but follows a distribution, according to the parton (or photon) distribution function (PDF) for the DY or PF processes.  

In the absence of a magnetic-moment parameter, $\kappa$, or for the unitary value $\kappa=1$ in the case of spin-1 monopoles studied in \cite{original}, the condition $\beta \ll 1$ would lead to strong suppression of the pertinent cross sections beyond the current experimental sensitivities, thereby rendering the limit $\beta \to 0$ experimentally irrelevant for placing bounds on monopole masses or magnetic charges.
Indeed, the various total cross sections discussed so far behave as follows, when $\beta \to 0$ (using the definition \eqref{etogb} of the magnetic fine structure constant):
\begin{alignat}{5}\label{reproRusakovich1b}
 \sigma_{\gamma\gamma\rightarrow M\overline{M}}^{S=1, \kappa=1} \quad  & \stackrel{\beta \to 0}{\simeq} && \quad 
\frac{19\, g^4}{8\pi\, s}\, \beta^5 \quad && \xrightarrow{\beta \to 0} && \quad 0 \qquad \text{(spin-1~PF)},  \nonumber \\
 \sigma^{S=1, \kappa=1}_{q\overline{q}\rightarrow M\overline{M}}   \quad  & \stackrel{\beta \to 0}{\simeq} && \quad 
\frac{135 \,  s \, \alpha _e \, g^2}{1728\,  M^4}\beta^5 \quad && \xrightarrow{\beta \to 0} && \quad 0 \qquad \text{(spin-1~DY)}, \nonumber \\
 \sigma_{\gamma\gamma\rightarrow M\overline{M}}^{S=\frac{1}{2}, \kappa=0} \quad  & \stackrel{\beta \to 0}{\simeq} && \quad  \frac{g^4}{4\pi\, s} \, \beta^5 
\quad && \xrightarrow{\beta \to 0} && \quad 0 \qquad \text{(spin-\half~PF)}, \nonumber \\
  \sigma^{S=\frac{1}{2}, \kappa=0}_{q\overline{q}\rightarrow M\overline{M}}   \quad  & \stackrel{\beta \to 0}{\simeq} && \quad  \frac{5\, \alpha_e \, g^2}{18\,  \, s}\, \beta^3 \quad && \xrightarrow{\beta \to 0} && \quad 0 \qquad \text{(spin-\half~DY)}, \nonumber \\
 \sigma^{S=0}_{\gamma\gamma\rightarrow \overline{M}M} \quad  & \stackrel{\beta \to 0}{\simeq} && \quad \frac{g^4}{4\pi\, s}\, \beta^5 \, && \xrightarrow{\beta \to 0} && \quad  0  \qquad \text{(spin-0~PF)},
\nonumber \\
  \sigma_{q\overline{q}\rightarrow M\overline{M}}^{S=0}  \quad & \stackrel{\beta \to 0}{\simeq} && \quad 
  \frac{5\,\alpha_{e}\, g^2}{108 \, s}\, \beta^5 \quad && \xrightarrow{\beta \to 0} && \quad  0  \qquad \text{(spin-0~DY)}. 
\end{alignat}

However, in the case of non-trivial and \emph{large} (dimensionless) magnetic-moment-related parameters $\kappa, \tilde \kappa$, relevant for the cases of vector and spinor monopoles, the situation changes drastically, as we shall now argue. 
To this end, we consider the limits 
\begin{equation}\label{limit}
\kappa \gg 1, \quad \tilde \kappa \gg 1, \quad \beta \ll 1, 
\end{equation}
with $\tilde \kappa$ defined in \eqref{ktilde},
but in such a way that the strength of the derivative magnetic-moment couplings, given in eq.~\eqref{FeynRuleS12eq} (see fig.~\ref{FeynRuleS12}) for spin-\half monopoles and on the left side of \eqref{spin1feyngraphs} (see fig.~\ref{FeynRulesFig}) for spin-1 monopoles is perturbatively small. Since the magnitude of the monopole momentum is proportional to $M \beta$, one expects that the condition of a perturbatively small derivative coupling for magnetic moment is guaranteed if, by order of magnitude, one has:
\begin{equation}\label{kbg}
g  \kappa^\prime  \beta^2  < 1  ~, 
\end{equation}
where $\kappa^\prime = \tilde\kappa$ for spin-\half monopole and $\kappa^\prime = \kappa$ for spin-1 monopole. 

For the spin-\half monopole, from \eqref{totxsec12} and \eqref{thisone}, one observes that in the limit \eqref{limit} and respecting \eqref{kbg}, upon requiring the absence of infrared divergences in the cross sections as $\beta \to 0$, and postulating that:
\begin{equation}\label{ktildelim}
(\tilde \kappa \beta g)^4  \beta \xlongequal{\stackrel{\beta \to 0}{\kappa \to \infty}}  |c^\prime_1|, \qquad c^\prime_1 = \text{finite constant},
\end{equation}
so that \eqref{kbg} is trivially satisfied, since $\tilde \kappa g \beta^2   = |c^\prime_1|^{\frac{1}{4}}  \beta^{\frac{3}{4}} \, \xrightarrow{\stackrel{\beta \to 0}{\kappa \to \infty}} 0$,
the dominant contributions to the PF and DY total cross sections  are   
given by 
\begin{equation}\label{totsec12lim}
\begin{split}
 \sigma^{S=\frac{1}{2}}_{\gamma\gamma\rightarrow M\overline{M}} &\sim\pi  \alpha_g^2(\beta) \beta \kappa^4 s = \frac{(\tilde \kappa \, g \, \beta)^4  \, \beta }{16 \pi M^4} s \xlongequal{\stackrel{{\beta \to 0}}{\tilde \kappa \to \infty}} \text{finite},
\end{split}
\end{equation}
and
\begin{equation}\label{totsecdy12lim}
\sigma^{S=\frac{1}{2}}_{q\overline{q}\rightarrow M\overline{M}}\sim\pi\alpha_e \alpha_g(\beta) \, \frac{10\beta\kappa^2}{9} = \frac{5 \alpha_e}{18M^2}  \, 
(\tilde \kappa \beta  g )^2 \beta   \xrightarrow{\stackrel{\beta \to 0}{\kappa \to \infty}} \, 0,
\end{equation}
respectively, where we used \eqref{etogb} and \eqref{ktilde}. Hence for slowly-moving spinor-monopoles, with velocity-dependent magnetic charge, and large magnetic moment parameters, it is the PF cross section which is the dominant one of relevance to collider experiments.

Similar results characterise the spin-1 monopole. Indeed, we observe that in the limit \eqref{limit}, \eqref{kbg}, the dominant contributions to the total cross sections for the PF (see \eqref{totxsec1k})  and DY (see \eqref{totxsecdyk}) processes are such that
\begin{equation}\label{totxsec1limit}
\begin{split}
 \sigma^{S=1}_{\gamma\gamma\rightarrow M\overline{M}}&\sim\pi  \alpha_g^2 \, \frac{29 \beta^5 \, \kappa^4}{4 s} = \frac{29}{64\pi s} \beta\,  \big(\kappa \beta g \big)^4 ,
\end{split}
\end{equation}
and 
\begin{equation}\label{dytotsec1lim}
\sigma^{S=1}_{q\overline{q}\rightarrow M\overline{M}}=\alpha _e \alpha _g(\beta)\pi \frac{40\beta^3\kappa^2}{27s} 
= \alpha_e \, \frac{10}{27s} \big(\kappa \beta g \big)^2  \beta^3,
\end{equation}
respectively, where we used \eqref{etogb}. 

We can see that, by requiring the \emph{absence of infrared} ($\beta \to 0$) \emph{divergences} in the total cross sections, one may consistently arrange that
the PF cross section \eqref{totxsec1limit} acquires a non-zero (finite) value as $\beta \to 0$, whilst the DY cross section \eqref{dytotsec1lim} vanishes in this limit:
\begin{eqnarray}\label{limits}
(\kappa \beta  g)^4  \beta & \xlongequal{\stackrel{\beta \to 0}{\kappa \to \infty}} &  |c_1|, \qquad c_1 = \text{finite constant}, \nonumber \\
\sigma^{S=1}_{\gamma\gamma\rightarrow M\overline{M}} 
& \xlongequal{\stackrel{\beta \to 0}{\kappa \to \infty}} &
\frac{29\, c_1}{64\, \pi \, s} ,
\nonumber \\
\sigma^{S=1}_{q\overline{q}\rightarrow M\overline{M}} 
& \xlongequal{\stackrel{\beta \to 0}{\kappa \to \infty}}&
\alpha_e \, \frac{10\, \sqrt{|c_1|}}{27\,s} \, \beta ^{\frac{5}{2}} \quad \xrightarrow{\stackrel{\beta \to 0}{\kappa \to \infty}}  \quad 0.
\end{eqnarray}
In such a limit, the quantity $\kappa g \beta^2  = |c_1|^{\frac{1}{4}} \, \beta^{\frac{3}{4}} \, \xrightarrow{\stackrel{\beta \to 0}{\kappa \to \infty}} \, 0$, so \eqref{kbg} is trivially satisfied, and thus  the perturbative nature of the magnetic moment coupling is guaranteed. 
Hence in this limiting case of velocity-dependent magnetic charge, large magnetic moment couplings and slowly moving vector monopoles, again the PF cross section is the dominant one relevant to searches in current colliders and can be relatively large (depending on the value of the phenomenological parameter $c_1$). This argument is successfully tested with simulated events in section~\ref{kappaMad}.

\section{\MAD Implementation \label{sec:mad}}

The \MAD generator~\cite{MG} is used to simulate the generation of monopoles. In this section, we briefly present the development of the \MAD Universal FeynRules Output (UFO) model~\cite{ufo} used to simulate different production mechanisms of monopole. This includes both the monopole velocity ($\beta$) dependent and independent photon-monopole-monopole coupling. Three different spin cases have been included:  spins 0, \half and 1.

\subsection{Monopole couplings}
\label{coupling}
In Dirac's model, the relation between the elementary electric charge $q_e$ and the basic magnetic charge $g$ is given in eq.~\eqref{diracrule2} in Gaussian units. However in \MAD, \Heav units are used, where \eqref{diracrule2} becomes
\begin{equation}
 q_e  g = 2\pi n \label{heavlo}, \quad n\in {\mathbb Z}.
\end{equation}
Hence, the unit of the magnetic charge is
\begin{equation}
 \gd =2\pi/q_e. \label{gd}
\end{equation}

The electromagnetic vertex in \Heav units simply becomes
\begin{equation}
 c_{em} = q_e. \label{cem}
\end{equation}

Similarly, the monopole-photon vertex becomes
\begin{equation}
 c_{mm} = g. \label{cmm}
\end{equation}

In \Heav units, electric charge $q_e$ is given by $\sqrt{4\pi\alpha}$ where $\alpha$ is the fine-structure constant. Hence, equation \eqref{gd} turns out to be 
\begin{equation}
 \gd = 2\pi/\sqrt{4\pi\alpha} = \sqrt{\pi/\alpha}. \label{gd2}
\end{equation}

In \eqref{gd2}, the monopole velocity $\beta$ is not used. But if we want to consider the monopole velocity dependent coupling, the value of \gd simply becomes $\beta\sqrt{\pi/\alpha}$. The velocity $\beta$ is expressed in~\eqref{btos} as a function of monopole mass $M$.

\subsection{Implementation of the monopole Lagrangians in $\textsc{MadGraph}$ \label{Implementation}}

In this section, the practical details on the use of \MAD~\cite{MG} to simulate the photon-fusion production mechanism of monopoles is described. The \MAD was downloaded and installed following the instructions given in~\cite{madgraphIn}. The general procedure to simulate a model with the help of \MAD is the following:
\begin{enumerate}
 \item Create a model\footnote{The code of the model will be publicly available in the \MAD web page~\cite{madgraphIn}.} with all the user defined fields, parameters and interactions. Lately, the use of the UFO format~\cite{ufo} is strongly encouraged for such models.
 \item In the \MAD command prompt, import that model.
 \item Generate the process which will be simulated using the {\tt generate} command and create an output folder.
 \item Fix the centre-of-mass energy, colliding particles, parton distribution functions in the run card.
 \item Fix the parameters (electric and magnetic charges, masses, etc) of the colliding and generated particles in the parameter card. 
 \item Launch the output folder in order to compile the model and create the Les Houches Event (LHE) files~\cite{lhe}. 
 \item These LHE files will be used to produce simulated results.
 \end{enumerate}

\subsection{Generating and validating the UFO models}\label{ufomodel}

To generate a UFO model from the Lagrangian, \Feyn~\cite{feyns}, an interface to \Math, was utilised. Here the parameters of a model (mass of a particle, spin, electric charge, magnetic charge, coupling constant, fermionic or bosonic field, etc) and the corresponding Lagrangians are written in a text file. From the Lagrangian, the UFO model is generated with the help of \Feyn.

The velocity $\beta$, defined in eq.~\eqref{btos}, is defined as a form factor in the generated UFO models. The instructions given in~\cite{formf}, specially the `Method 2: Fortran Way' were followed to get a proper form factor. To get the value of $\hat{s}$ inside the Fortran function, we used this formula (for elastic collision):
\begin{equation}
 \hat{s} = 2(P_1\cdot P_2)\label{shat}
\end{equation}
where $P_1$ and $P_2$ are the 4-momenta of the two colliding particles. 

For scalar monopoles, the inclusion in the simulation of the four-particle vertex shown in fig.~\ref{fig:seagull} in addition to the $u$- and $t$-channel, shown in  figs.~\ref{fig:uchan} and~\ref{fig:tchan} respectively, led to the necessary use of UFO model written as a \pyt object and abandon the rather older method in \fort code. The implementation of the four-vertex diagram proved to be non-trivial due to the $g^2$ coupling. The Lagrangian, which takes the form given in \eqref{spinzero}, is rewritten in a \Math format so that \Feyn can understand the variables. We created the text file containing all the information related to the field, mass, spin and charge, etc, following the instructions given in ref.~\cite{feyn2}.

To validate the \MAD UFO model for monopoles, we compare the cross sections for the photon-fusion process from the theoretical calculation derived in section~\ref{sec:spin} to those obtained from simulation. Since the theoretical calculations consider bare photon-to-photon scattering, we chose in \MAD the no-PDF option, i.e.\ we assume direct $\gamma\gamma$ collisions at $\sqrt{\sgg}=13~{\rm TeV}$. Also, the coupling used here is assumed to depend on $\beta$.

The cross-section values for spin-0 monopoles are shown in the first columns of table~\ref{tabcomb}. The UFO-model-over-theory ratio values, also shown in the fourth column of the table, are very close to unity. This clearly shows the validity of spin-0 monopole UFO model.

\begin{table}[ht!]
\begin{center}
\begin{tabular}{| c | c | c | c | c | c | c | c | c | c |}
\hline \hline
\multirow{3}{1cm}{ Mass (GeV)} & \multicolumn{3}{c|}{ Spin 0} & \multicolumn{3}{c|}{ Spin \half} & \multicolumn{3}{c|}{ Spin 1} \\
	\cline{2-10}
	 & \multicolumn{2}{c|}{$\ga,\;\;\sigma~{\rm (pb)}$}     &  Ratio     & \multicolumn{2}{c|}{$\ga,\;\;\sigma~{\rm (pb)}$}     &  Ratio    & \multicolumn{2}{c|}{$\ga,\;\;\sigma~{\rm (pb)}$}     &  Ratio         \\
	 \cline{2-3} \cline{5-6}  \cline{8-9}
           & UFO model                   & Theory                       &  UFO/th. & UFO model           & Theory            &  UFO/th. & UFO model                   & Theory                       &  UFO/th. \\
\hline
\hline           
1000 & $1.4493\times10^{4}$ & $1.4336\times10^{4}$ & $0.99$ & $1.364\times10^5$    & $1.358 \times 10^{5}$ & $1.004$ & $1.078\times10^{7}$ & $1.0781\times10^{7}$ & $0.999$ \\ 
2000 & $9.851\times10^{3}$ & $9.791\times10^{3}$ & $1.006$ & $8.341\times10^{4}$ & $8.2551\times10^{4}$ & $1.010$ & $2.277\times10^{6}$ & $2.2520\times10^{6}$ & $1.011$  \\ 
3000 & $5.685\times10^{3}$ & $5.640\times10^{3}$ & $1.007$ & $4.803\times10^{4}$ & $4.7554\times10^{4}$ & $1.010$ & $7.214\times10^{5}$ & $7.1290\times10^5$ & $1.012$    \\ 
4000 & $2847$ & $2810.5$ & $1.013$ & $2.251\times10^{4}$ & $2.2156\times10^{4}$ & $1.012$ & $2.275\times10^{5}$ & $2.2523\times10^5$ & $1.010$     \\ 
5000 & $1094$ & $1087$ & $1.006$ & $6362$       & $6331$ & $1.005$ & $5.256\times10^{4}$ & $5.1833\times10^4$ & $1.014$     \\ 
6000 & $117.8$ & $116.53$ & $1.011$ & $370$     & $365.5$ & $1.012$ & $3.034\times10^{3}$ & $3.014\times10^3$ & $1.007$  \\ 
\hline
\hline
\end{tabular}
\end{center}
\caption{Cross-section values obtained from theoretical calculations and from the \MAD UFO model at $\sqrt{\sgg}=13~{\rm TeV}$ without PDF for monopoles of spin 0, \half, 1 and a  $\beta$-dependent coupling through the photon-fusion production mechanism. The ratios simulation/theory prediction are also listed.}
\label{tabcomb}
\end{table}

In a similar fashion, spin-\half monopole Lagrangians~\eqref{12lag} are also rewritten in a \Math format. The magnetic-moment parameter $\tilde\kappa$ is also implemented in the model. No additional diagram was added to the $u/t$-channels already described in the UFO model. Again, the cross sections from theoretical calculations and \MAD UFO models (for no PDF) for spin-\half monopoles were compared, are shown in table~\ref{tabcomb}. The comparison clearly shows the validity of the \MAD UFO model for spin-\half monopoles.

Finally, the Lagrangian~\eqref{GaugeInvLag} for spin-1 monopoles is also written in \Math code. The possibility for choosing the value of the $\kappa$ parameter in~\eqref{GaugeInvLag} exists, yet for validation purposes, the value of $\xi$ is taken to be zero and the value of $\kappa$ is taken to be one. The cross sections for spin-1 monopoles from the theoretical calculations and \MAD UFO models (for no PDF), shown in table~\ref{tabcomb}, match which satisfactorily proved the validity of the \MAD UFO model for the spin-1 monopole. 

Apart from the photon fusion production mechanism, the UFO models were also tested for the DY production mechanism as well. The DY process for monopoles was already implemented in \MAD both for $\beta$-dependent and $\beta$-independent coupling using a \fort code setup for spin-0 and spin-\half monopoles. This setup was utilised by ATLAS~\cite{atlasmono1,atlasmono2} and MoEDAL~\cite{moedal,moedalplb} to interpret their search results in terms of monopoles under these assumptions. These \fort setups were rewritten in the context of this work as UFO models, following their PF counterparts, and they were extended to include the spin-1 case. The latter was used in the latest MoEDAL monopole-search analysis~\cite{moedalplb}. After validating the UFO models against their \fort implementations for scalar and fermionic monopoles, all spin cases were confronted by the theoretical predictions given in section~\ref{sec:spin}. Here again the \MAD UFO models produced satisfactorily close cross-section values with that predicted by the theory, as shown in table~\ref{tabDYcomb}.
\begin{table}[ht!]
\begin{center}
\begin{tabular}{| c | c | c | c | c | c | c | c | c | c |}
\hline \hline
\multirow{3}{1cm}{ Mass (GeV)} & \multicolumn{3}{c|}{ Spin 0} & \multicolumn{3}{c|}{ Spin \half} & \multicolumn{3}{c|}{ Spin 1} \\
	\cline{2-10}
	 & \multicolumn{2}{c|}{$\dy,\;\;\sigma~{\rm (pb)}$}     &  Ratio     & \multicolumn{2}{c|}{$\dy,\;\;\sigma~{\rm (pb)}$}     &  Ratio    & \multicolumn{2}{c|}{$\dy,\;\;\sigma~{\rm (pb)}$}     &  Ratio         \\
	 \cline{2-3} \cline{5-6}  \cline{8-9}
           & UFO model                   & Theory                       &  UFO/th. & UFO model           & Theory            &  UFO/th. & UFO model                   & Theory                       &  UFO/th. \\
            \hline
           \hline
1000 & 0.4223 & 0.4184 & 1.009 & 1.747 & 1.735 & 1.007 & 3362 & 3343.05 & 1.006 \\
2000 & 0.3484 & 0.3465 & 1.005 & 1.614 & 1.603 & 1.007 & 230.6 & 228.872 & 1.007 \\
3000 & 0.2463 & 0.2441 & 1.009 & 1.373 & 1.373 & 1.000 & 45.43 & 45.173 & 1.006 \\
4000 & 0.1361 & 0.1352 & 1.007 & 1.039 & 1.0352 & 1.004 & 11.38 & 11.3162 & 1.006 \\
5000 & 0.04724 & 0.0473 & 0.999 & 0.6029 & 0.601 & 1.003 & 2.299 & 2.282 & 1.007 \\
6000 & 0.003745 & 0.00373 & 1.004 & 0.1454 & 0.1442 & 1.008 & 0.1206 & 0.1196 & 1.008 \\
\hline
\hline
\end{tabular}
\end{center}
\caption{Cross-section values obtained from theoretical calculations and from the \MAD UFO model at $\sqrt{\sgg}=13~{\rm TeV}$ without PDF for monopoles of spin 0, \half, 1 and a  $\beta$-dependent coupling through the Drell-Yan production mechanism. The ratios simulation/theory prediction are also listed.}
\label{tabDYcomb}
\end{table}

\section{LHC phenomenology \label{sec:lhc}}

\subsection{Comparison between the photon fusion and the Drell-Yan production mechanisms}\label{sec:kinematics}

Apart from the total cross sections, it is important to study the angular distributions of the generated monopoles. This is of great interest to the interpretation of the monopole searches in collider experiments, given that the geometrical acceptance and efficiency of the detectors is not uniform as a function of the solid angle around the interaction point. The kinematic distributions for the direct $\gamma\gamma$ and $q\bar{q}$ scattering obtained with the UFO models were also compared against the calculated differential cross sections of section~\ref{sec:spin} and showed good agreement with respect to the pseudorapidity, $\eta$, and the transverse momentum, \pt, of the monopole. It is worth noting that the differential cross sections in section~\ref{sec:spin} are plotted for a specific value of $\beta\simeq0.986$, while in this section we consider a range of monopole velocities connected to the ratio of the selected monopole mass over the proton-proton collision energy, thus some differences in the PF-vs-DY comparison are expected. 
 
Here the kinematic distributions are compared between the photon-fusion ($\gamma\gamma$) and the Drell-Yan mechanisms. For this purpose, the $\beta$-dependent UFO monopole model was used in \MAD. Monopole-antimonopole pair events have been generated for proton-proton collisions at $\sqrt{s}=13$~ TeV, i.e.\ for the LHC Run-2 operating energy. The PDF was set to \texttt{NNPDF23}~\cite{nnpdf} at LO for the Drell-Yan and \texttt{LUXqed}~\cite{lux} for the photon-fusion mechanism. The latter choice is made due to the relatively small uncertainty in the photon distribution function in the proton provided by \texttt{LUXqed}~\cite{lux2}. The monopole magnetic charge is set to 1~\gd, yet the kinematic spectra are insensitive to this parameter. The distributions are normalised to the same number of events, in order to facilitate the shape comparison.

\begin{figure}[ht!]
\justify
\includegraphics[width=0.33\textwidth]{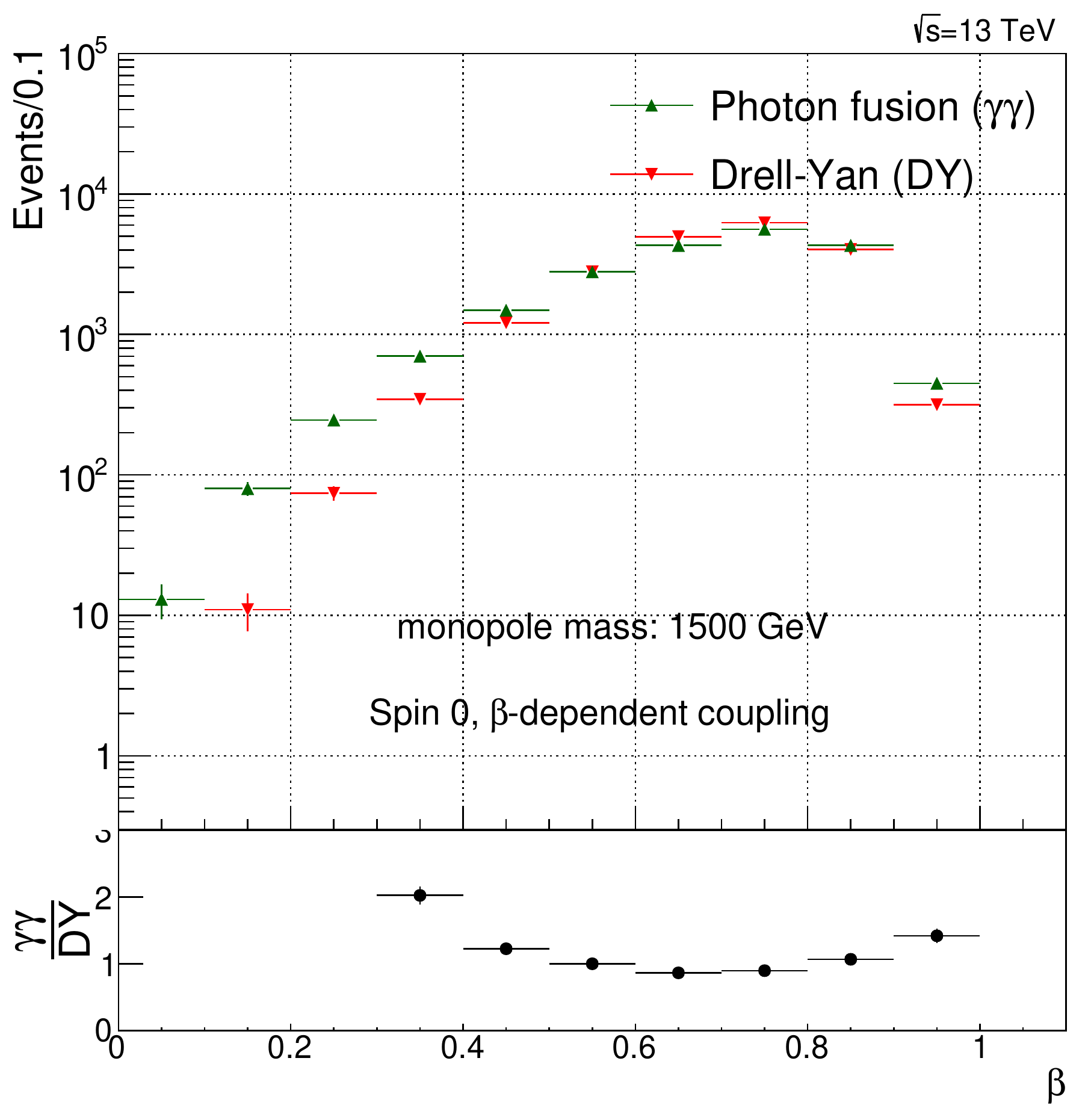}\hfill
\includegraphics[width=0.33\textwidth]{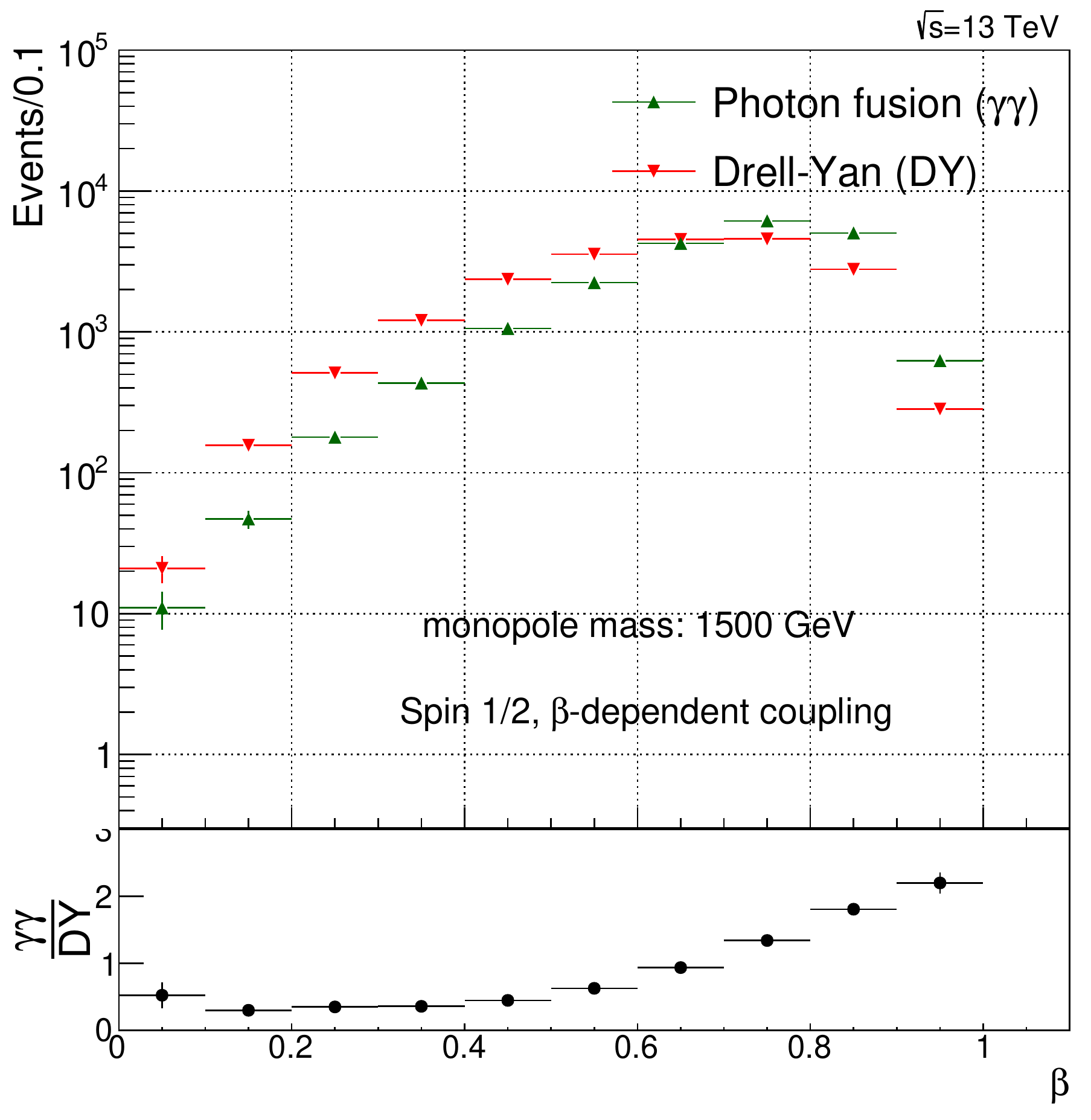}\hfill
\includegraphics[width=0.33\textwidth]{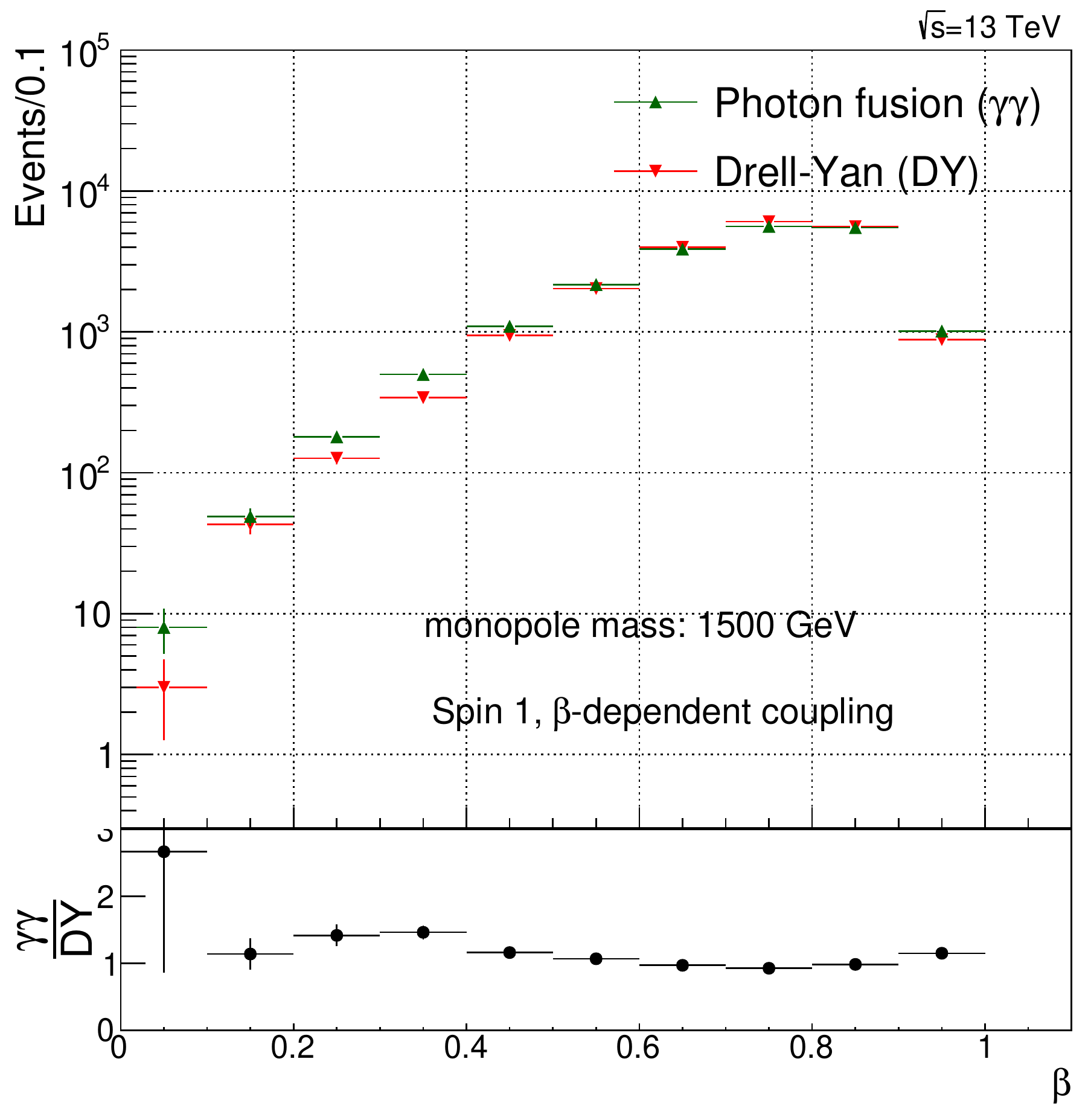}
\caption{The monopole velocity $\beta$ distributions of spin-0 (left), spin-\half (centre) and spin-1 monopoles (right) for both the photon-fusion and the Drell-Yan mechanisms at $\sqrt{s}=13$~TeV. The \texttt{LUXqed} and \texttt{NNPDF23} PDFs were used for the PF and the DY process, respectively.  }\label{fig:beta}
\end{figure}

The distributions of the monopole velocity, an important parameter for the detection of monopoles in experiments such as MoEDAL~\cite{moedal-review}, are depicted in fig.~\ref{fig:beta}. The velocity $\beta$, which is calculated in the centre-of-mass frame of the colliding protons, largely depends on the PDF of the photon (quark/antiquark) in the proton for the photon-fusion (Drell-Yan) process. For scalar monopoles, fig.~\ref{fig:beta} (left) shows that slower-moving monopoles are expected for PF than DY, an observation favourable for the discovery potential of MoEDAL NTDs, the latter being sensitive to low-$\beta$ monopoles. The comparison is reversed for fermionic monopoles, where PF yields faster monopoles than DY (fig.~\ref{fig:beta} (centre)). Last, as deduced from fig.~\ref{fig:beta} (right), the $\beta$ distributions for PF and DY are very similar. It should be remarked that the plotted $\beta$ expresses the monopole boost in the laboratory frame and, in the context of proton-proton collisions, it is distinct from the parameter $\beta$ defined in \eqref{btos} that enters into the monopole-photon coupling.

\begin{figure}[ht!]
\justify
\includegraphics[width=0.33\textwidth]{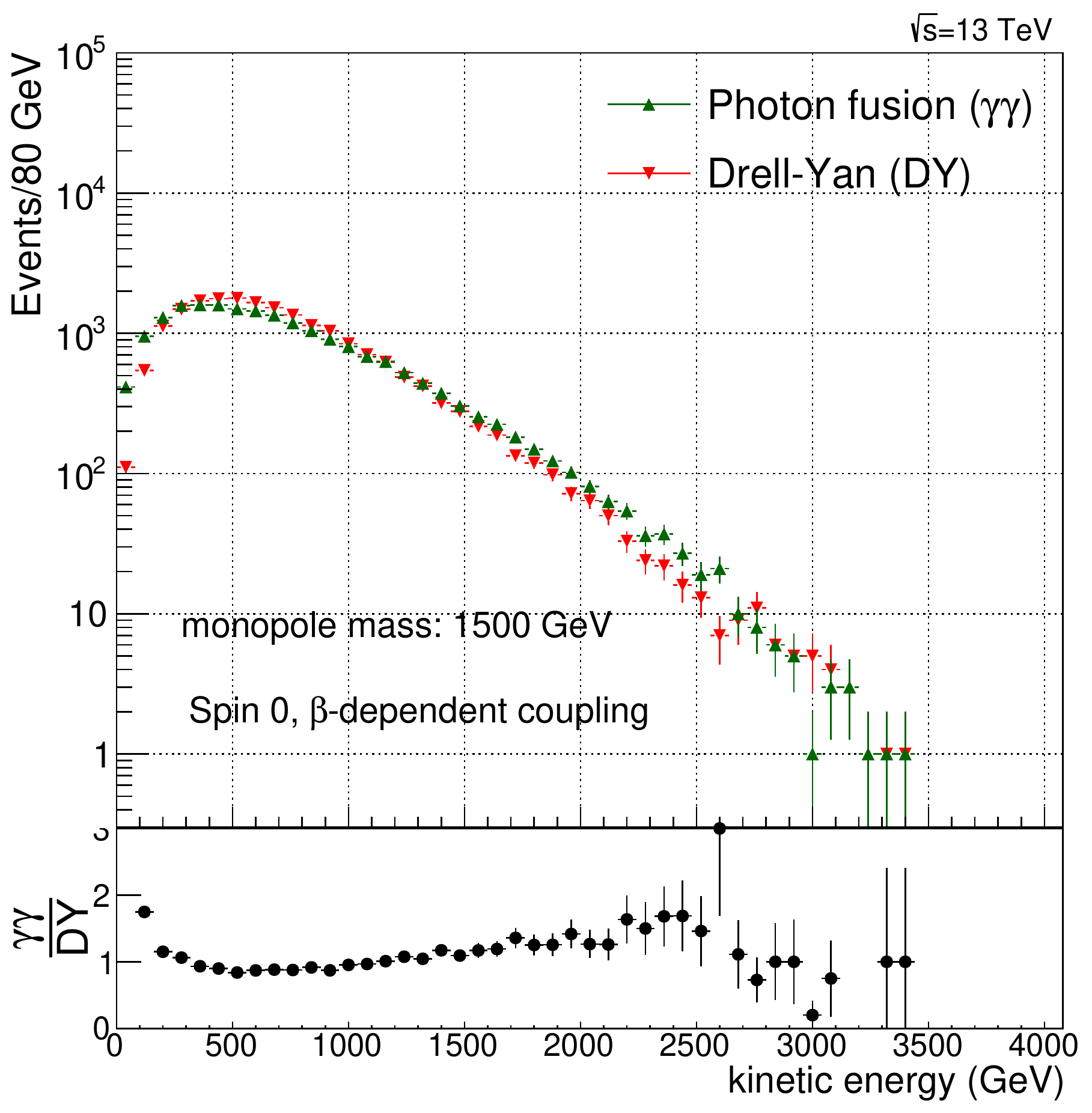}\hfill
\includegraphics[width=0.33\textwidth]{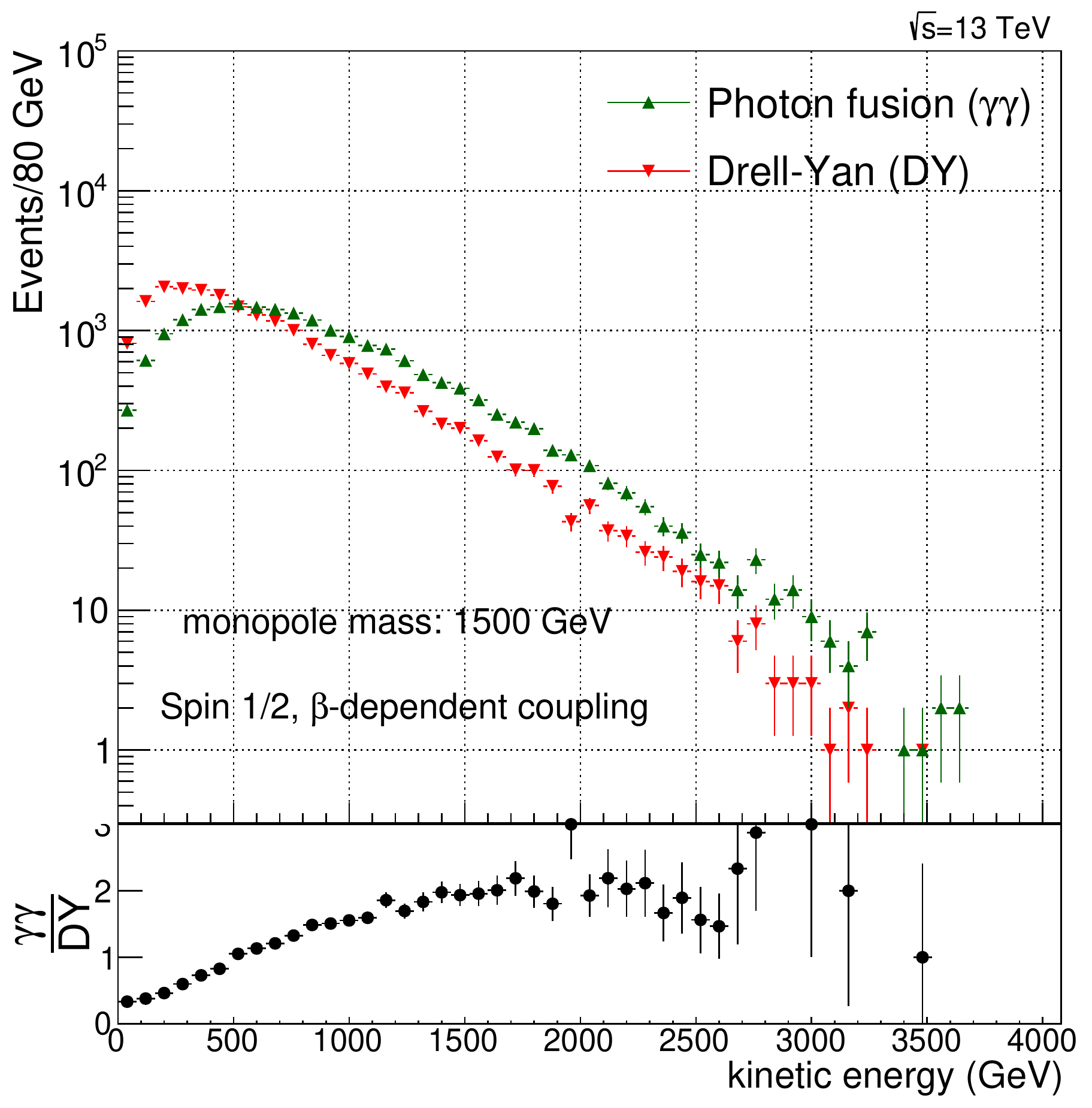}\hfill
\includegraphics[width=0.33\textwidth]{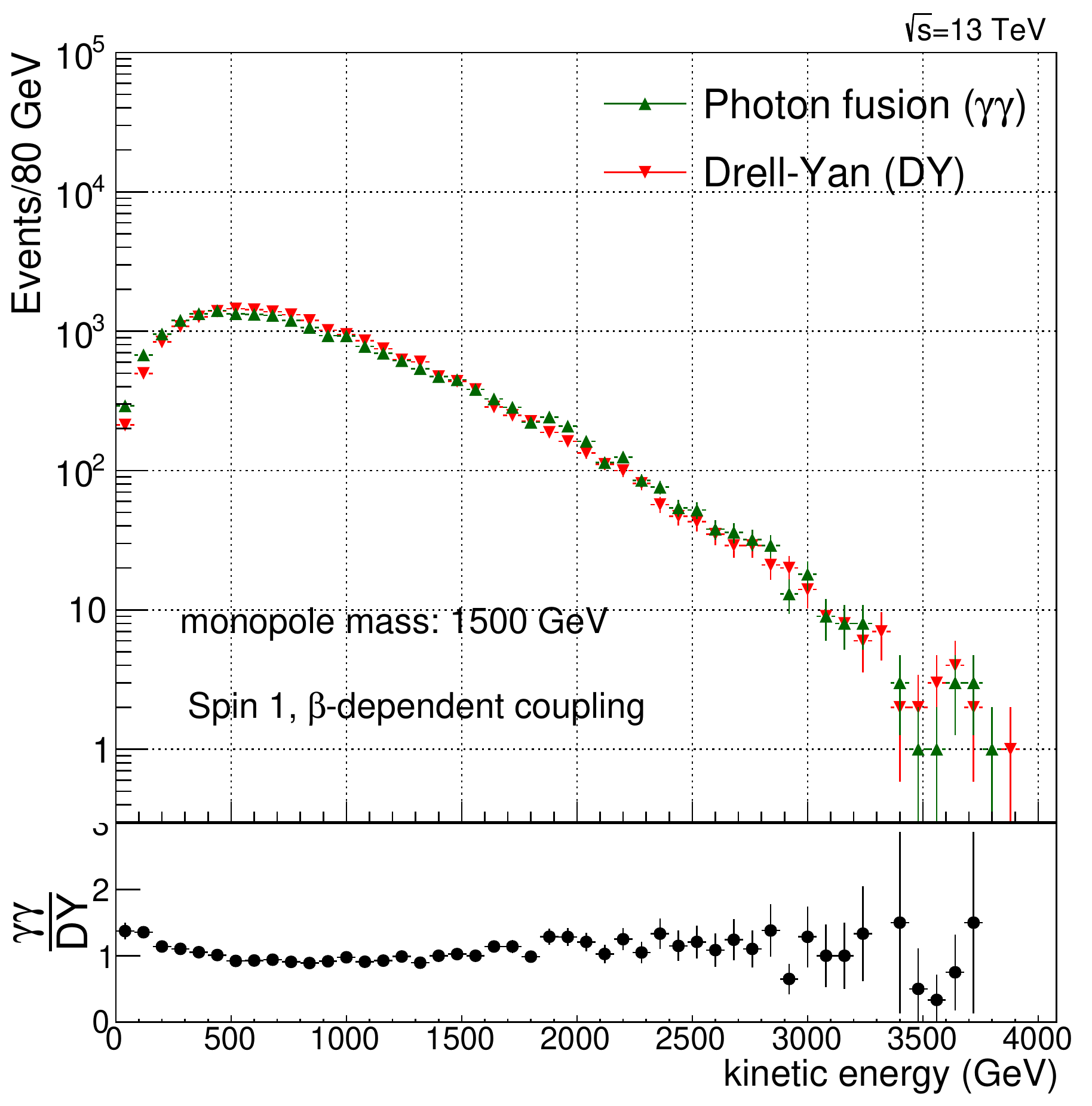}
\caption{The monopole kinetic energy distributions of spin-0 (left), spin-\half (centre) and spin-1 monopoles (right) for both the photon-fusion and the Drell-Yan mechanisms at $\sqrt{s}=13$~TeV and for $\tilde{\kappa}=0$. The \texttt{LUXqed} and \texttt{NNPDF23} PDFs were used for the PF and the DY process, respectively.  }\label{fig:ke}
\end{figure}

The kinetic energy\footnote{In the context of this work, the kinetic energy of a particle is defined as the scalar difference of its total energy and its mass.} spectra are shown in fig.~\ref{fig:ke}. We choose to show distributions of the kinetic energy because it is relevant for the monopole energy loss in the detector material, hence important for the detection efficiency. The kinetic-energy spectrum is slightly softer for PF than DY for scalar (left panel) and vector (right panel) monopoles, whereas it is significantly harder for fermions (central panel). This difference may be also due to the four-vertex diagram included in the bosonic monopole case. This observation is in agreement with the one made for $\beta$ previously.  We have also compared \MAD predictions for kinetic-energy and \pt distributions between with- and without-PDF cases, the latter also against analytical calculations (cf.\ section~\ref{sec:spin}) across different spins and production mechanisms. As expected, some features seen in the direct $\gamma\gamma$ or $q\bar{q}$ production are attenuated in the $pp$ production due to the sampling of different $\beta$ values in the latter as opposed to the fixed value in the former. 

\begin{figure}[ht!]
\justify
\includegraphics[width=0.33\textwidth]{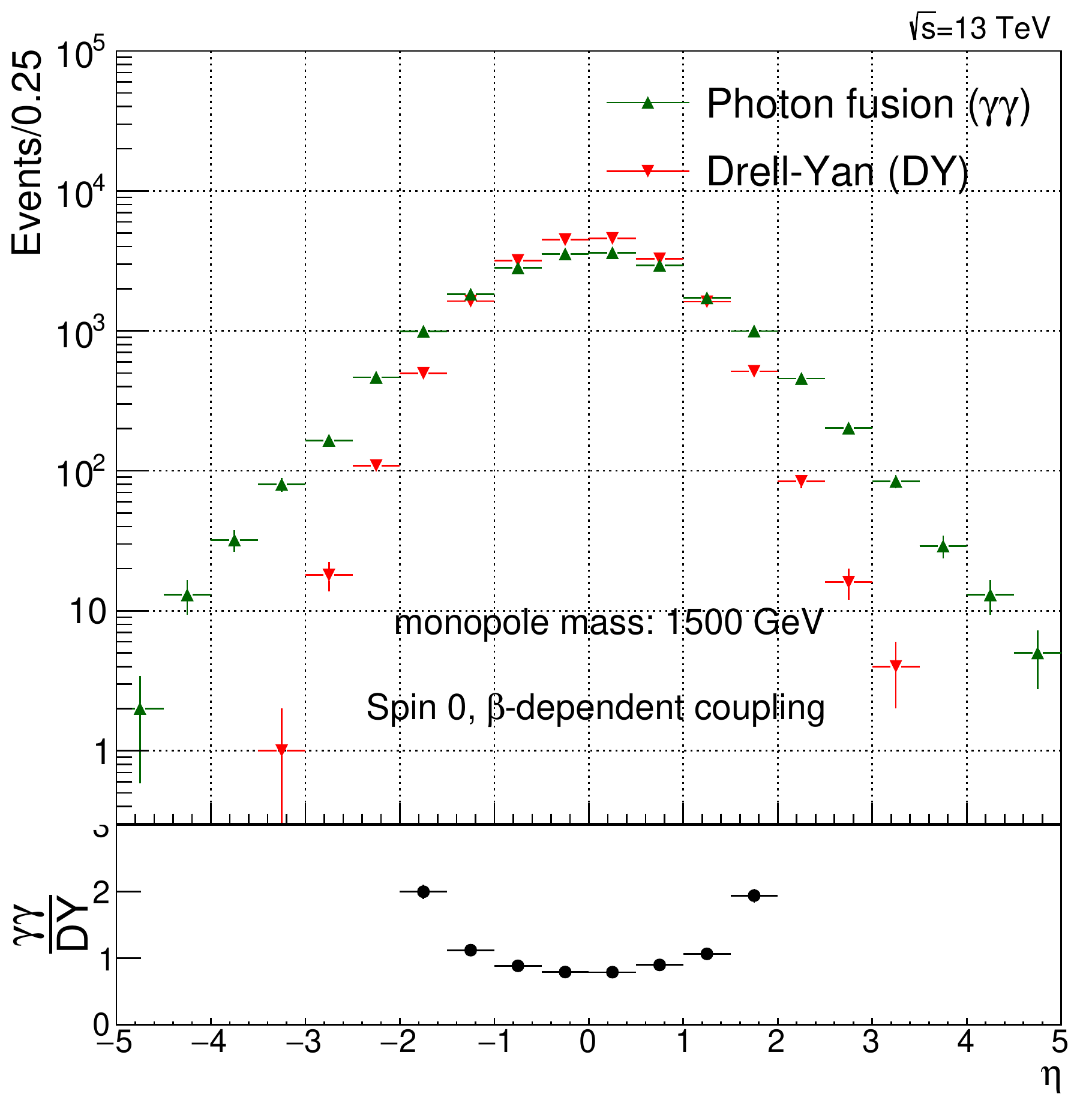}\hfill
\includegraphics[width=0.33\textwidth]{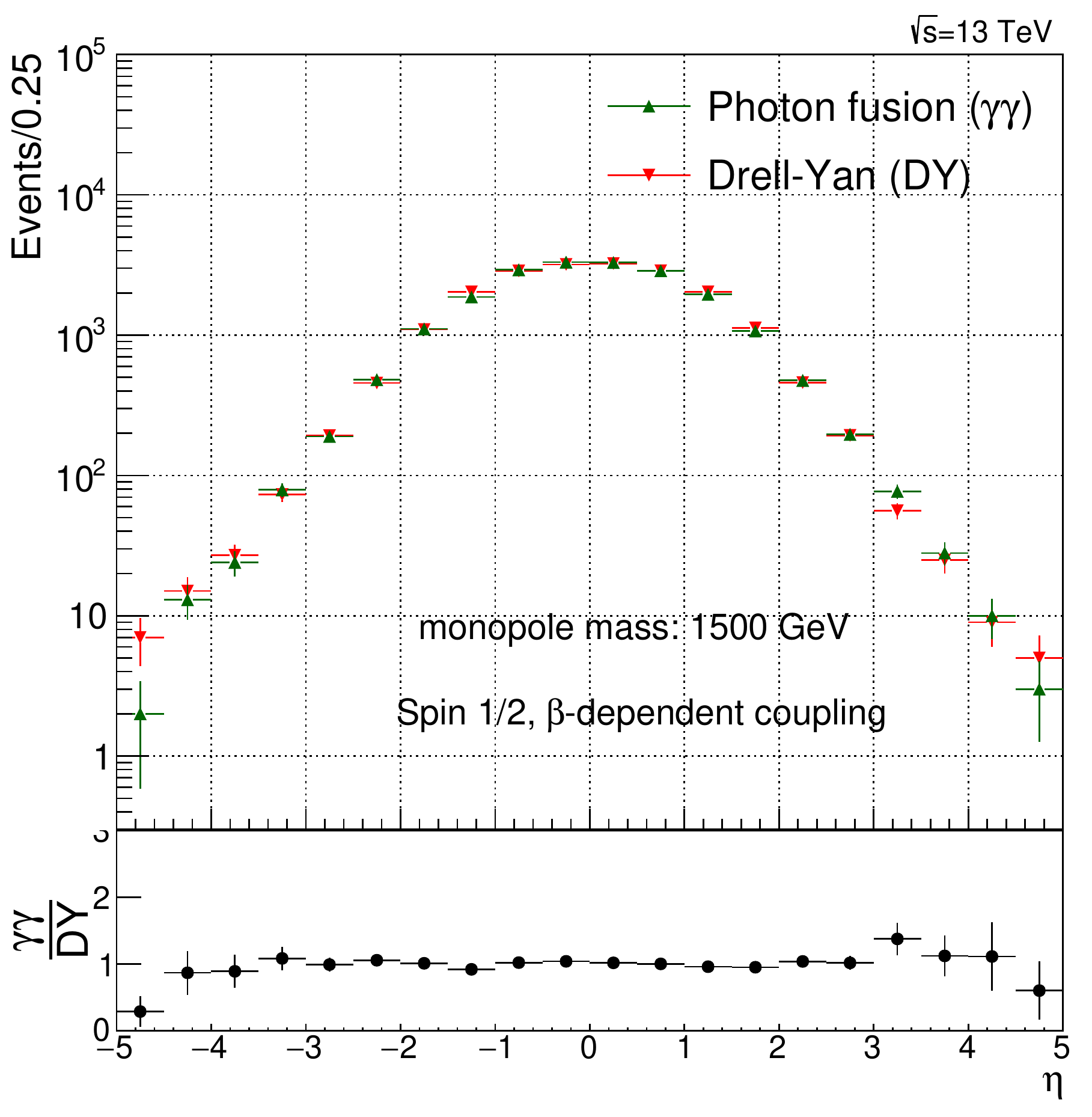}\hfill
\includegraphics[width=0.33\textwidth]{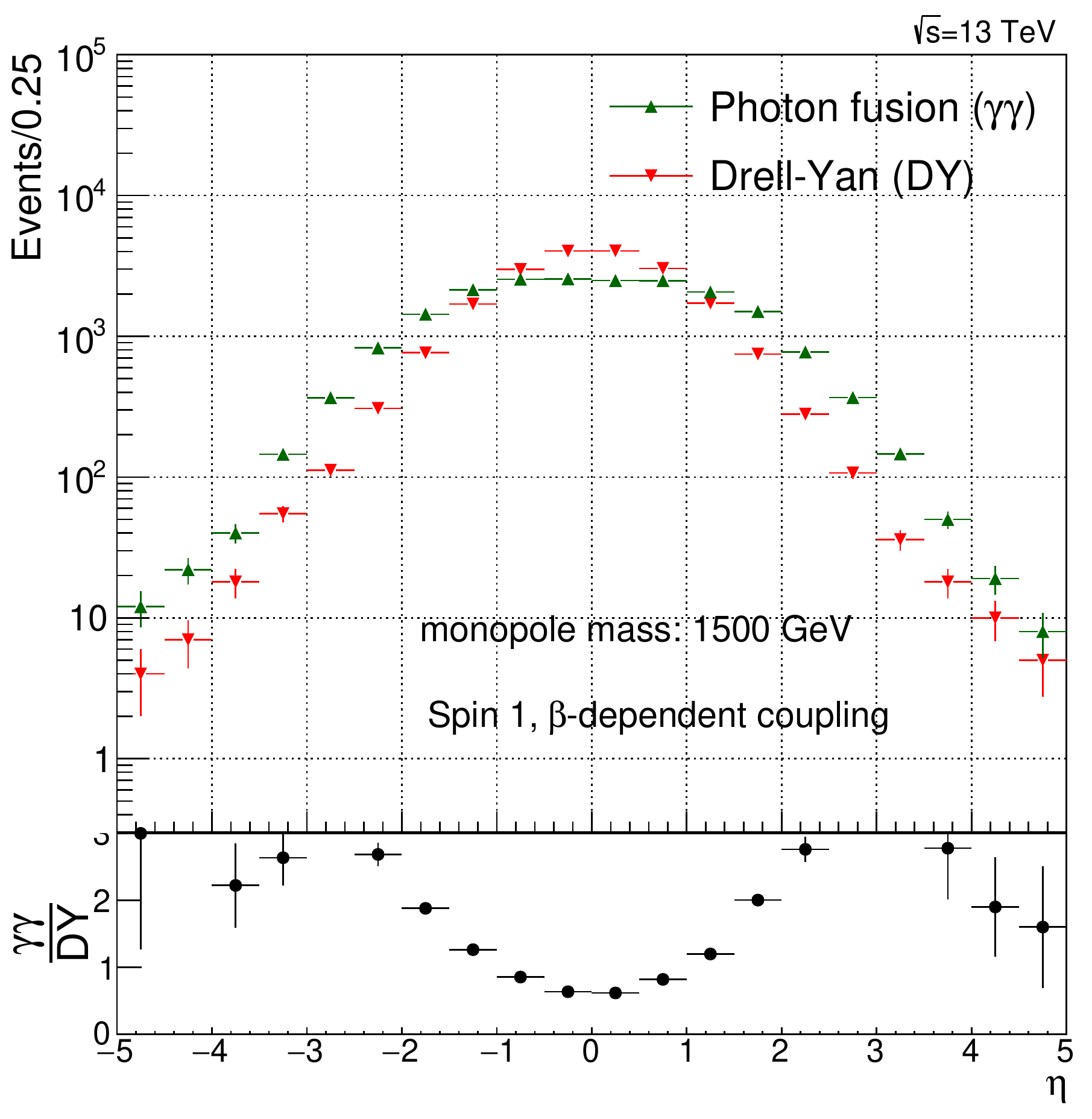}
\caption{The monopole pseudorapidity $\eta$ distributions of spin-0 (left), spin-\half (centre) and spin-1 monopoles (right) for both the photon-fusion and the Drell-Yan mechanisms at $\sqrt{s}=13$~TeV and for $\kappa=1$. The \texttt{LUXqed} and \texttt{NNPDF23} PDFs were used for the PF and the DY process, respectively.  }\label{fig:eta}
\end{figure}

As far as the pseudorapidity is concerned, its distributions are shown in fig.~\ref{fig:eta}. The spin-0 (left panel) and spin-1 (right panel) cases yield a more central production for DY than PF, whilst for spin-\half (central panel) the spectra are practically the same, although the one for PF is slightly more central. Again this behaviour of bosonic versus fermionic monopoles may be attributed to the (additional) four-vertex diagram for the bosons. In addition, the PF-versus-DY comparison of the three panels in fig.~\ref{fig:eta} is in agreement with their counterparts of figs.~\ref{PFdsigmaSpin0},~\ref{DYdsigmadetaSpin0} (scalar), figs.~\ref{LeeYangDsigDstuff12label},~\ref{LeeYangDsigDstuff12labeldy} (spinor), and figs.~\ref{LeeYangDsigDomega1pflabel}, ~\ref{LeeYangDsigDomega1label} (vector), respectively, as far as the production ``centrality'' is concerned. In the PF process, in particular, we observe that the depressions at $\eta\simeq 0$ for SM values of the $\kappa$ parameters for both spin-\half (see fig.~\ref{LeeYangDsigDstuff12label}) and spin-1 (see fig.~\ref{LeeYangDsigDomega1pflabel}) have been converted to flat tops when photon PDFs are also considered. This is normal taking into account the boost of the $\gamma\gamma$ centre-of-mass frame with respect to the $pp$ (laboratory) frame and the event-by-event variation of the monopole velocity $\beta$ that yields different event weight. 

\begin{figure}[ht!]
\justify
\includegraphics[width=0.33\textwidth]{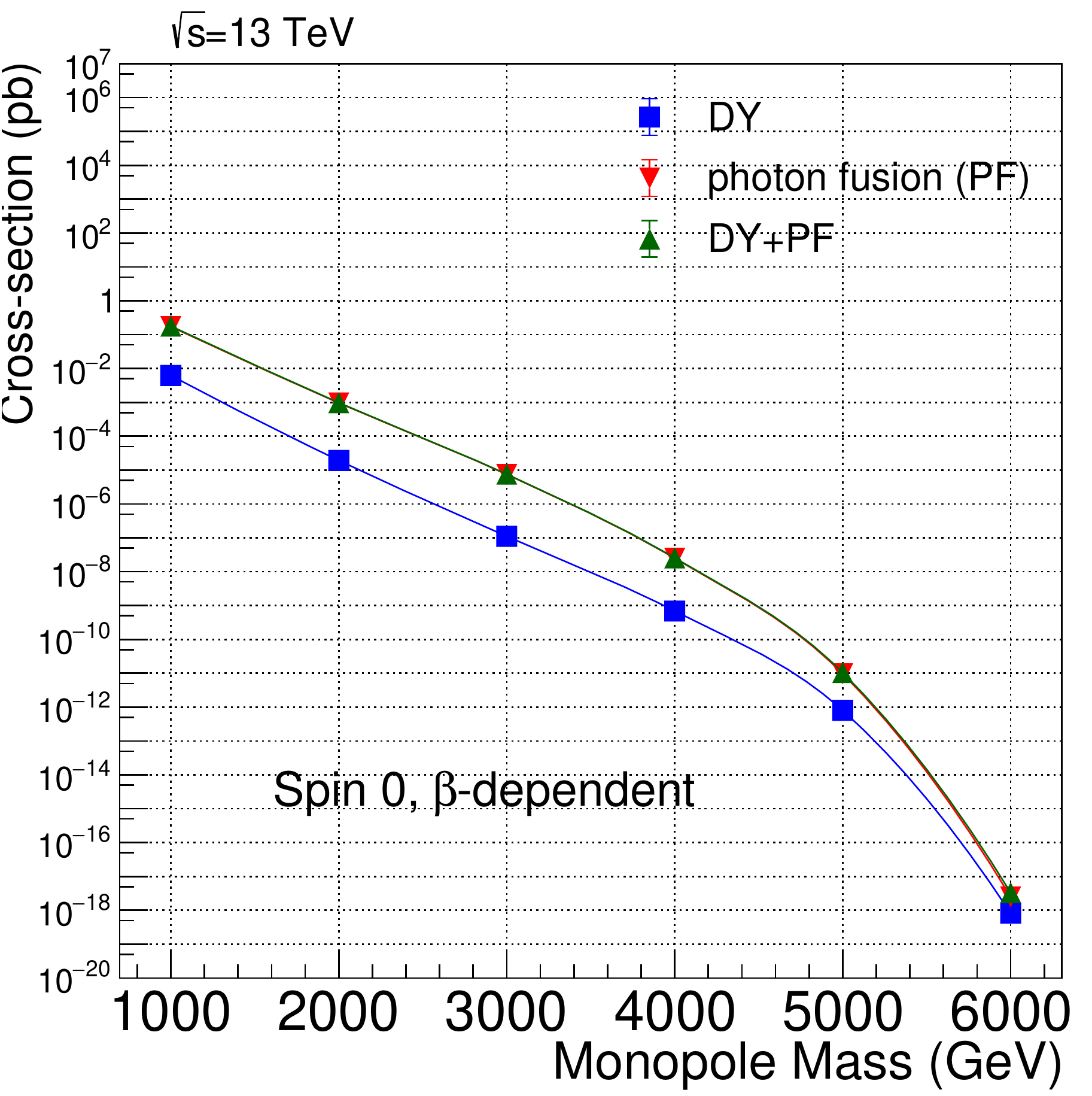}\hfill
\includegraphics[width=0.33\textwidth]{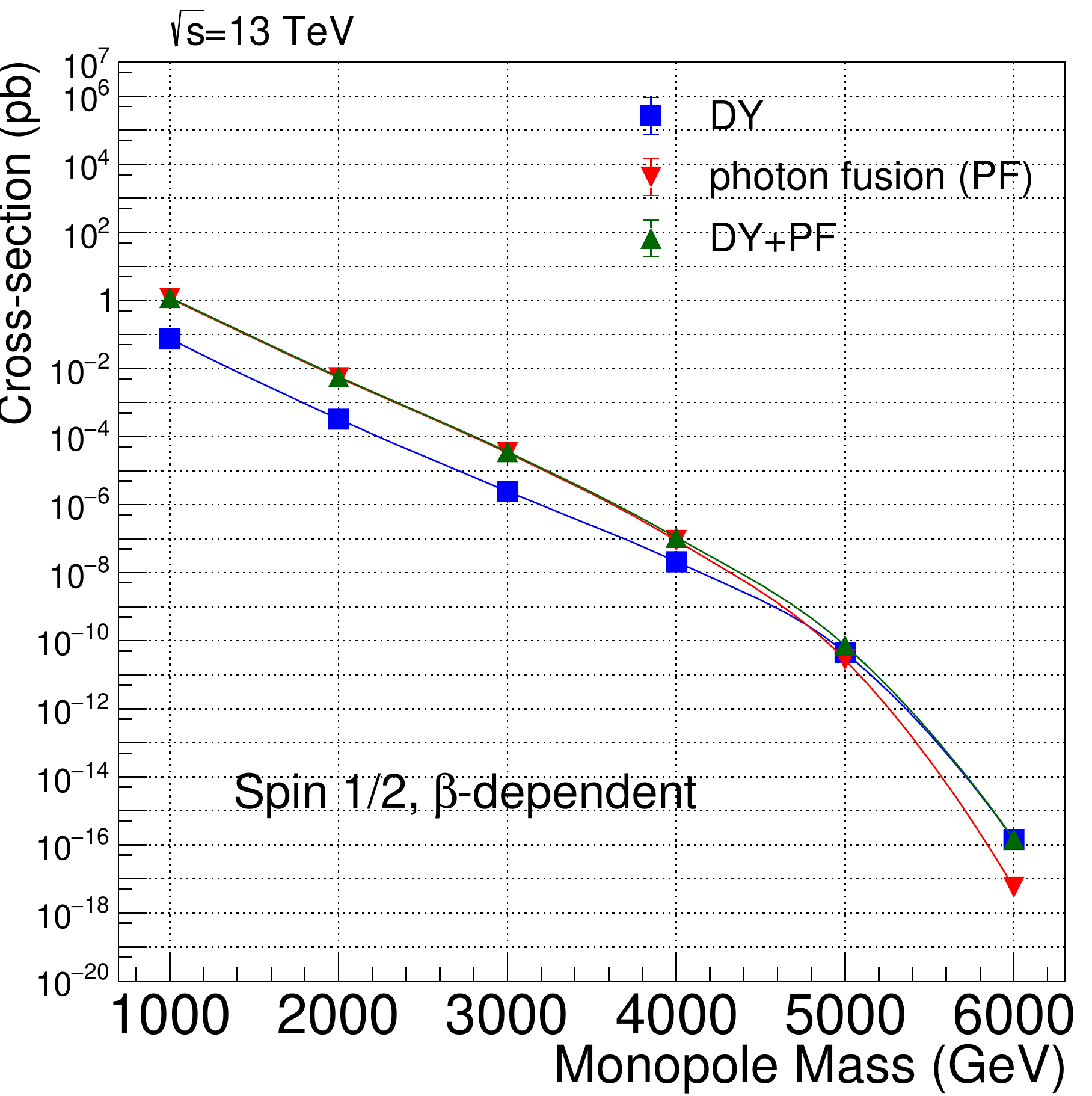}\hfill
\includegraphics[width=0.33\textwidth]{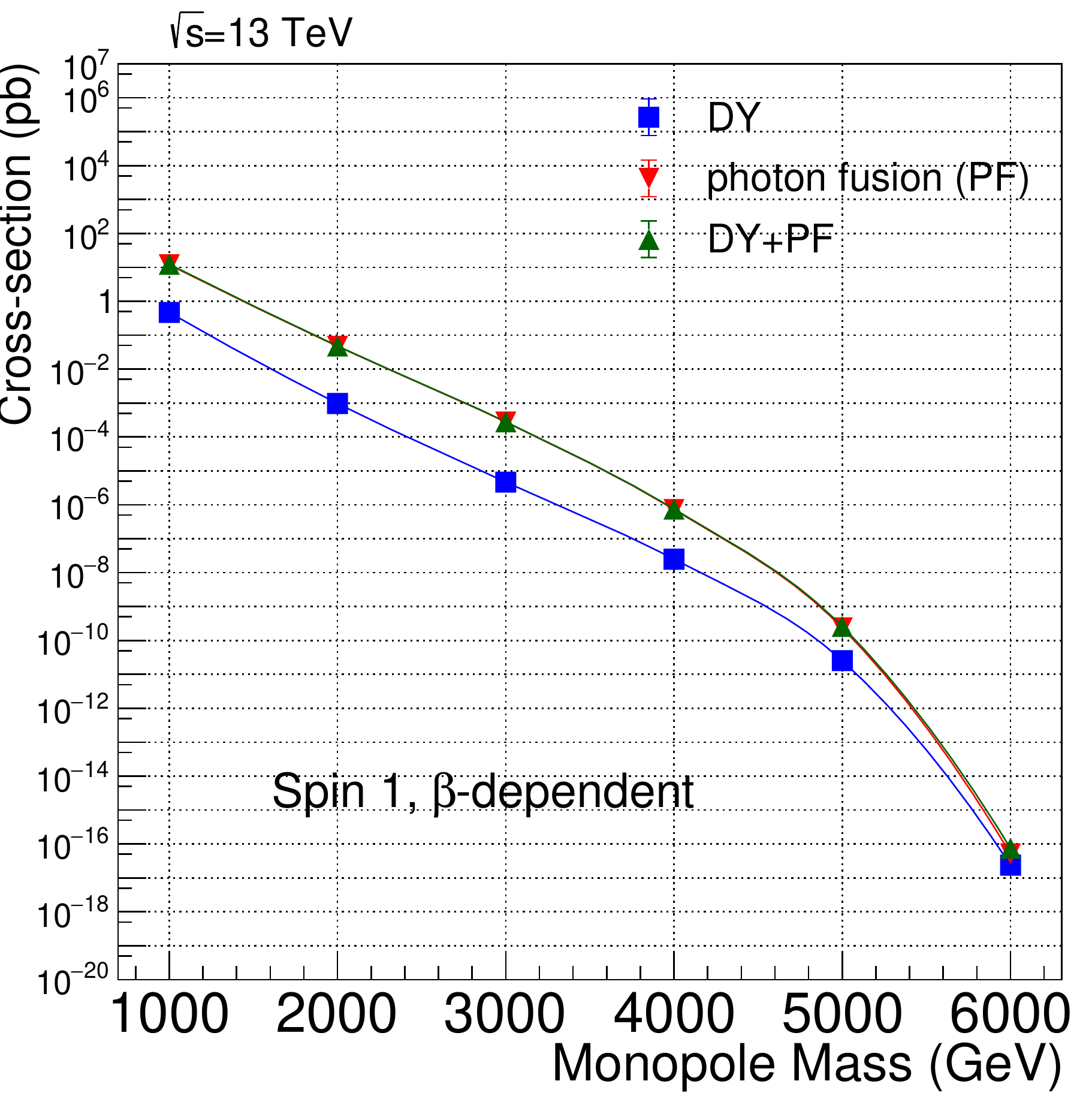}
\caption{Total cross section at $\sqrt{s}=13$~TeV for PF, DY and their sum versus the monopole mass. The \texttt{LUXqed} and \texttt{NNPDF23} PDFs were used for the PF and the DY process, respectively. For $S=\half$ and $S=1$, the SM values $\tilde{\kappa}=0$ and $\kappa=1$, respectively, are drawn, while there is no magnetic moment in the spin-0 case. }\label{fig:xsec-pdf}
\end{figure}

The total cross sections for the various spin cases, assuming SM magnetic-moment values for spin~\half and spin~1 and $\beta$-dependent coupling are drawn in fig.~\ref{fig:xsec-pdf} for photon fusion and Drell-Yan processes, as well as their sum. At tree level there is no interference between the PF and DY diagrams, so the sum of cross sections expresses the sum of the corresponding amplitudes. The PF mechanism is the dominant at the LHC energy of 13~TeV throughout the whole mass range of interest of $1\div 6~{\rm TeV}$ for the bosonic-monopole case. However if the monopole has spin~\half the PF dominates for masses up to $\sim 5~{\rm TeV}$, while DY takes over for $M\gtrsim5~{\rm TeV}$. This results underlines the importance that the photon-fusion production mechanism has for LHC without, however, overlooking the DY process.

\subsection{Perturbatively consistent limiting case of large $\kappa$ and small $\beta$ for photon fusion}\label{kappaMad}


In section~\ref{kappaThe}, the theoretical calculations show that in the perturbatively consistent limit of large $\kappa$ and small $\beta$, the cross sections are finite for both spin-\half  \eqref{totsec12lim} and spin-1 \eqref{limits} cases. In this section, we focus on this aspect of the photon-fusion production mechanism, since it dominates at LHC energies. We first put to test this theoretical claim utilising the \MAD implementation and later we discuss the kinematic distributions and comment on experimental aspects of a potential perturbatively-consistent search in colliders to follow in this context. 

\subsubsection{Spin-\half case}\label{spinhalfkappa}

\begin{figure}[ht!]
\justify
\includegraphics[width=0.33\textwidth]{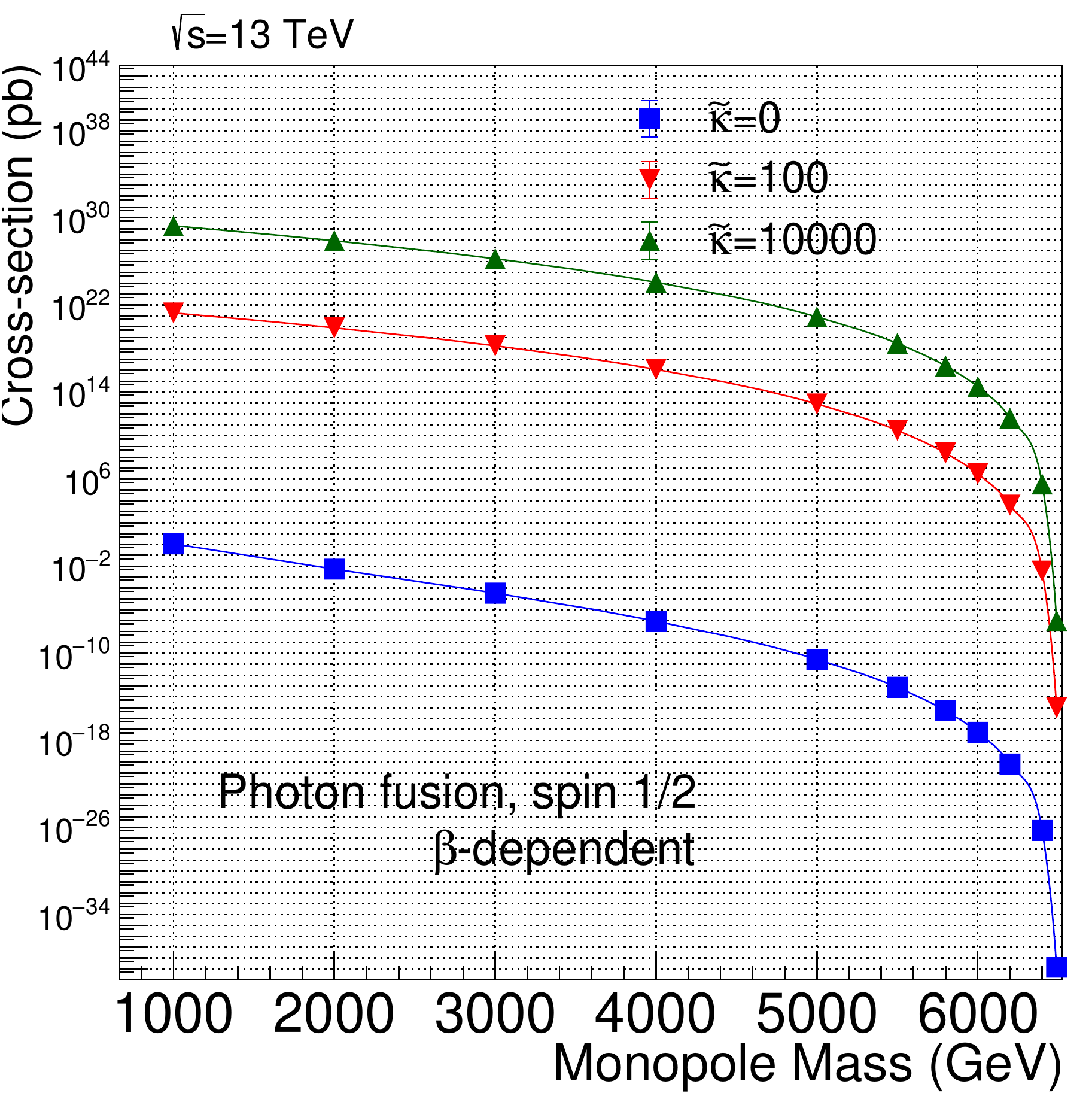}\hfill
\includegraphics[width=0.33\textwidth]{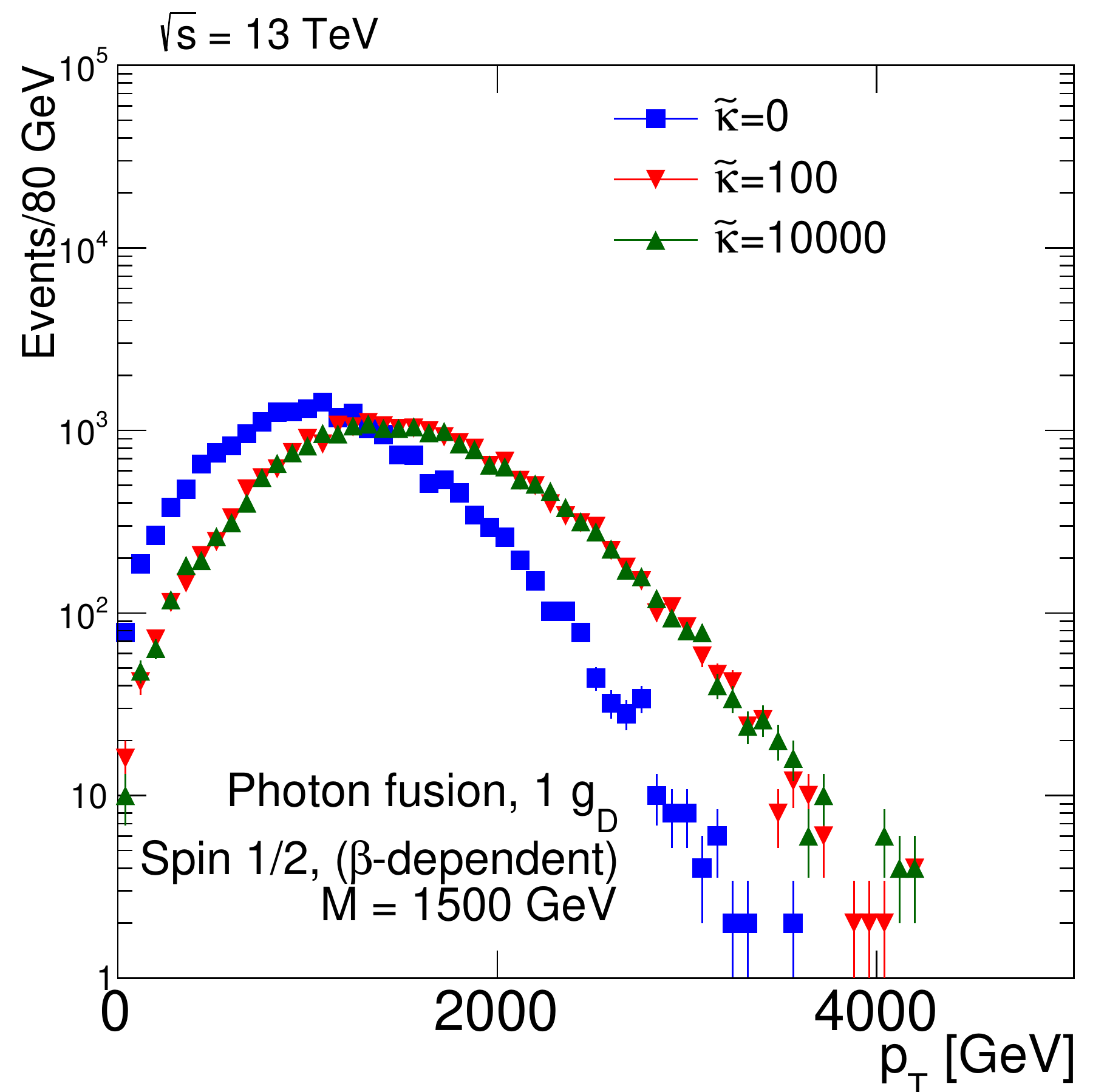}\hfill
\includegraphics[width=0.33\textwidth]{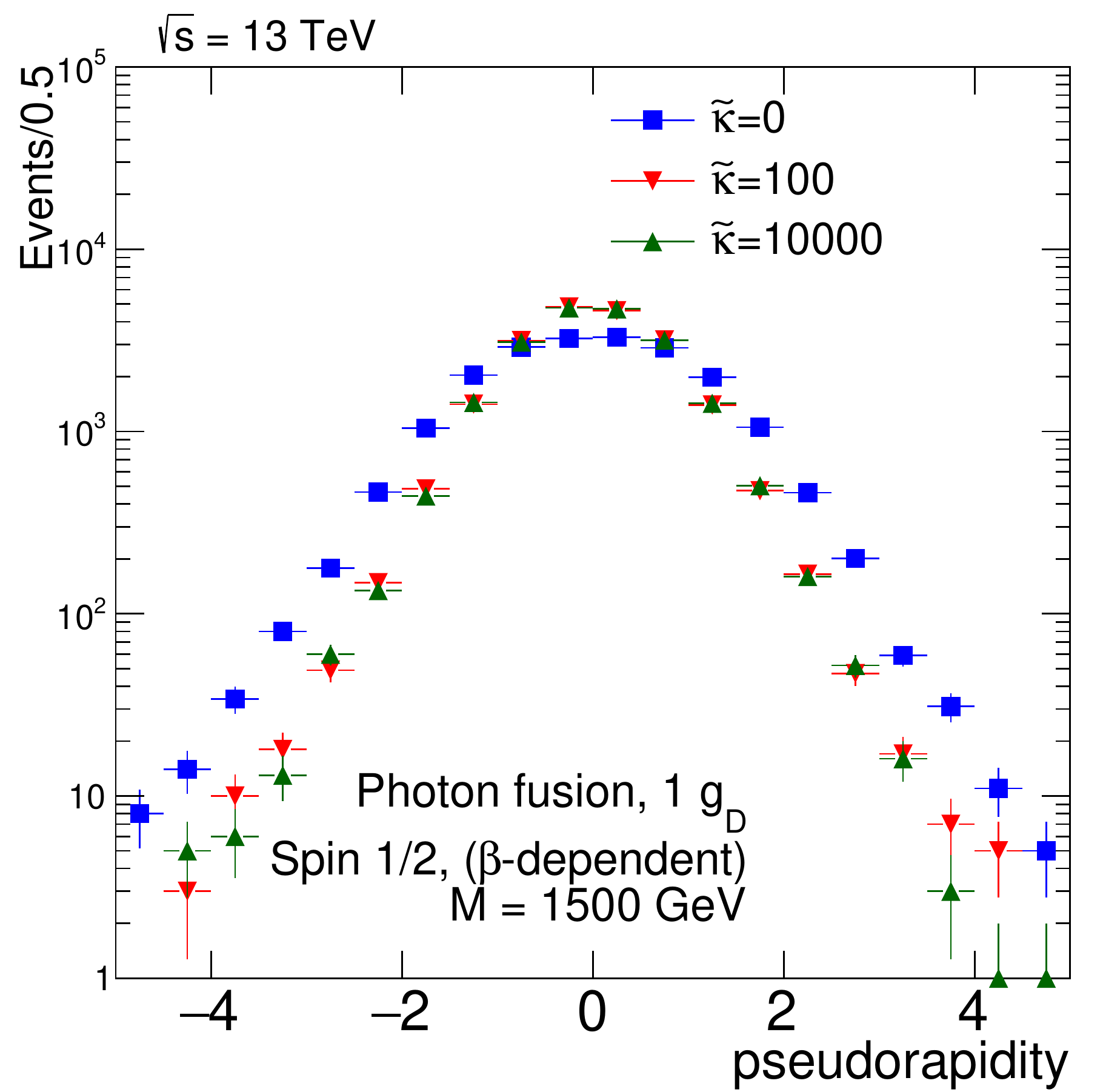}
\caption{Photon-fusion production at $\sqrt{s}=13$~TeV $pp$ collisions for spin-\half monopole, $\beta$-dependent coupling and various values of the $\tilde{\kappa}$ parameter: cross section versus monopole mass (left); \pt distribution for $M=1500~{\rm GeV}$ (centre); and $\eta$ distribution for $M=1500~{\rm GeV}$ (right). }\label{fig:spin12-kappa}
\end{figure}

For a spin-\half monopole, the dimensionless parameter $\tilde{\kappa}=\kappa M$, with $M$ the mass of the monopole, is varied from zero (the SM scenario) to 10,000 for $\gamma\gamma$ collisions at $\sqrt{\sgg}=13~{\rm TeV}$. The cross-section of the photon fusion process for $\tilde{\kappa}=0$ is going to zero very fast as $\beta\rightarrow 0$, as can be seen in the third column of table~\ref{tab:kappatilde}. However for non-zero $\tilde{\kappa}$, the cross-section values remain finite even if $\beta$ goes to zero, as expected, as becomes evident from the last row of table~\ref{tab:kappatilde}. The same conclusion is drawn from fig.~\ref{fig:spin12-kappa} (left), where the cross sections are plotted for $pp$ collisions at $\sqrt{s}=13$~TeV, i.e.\ with PDF.  For masses $M\gtrsim 6~{\rm TeV}$, the monopole production, although still rare, remains at detectable limits for the LHC experiments. 
\begin{table}[ht]
\begin{center}
\scalebox{0.93}{
\begin{tabular}{| c | c | c | c | c | c |}
\hline \hline
 Monopole   & $\beta$ & \multicolumn{4}{c|}{$\ga,\;\;\sigma~{\rm (pb)}$}          \\
 \cline{3-6}
mass (GeV) &         & $\tilde{\kappa}=0$    & $\tilde{\kappa}=10$       & $\tilde{\kappa}=100$  &  $\tilde{\kappa}=10,000$ \\
           \hline
           \hline
1000 & 0.9881 &  $1.37\times10^{5} \pm 4.6\times10^{2}$ & $1.639\times10^{24} \pm 3.3\times10^{21}$ & $1.639\times10^{28} \pm 3.3\times10^{25}$ & $1.639\times10^{36} \pm 3.3\times10^{33}$\\ 
2000 & 0.9515 &  $8.303\times10^{4} \pm 4.5\times10^{2}$ & $1.61\times10^{24} \pm 3.1\times10^{21}$ & $1.61\times10^{28} \pm 3.1\times10^{25}$ & $1.61\times10^{36} \pm 3.1\times10^{33}$\\ 
3000 & 0.8871 &  $4.78\times10^{4} \pm 3.5\times10^{2}$ & $1.356\times10^{24} \pm 2.5\times10^{21}$ & $1.356\times10^{28} \pm 2.5\times10^{25}$ & $1.356\times10^{36} \pm 2.5\times10^{33}$\\ 
4000 & 0.7882 &  $2.237\times10^{4} \pm 1.9\times10^{2}$ & $8.612\times10^{23} \pm 2.1\times10^{21}$ & $8.613\times10^{27} \pm 2.1\times10^{25}$ & $8.613\times10^{35} \pm 2.1\times10^{33}$\\ 
5000 & 0.639 &  $6396 \pm 61$ & $3.154\times10^{23} \pm 1.1\times10^{21}$ & $3.154\times10^{27} \pm 1.1\times10^{25}$ & $3.154\times10^{35} \pm 1.1\times10^{33}$\\ 
5500 & 0.5329 &  $2256 \pm 22$ & $1.247\times10^{23} \pm 4.5\times10^{20}$ & $1.247\times10^{27} \pm 4.5\times10^{24}$ & $1.247\times10^{35} \pm 4.5\times10^{32}$\\ 
5800 & 0.4514 &  $886.5 \pm 7.8$ & $5.28\times10^{22} \pm 2.5\times10^{20}$ & $5.28\times10^{26} \pm 2.5\times10^{24}$ & $5.28\times10^{34} \pm 2.5\times10^{32}$\\ 
6000 & 0.3846 &  $367.2 \pm 3$ & $2.294\times10^{22} \pm 7.6\times10^{19}$ & $2.294\times10^{26} \pm 7.6\times10^{23}$ & $2.294\times10^{34} \pm 7.6\times10^{31}$\\ 
6200 & 0.3003 &  $97.19 \pm 0.77$ & $6.43\times10^{21} \pm 3.3\times10^{19}$ & $6.43\times10^{25} \pm 3.3\times10^{23}$ & $6.43\times10^{33} \pm 3.3\times10^{31}$\\ 
6400 & 0.1747 &  $5.846 \pm 0.025$ & $4.065\times10^{20} \pm 1.5\times10^{18}$ & $4.065\times10^{24} \pm 1.5\times10^{22}$ & $4.065\times10^{32} \pm 1.5\times10^{30}$\\
6490 & $0.0554$ & $0.017\pm2.27\times 10^{-5}$ & $1.27\times10^{18}\pm 8.74\times10^{14}$ & $1.27\times10^{22}\pm 8.74\times10^{18}$ & $1.27\times10^{30}\pm 8.74\times10^{26}$ \\
\hline
\hline
\end{tabular}
}
\end{center}
\caption{Photon-fusion production cross sections at $\sqrt{\sgg}=13~{\rm TeV}$ for spin-\half monopole, $\beta$-dependent coupling and various values of the  $\tilde{\kappa}$ parameter.}
\label{tab:kappatilde}
\end{table}

The central and right-hand-side plots of fig.~\ref{fig:spin12-kappa} depict a comparison of the \pt and $\eta$ distributions, respectively, between the SM value $\tilde{\kappa}=0$ and much higher values up to  $\tilde{\kappa}=10^4$. The SM-like case is characterised by a distinguishably ``softer'' \pt spectrum and a less central angular distribution than the large-$\tilde{\kappa}$ case.\footnote{We should remark at this point, that in the absence of PDF, in the cases for the magnetic-moment parameter $\tilde{\kappa}\neq 0$, we observe a fast-increasing distribution of events at high \pt-values up to a cutoff of $\pt=\sqrt{\sgg}/2$, as expected from the non-unitarity of such cases.} The latter case, on the other hand, seems to converge to a common shape for the kinematic variables as $\tilde{\kappa}$ increases to very large values. This is not the case for $\tilde{\kappa}$ values distinct, yet near, the SM value, where the angular distributions are not considerably different, as shown in fig.~\ref{LeeYangDsigDstuff12label}. The common kinematics among large $\tilde{\kappa}$ values would greatly facilitate an experimental analysis targeting perturbatively reliable results. We note here that the DY process, which dominates the cross section for heavy monopoles \emph{for the SM magnetic-moment value} (see fig.~\ref{fig:xsec-pdf}), vanishes as $\tilde{\kappa}$ acquires large values as shown in section~\ref{kappaThe}, rendering the study of photon-fusion process sufficient at the perturbative-coupling limit.

\subsubsection{Spin-1 case}\label{spinonekappa}

\begin{figure}[ht!]
\justify
\includegraphics[width=0.33\textwidth]{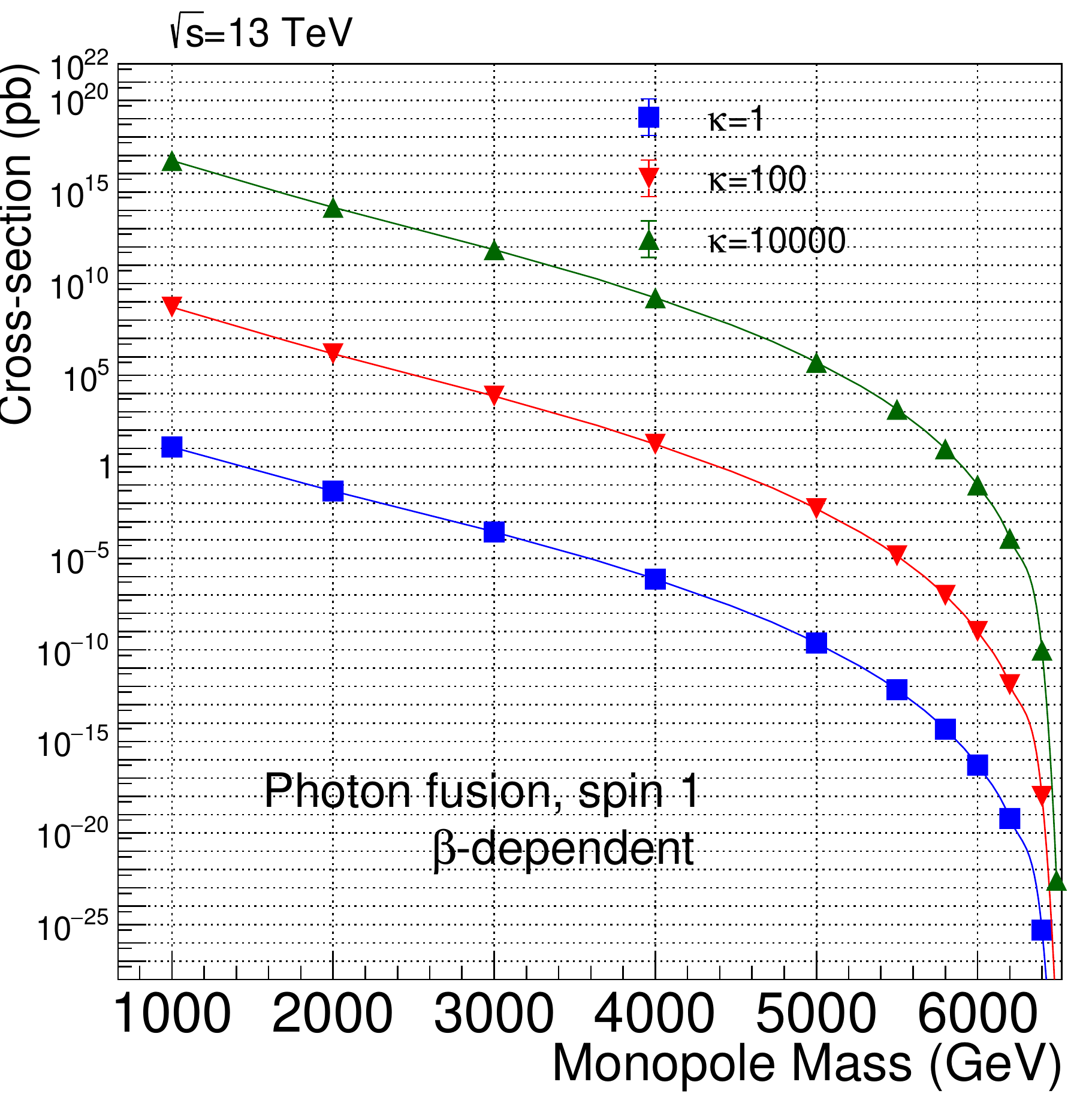}\hfill
\includegraphics[width=0.33\textwidth]{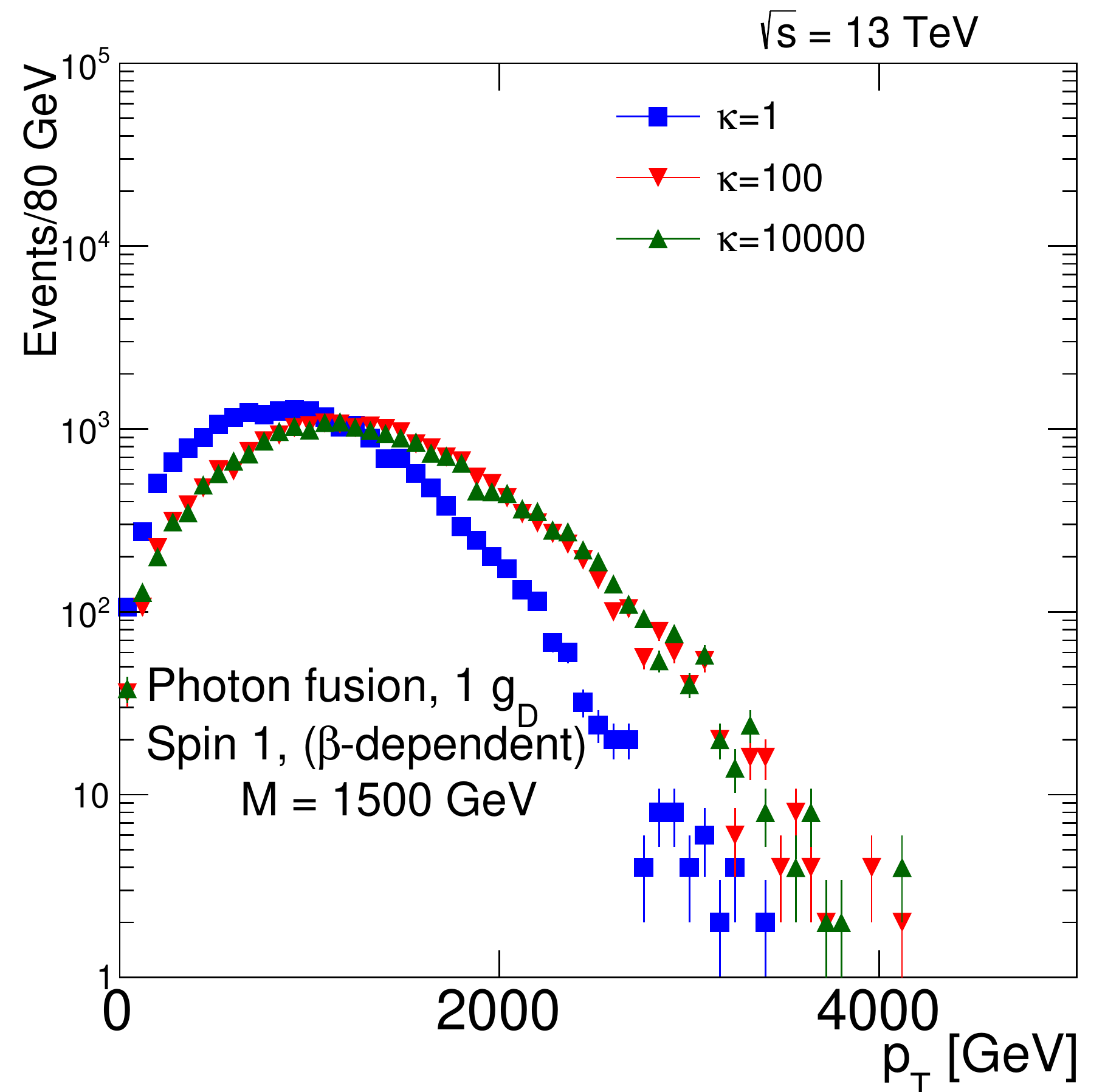}\hfill
\includegraphics[width=0.33\textwidth]{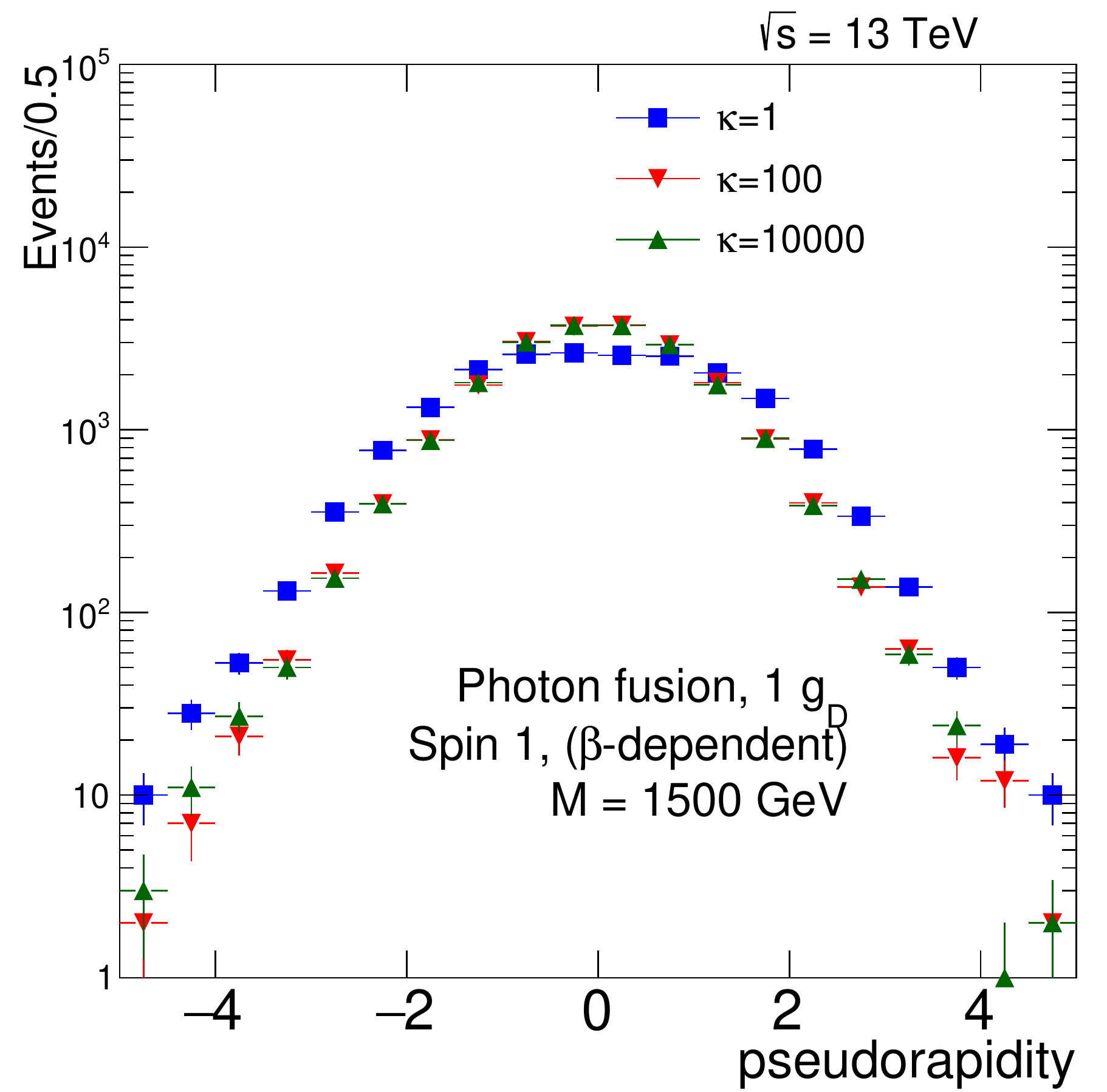}
\caption{Photon-fusion production at $\sqrt{s}=13$~TeV $pp$ collisions for spin-1 monopole, $\beta$-dependent coupling and various values of the $\kappa$ parameter: cross section versus monopole mass (left); \pt distribution for $M=1500~{\rm GeV}$ (centre); and $\eta$ distribution for $M=1500~{\rm GeV}$ (right).  }\label{fig:spin1-kappa}
\end{figure}

Repeating the same procedure this time for spin-1 monopoles, we vary the dimensionless parameter $\kappa$ from unity (the SM scenario) to 10,000 and we list the \ga cross sections in table~\ref{tab:kappa} for the photon-fusion process. Similar to the spin-\half monopole case, the cross section for $\kappa=1$, i.e.\ the SM scenario, is going to zero very fast as $\beta\rightarrow 0$. However, for $\kappa>1$, the cross section becomes finite even if $\beta$ goes to 0, as seen in the last row of table~\ref{tab:kappa}. This observation remains valid when the cross section for $pp$ collisions, instead of $\gamma\gamma$ scattering, is considered. Indeed, fig.~\ref{fig:spin1-kappa} shows that for large values of $\kappa$ and $M \simeq \sqrt{s}/2$, which is equivalent to $\beta \ll 1$, the cross section, although very small, remains finite.
\begin{table}[ht]
\begin{center}
\scalebox{1.05}{
\begin{tabular}{| c | c | c | c | c |}
\hline \hline
 Monopole   & $\beta$ & \multicolumn{3}{c|}{$\ga,\;\;\sigma~{\rm (pb)}$}          \\
 \cline{3-5}
mass (GeV) &              & $\kappa=1$           & $\kappa=100$  &  $\kappa=10,000$ \\
           \hline
           \hline
1000 & 0.9881 & $1.086\times10^{7} \pm 1.4\times10^{5}$ & $4.939\times10^{15} \pm 1\times10^{13}$ & $5.033\times10^{23} \pm 2.1\times10^{21}$ \\ 
2000 & 0.9515 & $2.275\times10^{6} \pm 1.6\times10^{4}$ & $2.844\times10^{14} \pm 4.9\times10^{11}$ & $2.879\times10^{22} \pm 9.8\times10^{19}$ \\ 
3000 & 0.8871 & $7.198\times10^{5} \pm 6.6\times10^{3}$ & $4.518\times10^{13} \pm 1.5\times10^{11}$ & $4.536\times10^{21} \pm 1.2\times10^{19}$ \\ 
4000 & 0.7882 & $2.273\times10^{5} \pm 2.2\times10^{3}$ & $9.079\times10^{12} \pm 2.7\times10^{10}$ & $9.002\times10^{20} \pm 3.2\times10^{18}$ \\ 
5000 & 0.639 & $5.232\times10^{4} \pm 4.9\times10^{2}$ & $1.513\times10^{12} \pm 9.2\times10^{9}$ & $1.5\times10^{20} \pm 9.3\times10^{17}$ \\ 
5500 & 0.5329 & $1.785\times10^{4} \pm 1.6\times10^{2}$ & $4.49\times10^{11} \pm 1.7\times10^{9}$ & $4.466\times10^{19} \pm 2.9\times10^{17}$ \\ 
5800 & 0.4514 & $7118 \pm 62$ & $1.658\times10^{11} \pm 1.1\times10^{9}$ & $1.624\times10^{19} \pm 8.4\times10^{16}$ \\ 
6000 & 0.3846 & $3025 \pm 24$ & $6.72\times10^{10} \pm 2.5\times10^{8}$ & $6.627\times10^{18} \pm 3.7\times10^{16}$ \\ 
6200 & 0.3003 & $836.9 \pm 6.3$ & $1.764\times10^{10} \pm 1\times10^{8}$ & $1.733\times10^{18} \pm 1\times10^{16}$ \\ 
6400 & 0.1747 & $53.42 \pm 0.23$ & $1.066\times10^{9} \pm 3.9\times10^{6}$ & $1.05\times10^{17} \pm 3.8\times10^{14}$ \\
6490 & $0.0554$ & $0.1694\pm0.00065$ & $3.293\times 10^{6} \pm 5.6\times 10^{3}$ & $3.244\times 10^{14}\pm 5.6\times 10^{11}$   \\
\hline
\hline
\end{tabular}
}
\end{center}
\caption{Photon-fusion production cross sections at $\sqrt{\sgg}=13~{\rm TeV}$ for spin-1 monopole, $\beta$-dependent coupling and various values of the $\kappa$ parameter.}
\label{tab:kappa}
\end{table}

In fig.~\ref{fig:spin1-kappa}, a comparison of the \pt (centre) and $\eta$ (right) distributions, between the SM value $\kappa=1$ and much higher values up to  $\kappa=10^4$ is given for spin~1. As for spin~\half, the large-$\kappa$ curves converge to a single shape independent of the actual $\kappa$ value, with the SM-case yielding different distributions from the large-$\kappa$ case.\footnote{As in the spin-\half-monopole case, we also observe here that, in the absence of PDF, in the non-unitary cases $\kappa\neq 1$, there is a fast-increasing distribution of events at high \pt-values up to a cutoff of $\pt=\sqrt{\sgg}/2$.} Therefore the $\eta$-distribution features shown in fig.~\ref{LeeYangDsigDomega1pflabel} without PDF are completely smoothed out for large $\kappa$ by folding with the photon distribution function in the proton. As discussed in section~\ref{spinhalfkappa} for fermionic monopoles, such an experimental analysis can be concentrated on the $\kappa$-dependence of the total cross section and the acceptance for very slow monopoles to provide perturbatively valid mass limits in case of non-observation of a monopole signal, since the kinematic distributions are almost $\kappa$-invariant in $pp$ collisions. The MoEDAL experiment~\cite{moedal-review}, in particular, being sensitive to slow monopoles can make the best out of this new approach in the interpretation of monopole-search results. 

\section{Conclusions \label{sec:concl}}

The work described in this article consists of two parts. 
In the first part, we dealt with the computation of differential and total cross sections for pair production of monopoles of  spin $S=0,~\half,~1$,  through either photon-fusion or Drell-Yan processes. We have employed duality arguments to justify an effective monopole-velocity-dependent magnetic charge in monopole-matter scattering processes. Based on this, we conjecture that such $\beta$-dependent magnetic charges might also characterise monopole production. 

A magnetic-moment term proportional to a new phenomenological parameter $\kappa$ is added to the effective Lagrangians describing the interactions of these monopoles with photons for spins \half and 1. The lack of unitarity and/or renormalisability is restored when the monopole effective theory adopts a SM form, that is when the bare magnetic-moment parameter takes on the values $\kappa =0$ for spin-\half monopoles, and $\kappa =1$ for spin-1 monopoles. However we remark that the lack of unitarity and renormalisability is not necessarily an issue, from an effective-field-theory point of view. Indeed, given that the microscopic high-energy (ultraviolet) completion of the monopole models considered above is unknown, one might not exclude the possibility of restoration of unitarity in extended theoretical frameworks, where new degrees of freedom at a high-energy scale might play a role. In this sense, we consider the spin-1 monopole as a potentially viable phenomenological case worthy of further exploration. 

The motivation behind the magnetic-moment introduction is to enrich the monopole phenomenology with the (undefined) correction terms to the monopole magnetic moment to be treated as free parameters potentially departing from the ones prescribed for the electron or $W^{\pm}$ bosons in the SM. Lacking a fundamental microscopic theory of magnetic poles, such an addition appears reasonable. This creates a dependence of the scattering amplitudes of processes on this parameter, which is passed on to the total cross sections and, in some cases, to kinematic distributions.  Therefore the parameter $\kappa$ is proposed as a tool for monopole searches which can be tuned to explore different models. 

Moreover, even more intriguing is the possibility to use the parameter $\kappa$ in conjunction with the monopole velocity $\beta$ to achieve a perturbative treatment of the monopole-photon coupling. Indeed, in general the large value of the magnetic charge prevents any perturbative treatment of the monopole interactions limiting us to a necessary truncation of the Feynman-like diagrams at the tree level. By limiting the discussion to very slow ($\beta \ll 1$) monopoles, the perturbativity is guaranteed, however, at the expense of a vanishing cross section. Nonetheless it turns out that the photon-fusion cross section remains finite \emph{and} the coupling is perturbative at the formal limits $\kappa\to\infty$ and $\beta\to 0$. This ascertainment opens up the possibility to interpret the cross-section bounds set in collider experiments, such as MoEDAL, in a proper way, thus yielding sensible monopole-mass limits. 

In the second part of this article, a complete implementation in \MAD of the monopole production is performed both for the photon-fusion and the Drell-Yan processes, also including the magnetic-moment terms. The UFO models were successfully validated by comparing cross-section values obtained by the theoretical calculations and the \MAD UFO models. Kinematic distributions, relevant for experimental analyses, were compared between the photon-fusion and the Drell-Yan production mechanism of spins 0, \half  and 1 monopoles. This tool will allow to probe for the first time the ---dominant at LHC energies--- photon-fusion monopole production. Furthermore, the experimental aspects of a perturbatively valid monopole search for large values of the magnetic-moment parameters and slow-moving monopoles have also been outlined, based on these kinematic distributions.

 \section*{Acknowledgements}

We would like to thank J.R.~Ellis, V.~Vento, J.~Bernabeu and D.V.~Shoukavy for discussions on theoretical issues, Wendy Taylor, Manoj Kumar Mondal and Olivier Mattelaer for discussions on the \MAD implementation, and fellow colleagues from the MoEDAL-LHC Collaboration for their interest. 
The work of S.B.\ is supported by an STFC (UK) doctoral studentship. The work of N.E.M.\ is partially supported by STFC (UK) under the research grant ST/P000258/1. N.E.M.\ also acknowledges a scientific associateship (``\emph{Doctor Vinculado}'') at IFIC-CSIC-Valencia Univ.\ (Spain).  V.A.M.\ and A.S.\ acknowledge support by the Generalitat Valenciana (GV) through the MoEDAL-supporting agreements, by the Spanish MINECO under the project FPA2015-65652-C4-1-R and by the Severo Ochoa Excellence Centre Project SEV-2014-0398.  V.A.M.\ acknowledges support by the GV Excellence Project PROMETEO-II/2017/033 and by a 2017 Leonardo Grant for Researchers and Cultural Creators, BBVA Foundation. 

\appendix

\section{Basic definitions}\label{app:basic}

Listed in this appendix are the standard identities for sums and traces over vector boson polarisation and spinor spin states, as well as basic properties of Dirac $\gamma$-matrices used in the text to derive the various scattering amplitudes. The variables related to the differential cross section of a two-particle scattering are also defined.

Our notation and conventions are:

Space-time metric  is four-dimensional, Minkowski flat: 
\begin{equation}
g^{\mu\nu} = \eta^{\mu\nu} = {\rm Diag}\Big(+1, -1, -1, -1 \Big).
\end{equation} 

Dirac $\gamma$-matrices satisfy the Clifford algebra:
\begin{equation}
\{ \gamma^\mu, \gamma^\nu \} = 2 g^{\mu\nu} ~, 
\end{equation}
where $\{ A, B \} = AB + BA$ denotes the anticommutator. 

The sum over photon polarisation states is given by:
\begin{equation}\label{sumphot}
\sum_{\lambda}\varepsilon_{\lambda}^{\mu}\varepsilon_{\lambda}^{\nu*}=-g^{\mu\nu}.
\end{equation}

The sum over massive (of mass $M$) vector meson polarisation states reads: 
\begin{equation}
\sum_{\kappa} \sum_{\kappa'} \, \Big[\Upsilon_{\kappa}^{\rho}\Upsilon_{\kappa'}^{\sigma*}\Upsilon_{\kappa}^{*\rho'}\Upsilon_{\kappa'}^{\sigma'}\Big]= \left(-g^{\rho\rho'}+\frac{p_1^{\rho}p_1^{\rho'}}{M^2}\right) \left(-g^{\sigma\sigma'}+\frac{p_2^{\sigma}p_2^{\sigma'}}{M^2}\right).\label{evaltrace1}
\end{equation}

The spinor sum rules and properties of Dirac's $\gamma$ matrices are: 
\begin{eqnarray}
	&& \sum_{s}u_{s}\overline{u}_{s}=(\slashed{p}_{1}+m), \qquad \sum_{r}v_{r}\overline{v}_{r}=(\slashed{p}_{2}-m)\label{spinsum}, \nonumber \\
	&& \Tr[{\rm odd~number~of~\gamma~matrices}] = 0, \nonumber \\ 
	&& \Tr[\gamma^\mu\, \gamma^\nu] =  4\, g^{\mu\nu}, \qquad \Tr[\gamma^{\mu}\gamma^{\nu}\gamma^{\sigma}\gamma^{\rho}]=4g^{\mu\nu}g^{\sigma\rho}-4g^{\mu\sigma}g^{\nu\rho}+4g^{\mu\rho}g^{\nu\sigma},
\end{eqnarray}
from which we obtain for the trace over spinor polarisations: 
 \begin{equation}
 \begin{split}
\Tr\Big[(u_{\alpha}&\gamma^{\mu}\overline{v}_{\beta})(v_{\beta}\gamma^{\mu'}\overline{u}_{\alpha})\Big]
=\Tr\Big[\left(\slashed{ q_1}+m\right)\gamma^{\mu }\left(\slashed{  q_2}-m\right)\gamma ^{\mu '}\Big]  \\
&=q_{1\rho}q_{2\sigma}(4g^{\mu\sigma}g^{\mu'\rho}-4g^{\mu\mu'}g^{\sigma\rho}+4g^{\mu\rho}g^{\mu'\sigma}) - 4m^2 \, g^{\mu\mu '}
\label{evaltrace2}
\end{split}
\end{equation}

\begin{enumerate}

\item The angular differential cross section distribution $\frac{\dd\sigma}{\dd\Omega}$ is defined as usual
\begin{equation}\label{Xsecdeff}
	\frac{\dd\sigma}{\dd\Omega} =\frac{1}{64\pi^{2} s}\frac{|\vec{p}_1|}{|\vec{q}_1|}|\overline{\mathcal{M}}|^{2}~,
\end{equation}
where $|\overline{\mathcal{M}}|^{2}$ is the squared matrix amplitude, averaged over initial spin or polarisation states (depending on the process) and summed over final ones, with the matrix element $\mathcal{M}$ denoting the sum of the matrix amplitudes of all the relevant processes contributing to the interaction. This expression is evaluated in terms of the Mandelstam variables defined as
\begin{equation}
\begin{split}
	t & = (q_{1}-p_{1})^2 = m_q^2 + m_p^2 -2q_{1}p_{1}\\
	& = (-q_{2}+p_{2})^2 = m_q^2 + m_p^2-2q_{2}p_{2},\\
	u & = (q_{1}-p_{2})^2 = m_q^2 + m_p^2-2q_{1}p_{2}\\
	& = (-q_{2}+p_{1})^2 = m_q^2 + m_p^2-2q_{2}p_{1},\\
	s & = (p_1+p_2)^2= (q_1+q_2)^2,  \\
	 s & + t + u = \sum_{i=1}^4 m_i^2,
	\label{Mandelstam}
\end{split}
\end{equation}
where $m_i$ are the masses of initial and final state particles $i$ for which $q^2=m^2_q$ and $p^2=m^2_p$ as defined in fig.~\ref{geometry}. 

\item The kinematics of the scattering of two initial-state particles is shown in fig.~\ref{geometry} in the centre-of-mass frame. These states with momenta $q_1, q_2$ interact with the same energy and with three momentum vectors that have opposite directions and equal magnitudes. In this frame, the energies $E_{q_1} = E_{q_2}$ and $E_{p_1}=E_{p_2}$. Finally, it is worth noting that the $s$-channel exchange energy $s=(p_1+p_2)^{2}=(q_1+q_2)^{2}$ demands that $E_q=E_p$. Using the geometry of the system, $|\overline{\mathcal{M}}|^{2}$ is reduced to its final form as a function of $\theta$, defined as the angle between the axes of propagation of the initial state particles and the scattered monopoles (see fig.~\ref{geometry}). The Mandelstam variables in this frame take the form $t = m_q^2 + m_p^2-\frac{s}{2}(1-\beta_{p}\beta_{q}\cos\theta)$ and $u =  m_q^2 + m_p^2-\frac{s}{2}(1+\beta_{p}\beta_{q}\cos\theta)$.
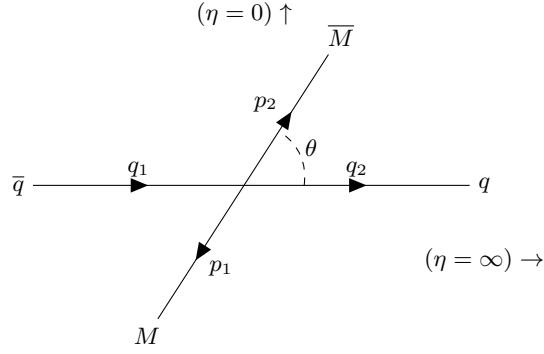
\begin{figure}[H]
\centering
\begin{tikzpicture}
  \begin{feynman}
  	\vertex (a1) {\(\overline{q}\)};
	\vertex[right=3cm of a1] (a2);
	\vertex[right=3cm of a2] (a3) {\(q\)};
	\vertex[below=1cm of a3] (c1) {\((\eta=\infty)\rightarrow\)};
	\vertex[above=2cm of a2] (c2) {\((\eta=0)\uparrow\)};
	\vertex[below=2cm of a2] (a6);
	\vertex[left=1cm of a6] (a7)  {\(M\)};
	\vertex[above=2cm of a2] (a4);
	\vertex[right=1cm of a4] (a5)  {\(\overline{M}\)};
	\vertex[below=1.5cm of a5] (a8);
	\vertex[left=0.2cm of a8] (a9) {\(\theta\)};
	\vertex[right=0.8cm of a2] (b1);
	\vertex[above=0.7cm of a2] (b3);
	\vertex[right=0.5cm of b3] (b2);
	\diagram* {
       		(a2) -- [fermion, edge label=\(p_{1}\)] (a7),
		(a2) -- [fermion, edge label=\(p_{2}\)] (a5),
		(a1) -- [fermion, edge label=\(q_{1}\)] (a2) -- [fermion, edge label=\(q_{2}\)] (a3),
		(b1) -- [scalar, bend right] (b2)
      };
\end{feynman}
\end{tikzpicture}
\caption{Definition of coordinates pertaining to the particle-antiparticle pair production from the interaction of two initial-state particles in the centre-of-mass frame. The process has cylindrical symmetry in the longitudinal direction ($q_i$), perpendicular to which the azimuth $\phi$ is defined. The scattering angle $\theta$ and pseudorapidity $\eta$ of the (final-state) $M$ and $\overline{M}$ particles is also defined. }
\label{geometry}
\end{figure}

\item At this point, $|\overline{\mathcal{M}}|^{2}$ is inserted into the expression for the angular cross section distribution in~\eqref{Xsecdeff}. The behaviour of the cross section can be studied as a function of kinematic variables other than $\theta$ by a change of variable. Specifically, experimental monopole searches are usually interested in expressing the pertinent cross sections in terms of the pseudorapidity $\eta$, defined as
\begin{align}\label{rapidity}
	 \eta \equiv -\ln\tan\frac{\theta}{2}~. 
\end{align}
The latter represents an angular coordinate relative to the beam axis as identified in fig.~\ref{geometry}, or alternatively the significance of the longitudinal boost relative to the total momentum. It is also the high energy limit of the rapidity $y$, which is the relativistic, i.e.\ Lorentz-invariant, realisation of velocity. It becomes the new kinematic variable through the coordinate transformation 
\begin{align}\label{drapidity}
	 \frac{1}{\cosh^{2}\eta}\dd\eta=\frac{1}{2\pi} \dd\Omega~. 
\end{align}
These cross section distributions are plotted in an effort to understand the behaviour of the particles scattering from these interactions according to various phenomenological models. In view of~\eqref{etogb}, the $\beta$ dependence of the coupling acts as a scaling factor on the distributions. In the first part of this work, the distributions in the $\beta$-independent-magnetic-coupling case are shown explicitly for brevity; those in the $\beta$-dependent case can be recovered by multiplying the vertical axes by a factor $\beta$. 

\item Finally, the total cross section $\sigma$ is evaluated as the definite integral of \eqref{Xsecdeff} over the solid angle $\dd\Omega=\dd\phi \, \dd\!\cos\theta$. The longitudinal angle $\phi$ spans from $\phi_0=0$ to $\phi_1=2\pi$, while the integration limits on the scattering angle are $\cos\theta_0=-1$ and $\cos\theta_1=1$.

\end{enumerate} 

\section{Detailed cross-section calculation}\label{app:xsection}

This appendix gives a more detailed discussion on how the analytic expressions for cross section distributions were evaluated. We outline below the basic steps leading to the evaluation of the pertinent cross sections for monopole production. 

\subsection{Spin-0 monopole cross section}

Scalar monopole production by PF manifests itself through three processes: a $t$-channel process, a $u$-channel process and a 4-point interaction as shown in fig.~\ref{Spinzerographs}. In fig.~\ref{FeynmanRulesSpinZero}, $\epsilon_{\lambda}(q)$ is the polarisation vector of the photon and the magnetic charge $g(\beta)$ is linearly dependent on the monopole boost, $\beta=|\vec{p}| / E_p$. The amplitudes are derived from the Feynman rules of \eqref{FeynRulesSpinZero}:

\begin{figure}[ht!]
\justify
\begin{minipage}[t]{0.5\textwidth}
\centering
\begin{tikzpicture}
  \begin{feynman}
  	\vertex (a1) {\(\mathcal{A}_{\rho},\epsilon_{\lambda}\)};
	\vertex[below=2cm of a1] (a2);
	\vertex[left=1.75cm of a2] (a3);
	\vertex[below=1cm of a3] (a4) {\(M\)};
	\vertex[right=1.75cm of a2] (a5);
	\vertex[below=1cm of a5] (a6) {\(M\)};
	
	\diagram* {
       		(a4) -- [scalar, edge label=\(p_{1_{\mu}}\)] (a2) -- [scalar, edge label=\(p_{2_{\nu}}\)] (a6),
		(a1) -- [boson, edge label=\(q_{\rho}\)] (a2),
      };
\end{feynman}
\end{tikzpicture}
\end{minipage}
~
\begin{minipage}[t]{0.5\textwidth}
\centering
\begin{tikzpicture}
  \begin{feynman}
  	\vertex (a2);
	\vertex[left=1.75cm of a2] (a3);
	\vertex[below=1cm of a3] (a4) {\(M\)};
	\vertex[right=1.75cm of a2] (a5);
	\vertex[below=1cm of a5] (a6) {\(M\)};
	\vertex[above=1cm of a3] (a8) {\(\mathcal{A}_{\rho},\epsilon_{\lambda}\)};
	\vertex[above=1cm of a5] (a7) {\(\mathcal{A}_{\sigma},\epsilon_{\lambda'}\)};
	\diagram* {
       		(a4) -- [scalar, edge label=\(p_{1_{\mu}}\)] (a2) -- [scalar, edge label=\(p_{2_{\nu}}\)] (a6),
		(a8) -- [boson, edge label=\(q_{1_{\rho}}\)] (a2) -- [boson, edge label=\(q_{2_{\sigma}}\)] (a7)
      };
\end{feynman}
\end{tikzpicture}
\end{minipage}
\begin{equation} \label{FeynRulesSpinZero}
\propto -ig(\beta) (p_{1}+p_{2})_{\mu} \qquad \qquad \qquad \qquad \qquad \qquad \qquad \qquad \qquad \qquad
\propto 2ig^{2}(\beta)g_{\mu\nu}
\end{equation}
\caption{In SQED, spin-$0$ scalar bosons (dashed lines) interact with photons (wavy lines) according to the vertices in the figure. The corresponding Feynman rules are given in \eqref{FeynRulesSpinZero}. }
\label{FeynmanRulesSpinZero}
\end{figure}

\begin{subequations} \label{MatAmplSpinZero}
	\begin{align}
	\mathcal{M}_{t} & =\varepsilon_{\lambda^{'}}^{\nu} ig(\beta) (-k_{\nu}+p_{2\nu}) \frac{i}{k^{2}-M^{2}} ig(\beta) (-k_{\mu}-p_{1\mu})\varepsilon_{\lambda}^{\mu} \label{MatAmplSpinZeroa} \\ 
	\mathcal{M}_{u} & =\varepsilon_{\lambda}^{\nu} ig(\beta) (-\widetilde{k_{\nu}}+p_{2\nu}) \frac{i}{\widetilde{k^{2}}-M^{2}} ig(\beta) (-p_{1\mu}-\widetilde{k_{\mu}})\varepsilon_{\lambda^{'}}^{\mu} \label{MatAmplSpinZerob}  \\
	\mathcal{M}_{4} & = 2ig^{2}(\beta)g_{\mu\nu}\varepsilon_{\lambda}^{\mu} \varepsilon_{\lambda^{'}}^{\nu}. \label{MatAmplSpinZeroc}
	\end{align}
\end{subequations}
As usual, the spin-0 monopole has mass $M$ and momenta $p_{1}$ and $p_{2}$, the photons have polarisation vectors $\varepsilon_{\lambda}$ and $\varepsilon_{\lambda^{'}}$ and momenta $q_{1}$ and $q_{2}$. These are indicated in fig.~\ref{Spinzerographs} where $p_{i_{\mu}}^{2}=M^{2}$ and $q_{i_{\mu}}^{2}=0$. $k$ and $\tilde{k}$ are the $t$- and $u$-channel exchange momenta, respectively.

The matrix amplitudes are reduced to the following forms in order to remove the $k$ and $\tilde{k}$ dependences.
\begin{eqnarray}\label{RedMatAmplSpinZero}
\notag	\mathcal{M}_{t} &=& -ig^{2}(\beta) \varepsilon_{\lambda^{'}}^{\nu} (2p_{2\nu}-q_{2\nu}) \frac{1}{2p_{1}q_{1}} (2p_{1\mu}-q_{1\mu})\varepsilon_{\lambda}^{\mu} \\ 
\notag	\mathcal{M}_{u} &=& -ig^{2}(\beta) \varepsilon_{\lambda}^{\nu} (2p_{2\nu}-q_{1\nu}) \frac{1}{2p_{2}q_{1}} (2p_{1\mu}-q_{2\mu})\varepsilon_{\lambda^{'}}^{\mu}  \\
	\mathcal{M}_{4} &=& 2ig^{2}(\beta)g_{\mu\nu}\varepsilon_{\lambda}^{\mu} \varepsilon_{\lambda^{'}}^{\nu} 
	\end{eqnarray}
The total matrix amplitude $\mathcal{M}_{PF}=\mathcal{M}_{t}+\mathcal{M}_{u}+\mathcal{M}_{4}$ is squared, summed over final states and averaged over photon polarisation states.  As a non-trivial but quite useful example of how the rules in appendix~\ref{app:basic} are used, the next paragraphs detail the analytical procedure involved in calculating $|\overline{\mathcal{M}}|^2_{PF}$ from the matrix amplitudes in eqs.~\eqref{RedMatAmplSpinZero}. 

\begin{eqnarray}\label{mtmt}
 \sum_{pol}\mathcal{M}_{t}\mathcal{M}_{t}^{*}  & = & \sum_{\lambda,\lambda^{'}}[-ig^{2}(\beta) \varepsilon_{\lambda^{'}}^{\nu} (2p_{2\nu}-q_{2\nu}) \frac{1}{2p_{1}q_{1}} (2p_{1\mu}-q_{1\mu})\varepsilon_{\lambda}^{\mu}]\cdot[ie^{2} \varepsilon_{\lambda}^{\sigma*} (2p_{1\sigma}-q_{1\sigma}) \frac{1}{2p_{1}q_{1}} (2p_{2\rho}-q_{2\rho})\varepsilon_{\lambda^{'}} ^{\rho*}] \nonumber \\ \nonumber\\
& = & g^{4}(\beta) (2p_{2\nu}-q_{2\nu})^{2} \frac{1}{(2p_{1}q_{1})^{2}} (2p_{1\mu}-q_{1\mu})^{2}  = g^{4}(\beta) \Big[\frac{4M^{4}}{(p_{1}q_{1})^{2}}+4-\frac{8M^{2}}{(p_{1}q_{1})}\Big], \nonumber \\ \nonumber\\
 \sum_{pol}\mathcal{M}_{u}\mathcal{M}_{u}^{*} & = & g^{4}(\beta) \Big[\frac{4M^{4}}{(p_{2}q_{1})^{2}}+4-\frac{8M^{2}}{(p_{2}q_{1})}\Big], \nonumber \\ \nonumber\\
 \sum_{pol}\mathcal{M}_{4}\mathcal{M}_{4}^{*} & = & \sum_{\lambda,\lambda^{'}} 2ig^{2}(\beta)g_{\mu\nu}\varepsilon_{\lambda}^{\mu} \varepsilon_{\lambda^{'}}^{\nu} \cdot (-2i)e^{2}g_{\sigma\rho}\varepsilon_{\lambda}^{\sigma*} \varepsilon_{\lambda^{'}}^{\rho*} \nonumber \\ \nonumber\\
 & = & 4 g_{\sigma\mu}g^{\sigma\mu}  = 16,  \nonumber \\ \nonumber\\
 2\sum_{pol}\mathcal{M}_{t}\mathcal{M}_{u}^{*} & = & 2\sum_{\lambda,\lambda^{'}}[-ig^{2}(\beta) \varepsilon_{\lambda^{'}}^{\nu} (2p_{2\nu}-q_{2\nu}) \frac{1}{2p_{1}q_{1}} (2p_{1\mu}-q_{1\mu})\varepsilon_{\lambda}^{\mu}]\cdot[ie^{2} \varepsilon_{\lambda^{'}}^{\rho*} (2p_{1\rho}-q_{2\rho}) \frac{1}{2p_{2}q_{1}} (2p_{2\sigma}-q_{1\sigma})\varepsilon_{\lambda}^{\sigma*}] \nonumber  \\ \nonumber\\
& = & \frac{1}{(p_{1}q_{1})(p_{2}q_{1})} (2(p_{2}q_{1})^{2}+2(p_{1}q_{1})^{2}+4(p_{2}q_{1})(p_{1}q_{1})+8M^{4}-8M^{2}(p_{2}q_{1}+p_{1}q_{1})) \nonumber \\ \nonumber\\
 2\sum_{pol}\mathcal{M}_{t}\mathcal{M}_{4}^{*} & = & 2\sum_{\lambda,\lambda^{'}}[-ie^{2} \varepsilon_{\lambda^{'}}^{\nu} (2p_{2\nu}-q_{2\nu}) \frac{1}{2p_{1}q_{1}} (2p_{1\mu}-q_{1\mu})\varepsilon_{\lambda}^{\mu}]\cdot [-2ig^{2}(\beta)g_{\sigma\rho}\varepsilon_{\lambda}^{\sigma*} \varepsilon_{\lambda^{'}}^{\rho*}] 
= -10 + \frac{8M^{2}}{p_{1}q_{1}}-\frac{2p_{2}q_{1}}{p_{1}q_{1}} \nonumber \\ \nonumber\\
 2\sum_{pol}\mathcal{M}_{u}\mathcal{M}_{4}^{*} & = & 2\sum_{\lambda,\lambda^{'}}[-ie^{2} \varepsilon_{\lambda}^{\nu} (2p_{2\nu}-q_{1\nu}) \frac{1}{2p_{2}q_{1}} (2p_{1\mu}-q_{2\mu})\varepsilon_{\lambda^{'}}^{\mu}]\cdot [-2ig^{2}(\beta)g_{\sigma\rho}\varepsilon_{\lambda}^{\sigma*} \varepsilon_{\lambda^{'}}^{\rho*}] =-10 + \frac{8M^{2}}{p_{2}q_{1}}-\frac{2p_{1}q_{1}}{p_{2}q_{1}}. \label{mum4}
\end{eqnarray}

Hence, using the Mandelstam variables \eqref{Mandelstam}, one obtains  
\begin{align}
\notag	|\overline{\mathcal{M}}|^{2}_{PF} & =\frac{2!}{4}\sum_{pol}\Big[\mathcal{M}_{t}\mathcal{M}_{t}^{*}+\mathcal{M}_{u}\mathcal{M}_{u}^{*}+\mathcal{M}_{4}\mathcal{M}_{4}^{*}+2\mathcal{M}_{t}\mathcal{M}_{4}^{*}+2\mathcal{M}_{u}\mathcal{M}_{4}^{*}+2\mathcal{M}_{t}\mathcal{M}_{u}^{*}\Big] \\
\notag	& = 2g^{4}(\beta) \Big\{1+\Big[1-\Big(\frac{m^{2}(p_{1}q_{1}+p_{2}q_{1})}{(p_{1}q_{1})(p_{2}q_{1})}\Big)\Big]^{2}\Big\} = 2g^{4}(\beta)\Big\{1+\Big[1-\Big(\frac{2M^{2}\sgg}{(t-M^{2})(u-M^{2})}\Big)\Big]^{2}\Big\} \\
	& = 2g^{4}(\beta)\Big\{ 2 + \frac{4M^{4}}{(t-M^{2})^{2}} +  \frac{4M^{4}}{(M^{2}-\sgg-t)^{2}} + \frac{4M^{2}}{(t-M^{2})} + \frac{4M^{2}}{(M^{2}-\sgg-t)}+\frac{8M^{4}}{(t-M^{2})(M^{2}-\sgg-t)} \Big\}, \label{TotMatEl}
\end{align}
where the $2!$ is the symmetry factor of the final states.  In terms of the angle $\theta$, using the definition of the Mandelstam variable $t$~\eqref{Mandelstam}, which in this case reads:
\begin{equation}\label{Mandt}
t  = M^{2} - 2E^{2} + 2\vec{q}_1\cdot\vec{p}_1= M^{2} - \frac{\sqq}{2}(1 - \beta \cos{\theta}), 
\end{equation}
the expression is reduced to:
\begin{align}
	|\overline{\mathcal{M}}|^{2}_{PF} & = 2g^{4}(\beta) \Big\{1+\Big[1-\Big(\frac{8M^{2}}{\sgg(1-\beta^{2}\cos^{2}\theta)}\Big)\Big]^{2}\Big\}. \label{TotMatElOmega}
\end{align}
where $\beta$ is defined in~\eqref{btos}. 

 In the centre-of-mass frame, $\sgg=4E_{q_{1}}E_{q_{2}}$ and the photon three-momenta $\vec{q}_1=\vec{q_{2}}$ are defined as $|\vec{q}|=E_{q}$ for any massless photon. It follows from this that the $s$-channel exchange energy $\sgg=(p_1+p_2)^2=(q_1+q_2)^2$ requires that $E_q=E_p$. 
 As discussed in section~\ref{sectionAmtoKin}, this squared matrix amplitude is used to evaluated the kinematic distributions in terms of the angle $\theta$ and the pseudorapidity $\eta$.
 The total cross section in the centre-of-mass frame is obtained by integrating \eqref{TotMatElOmega} over the solid angle $\dd\Omega = \dd\phi\, \dd\!\cos\theta$:
\begin{equation} \label{dsdO}
	\sigma^{S=0}_{\gamma\gamma\rightarrow \overline{M}M} =\frac{1}{64\pi^{2} \sgg}\frac{|\vec{p}_1|}{|\vec{q}_1|}\int_0^{2\pi}\, \int_{-\pi}^0\, |\overline{\mathcal{M}}|^{2}_{PH} \, \dd\phi\, \dd\!\cos\theta.
\end{equation} 

\begin{figure}[ht!]
\justify
\begin{minipage}[t]{0.5\textwidth}
\centering
\begin{tikzpicture}
  \begin{feynman}
  	\vertex (a1) {\(\mathcal{A}_{\mu}\)};
	\vertex[right=2cm of a1] (a2);
	\vertex[above=1.75cm of a2] (a3);
	\vertex[right=1cm of a3] (a4) {\(M\)};
	\vertex[below=1.75cm of a2] (a5);
	\vertex[right=1cm of a5] (a6) {\(\overline{M}\)};
	\vertex[right=3cm of a2] (b1) {\((a)\)};

	\diagram* {
       		(a6) -- [scalar, edge label=\(p_{2\sigma}\)] (a2) -- [scalar, edge label=\(p_{1\rho}\)] (a4),
		(a1) -- [boson, edge label=\(k_{\mu}\)] (a2),
      };
\end{feynman}
\end{tikzpicture}
\end{minipage}
\begin{minipage}[t]{0.5\textwidth}
\centering
\begin{tikzpicture}
  \begin{feynman}
  	\vertex (a1) {\(\mathcal{A}_{\mu}\)};
	\vertex[right=2cm of a1] (a2);
	\vertex[above=1.75cm of a2] (a3);
	\vertex[right=1cm of a3] (a4) {\(q\)};
	\vertex[below=1.75cm of a2] (a5);
	\vertex[right=1cm of a5] (a6) {\(\overline{q}\)};
	\vertex[right=3cm of a2] (b1) {\((b)\)};
	
	\diagram* {
       		(a6) -- [fermion, edge label=\(q_{2\sigma}\)] (a2) -- [fermion, edge label=\(q_{1\rho}\)] (a4),
		(a1) -- [boson, edge label=\(k_{\mu}\)] (a2),
      };
\end{feynman}
\end{tikzpicture}
\end{minipage}
\begin{equation} \label{vertexScalar}
\propto -ig(\beta) (p_{1}+p_{2})_{\mu} \;\;\;\;\;\;\;\;\;\;\;\;\;\;\;\;\;\;\;\;\;\;\;\;\;\;\;\;\;\;\;\;\;\;\;\;\;\;\;\;\;\;\;\;\;\;\;\;\;\;\;\;\;\;\;\;\;\;\;\;\;\;\;\;\;\;\;\propto-iQe \gamma^{\mu}\delta_{ij}
\end{equation}
\caption{(a) Three-point vertex representing the tree-level interaction of scalar monopoles (dashed lines) with photons (wavy lines). (b) The quark-photon three point vertex as in ordinary QED. The corresponding Feynman rules are given in \eqref{vertexScalar}, where $g(\beta)$ is the (in general $\beta$-dependent) monopole magnetic charge.}\label{vetexScalara}
\end{figure}
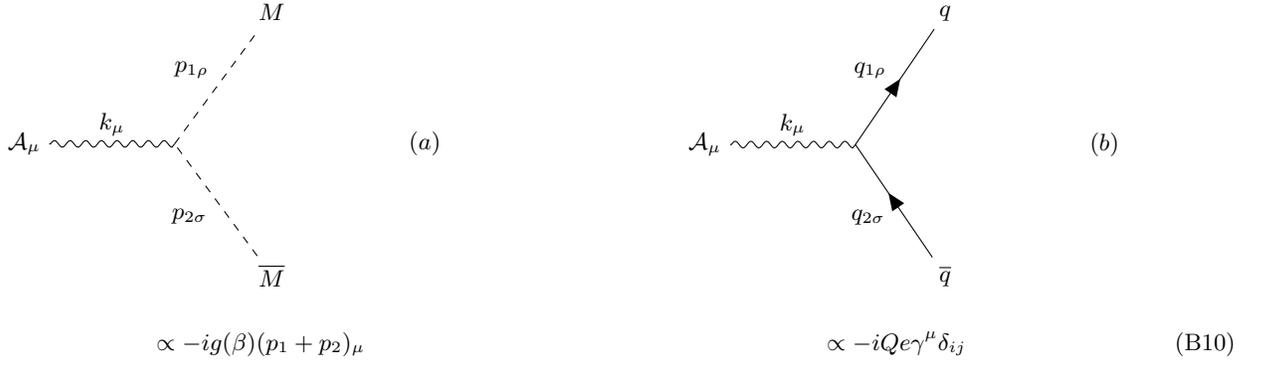

Scalar monopole pairs produced by the DY digram shown in fig.~\ref{FeynScalar} have kinematic distributions the analytic forms of which are derived analogously to the PF case. The total matrix amplitude, derived from the Feynman rules in fig.~\ref{vetexScalara}, is given by 
\begin{equation}\label{DrellYanMatScalar}
\mathcal{M}_{DY}=u_{\alpha i}(-iQe\gamma_{\mu}\delta_{ij})\overline{v}_{\beta j}(\frac{-ig^{\mu\nu}}{k^{2}})(-ig(\beta))(p_{1\nu}-p_{2\nu}),
\end{equation}
where $u_{\alpha i}, v_{\beta j}, \overline{u}_{\alpha i}, \overline{v}_{\beta j}$ are the quark fermionic spinors with $\delta_{ij}$ the colour index factor and $g^{\mu\nu}$ is the Minkowski space-time metric tensor. The quarks have a mass $m$ each and are characterised by momentum 4-vectors $q_{1\mu}$ and $q_{2\mu}$, where $q_{1,2}^2=m^2$. Similarly, the scalar monopoles have mass $M$ and momentum 4-vectors $p_{1\mu}$ and $p_{2\mu}$, where, on shell, $p_{1,2}^2=M^2$. The quark masses are negligible compared to a monopole mass. The quantity $e$ represents the positron charge and $Q$ is the charge fraction relevant for quarks, $Q=\frac{1}{3},\frac{2}{3}$.

The squared matrix amplitude, averaged over quark spins and colours, $|\overline{\mathcal{M}^{2}}|_{DY}$ is then computed:
\begin{eqnarray}
\notag	 |\overline{\mathcal{M}^{2}}|_{DY} 	& =\ & \sum_Q(\frac{3Q^2e^2g^2(\beta)}{k^4})\frac{2!}{4}\frac{1}{3} \Tr\Big[(u_{\alpha i}\gamma_{\mu}\delta_{ij}\overline{v}_{\beta j})(g^{\mu\nu})(p_{1\nu}-p_{2\nu})(p_{1\nu'}-p_{2\nu'})(g^{\nu'\mu'})(v_{\beta j'}\gamma_{\mu'}\delta_{j'i'}\overline{u}_{\alpha i'})\Big]  \\
 \notag								& = & \sum_Q(\frac{3Q^2e^2g^2(\beta)}{k^4})\frac{1}{3} \Tr[\delta_{i'i}\delta_{ij}\delta_{jj'}\delta_{j'i'}]\frac{1}{2} \Tr\Big[(\slashed{q_{1}}+m)\gamma_{\mu}(\slashed{q_{2}}-m)\gamma_{\mu'}\Big](p_1^{\mu}-p_2^{\mu})(p_1^{\mu'}-p_2^{\mu'}) \\
								& =  &(\frac{5}{9}\frac{3e^2g^2(\beta)}{2k^4})\Big[q_1^{\sigma}q_{2}^{\rho}(4g_{\sigma\mu}g_{\rho\mu'}-4 g_{\sigma\rho} g_{\mu\mu'}+4 g_{\sigma\mu'} g_{\rho\mu})-4m^2\, g_{\mu\mu'}\Big](p_1^{\mu}-p_2^{\mu})(p_1^{\mu'}-p_2^{\mu'}).  \label{DrellYanSquMatScalarc}
\end{eqnarray}
The $2!$, $\frac{1}{4}\frac{1}{3}$ factors account for the symmetry factor of the final states, and the averaging over spins and colour states. The factor of $3$ includes the contribution from 3 quark flavours. Finally, the sum over quark charges adds a factor of $\frac{5}{9}$ to the result. 
The rest of the expression was found using the spin sum rules and properties of $\gamma$-matrices 
listed in appendix~\ref{app:basic} (see eqs.~\eqref{spinsum}), \eqref{evaltrace2}).

Using the relationship between the Mandelstam variables \eqref{Mandelstam} and the scattering angle between the trajectories of the incoming quarks and outgoing scalar bosons as depicted in fig.~\ref{geometry}, 
\begin{align}
t & =  m^2+M^2-\frac{\sqq}{2}(1-\beta_{q}\beta_{p}\cos(\theta)) \\
u & =  m^2+M^2-\frac{\sqq}{2}(1+\beta_{q}\beta_{p}\cos(\theta)),
\end{align}
with $\beta_{q} = |q|/E_q $ and $\beta_{p} = |p|/E_p$, \eqref{DrellYanSquMatScalarc} simplifies  to: 
\begin{subequations}
	\begin{align}
	 |\overline{\mathcal{M}^{2}}|_{DY}	& =\frac{15}{9}(\frac{-e^2g^2(\beta_p)}{\sqq^2})((-t+u)^2+4M^2-\sqq^2)\label{finalMatrixScalara} \\
 								& =\frac{5}{3} e^2g^2(\beta_p)\beta_{p}^2(1-\beta_{q}^2\cos(\theta)^2)\label{finalMatrixScalarb},
	\end{align}
\end{subequations}
which is used to compute the differential cross section.
\begin{align}\label{Xsecdeff0}
\frac{\dd\sigma_{q\overline{q}\rightarrow M\overline{M}}^{S=0}}{\dd\Omega} & =\frac{1}{64\pi^{2} \sqq}\frac{|\vec{p}_1|}{|\vec{q}_1|}|\overline{\mathcal{M}}|^{2}_{DY}
	 =\frac{1}{64\pi^{2} \sqq}\frac{\beta_{p}}{\beta_{q}}|\overline{\mathcal{M}}|^{2}_{DY}.
\end{align}
 The quark masses are assumed negligible at this point, hence $\beta_q=1$ and the subscript is dropped on $\beta_p$. The total cross section is obtained by integrating over the solid angle as in the photon fusion case.

\subsection{Spin-\half monopole cross section}

Starting with pair production by PF, the kinematic distributions for spin-\half monopoles are derived. The total matrix amplitude for the process is the sum of a $t$- and a $u$-channel process, shown in fig.~\ref{spinhalfgraphs}, $\mathcal{M}_{PH}=\mathcal{M}_t+\mathcal{M}_u$. Each amplitude is derived from the Feynman rule \eqref{FeynRuleS12eq}, shown in fig.~\ref{FeynRuleS12}. The fermion in the latter may represent a spin-\half monopole with magnetic moment $\kappa \ne 0$, upon dualising the theory by replacing the electric charge by a (in general $\beta$-dependent) magnetic charge $g(\beta)$. The amplitude in the latter case is given in eq.~\eqref{FeynRuleS12eq} where the influence of the added magnetic moment term is visible. The term $[\gamma^{\mu},\gamma^{\nu}]$ is a commutator of $\gamma$ matrices and $q_{\mu}$ is the photon four-momentum.
\begin{subequations} \label{matrixel}
	\begin{align}
	\mathcal{M}_{t}&= -ig(\beta)^{2} \varepsilon_{\lambda\mu} \overline{u}_{s}(\gamma^{\mu}+\frac{1}{2}\kappa q_{1\sigma}[\gamma^{\sigma},\gamma^{\mu}]) \frac{(\slashed{k}+M)}{k^{2}-M^{2}}(\gamma^{\nu}+\frac{1}{2}\kappa q_{2\sigma}[\gamma^{\sigma},\gamma^{\nu}]) v_{r}\varepsilon_{\lambda^{'}\nu} \\ 
	\mathcal{M}_{u}&=-ig(\beta)^{2} \varepsilon_{\lambda^{'}\mu} \overline{u}_{s}(\gamma^{\mu}+\frac{1}{2}\kappa q_{2\sigma}[\gamma^{\sigma},\gamma^{\mu}])  \frac{(\tilde{\slashed{k}}+M)}{\tilde{k}^{2}-M^{2}}(\gamma^{\nu}+\frac{1}{2}\kappa q_{1\sigma}[\gamma^{\sigma},\gamma^{\nu}])v_{r}\varepsilon_{\lambda\nu}
	\end{align}
\end{subequations}

\begin{figure}[ht!]
\centering
\begin{tikzpicture}
  \begin{feynman}
  	\vertex (a1) {\(\mathcal{A}_{\mu}\)};
	\vertex[below=2cm of a1] (a2);
	\vertex[left=1.75cm of a2] (a3);
	\vertex[below=1cm of a3] (a4) {\(\psi\)};
	\vertex[right=1.75cm of a2] (a5);
	\vertex[below=1cm of a5] (a6) {\(\psi\)};
	
	\diagram* {
       		(a4) -- [fermion, edge label=\(p_{1\mu}\)] (a2) -- [fermion, edge label=\(p_{2\mu}\)] (a6),
		(a1) -- [boson, edge label=\(q_{\mu}\)] (a2),
      };
\end{feynman}
\end{tikzpicture}
\begin{equation} \label{FeynRuleS12eq}
\propto -ig(\beta)\gamma^{\nu}-i\kappa\frac{1}{2} g(\beta)q_{\mu}[\gamma^{\mu},\gamma^{\nu}]
\end{equation}
\caption{Interaction vertex representing the only coupling of a spin-\half fermion (continuous line), which may be a monopole, and a photon (wavy line)  in $U(1)$ gauge invariant quantum-electrodynamics. The corresponding Feynman rule is given in \eqref{FeynRuleS12eq}.}
\label{FeynRuleS12}
\end{figure}

It is squared, averaged over photon polarisation states, and summed over final-state monopole spins. The expression becomes:
\begin{eqnarray}\label{TotalSquaredMatrixAmpSpin12}
|\overline{\mathcal{M}}|^{2}_{PH}&=& \frac{1}{4}\sum_{\lambda\lambda'}\sum_{\alpha\alpha'}\Big[\mathcal{M}_{t}\mathcal{M}_{t}^{*}+\mathcal{M}_{u}\mathcal{M}_{u}^{*}+2\mathcal{M}_{t}\mathcal{M}_{u}^{*}\Big]   \nonumber \\ 
&=&\frac{1}{4} \sum_{\lambda\lambda'}\epsilon^{*\nu}_{\lambda'}\epsilon^{\mu}_{\lambda}\epsilon^{\nu'}_{\lambda'}\epsilon^{*\mu'}_{\lambda}\sum_{\alpha\alpha'}\Big[\overline{u}_{\alpha}\Gamma_{\mu}\Delta_{f}\Gamma_{\nu}v_{\alpha'}+\overline{u}_{\alpha}\Gamma_{\nu}\Delta_{f}\Gamma_{\mu} v_{\alpha'}\Big]\Big[u_{\alpha}\Gamma^*_{\mu'}\Delta_{f}\Gamma^*_{\nu'}\overline{v}_{\alpha'}+u_{\alpha}\Gamma^*_{\nu'}\Delta_{f}\Gamma^*_{\mu'} \overline{v}_{\alpha'}\Big]^{T} \nonumber \\
	&= &\frac{1}{4} (-g^{\mu\mu'})(-g^{\nu\nu'}) \Tr\Big[\Big(\overline{u}_{\alpha}\Gamma_{\mu}\Delta_{f}\Gamma_{\nu}v_{\alpha'}+\overline{u}_{\alpha}\Gamma_{\nu}\Delta_{f}\Gamma_{\mu} v_{\alpha'}\Big)\Big(u_{\alpha}\Gamma^*_{\mu'}\Delta_{f}\Gamma^*_{\nu'}\overline{v}_{\alpha'}+u_{\alpha}\Gamma^*_{\nu'}\Delta_{f}\Gamma^*_{\mu'} \overline{v}_{\alpha'}\Big)^{T}\Big],
\end{eqnarray}
where $u_{\alpha}$ and $v_{\alpha'}$ denote the monopole spinors, $\Gamma_{\nu}$ represents the $\kappa$ (or $\tilde{\kappa}$) dependent coupling at the vertices and $\Delta_{f}$ is the fermionic monopole propagator. The necessary sums and traces are listed in appendix~\ref{app:basic} (see eqs.~\eqref{evaltrace1}, \eqref{spinsum} and \eqref{evaltrace2}). The expression is reduced in \Math and the differential cross section is evaluated:
\begin{equation}
	\frac{\dd\sigma^{S=\frac{1}{2}}_{\gamma\gamma\rightarrow\overline{M}M}}{\dd\Omega} =\frac{1}{64\pi^{2} \sgg}\beta|\overline{\mathcal{M}}|^{2}_{PH}
\end{equation}
where the monopole boost is $\beta=|\vec{p}| / E_p$ and the photon momentum is $|\vec{q}|=E_q$, recognising also that $E_q=E_p$ in the centre-of-mass frame. It is expressed in terms of the scattering angle $\theta$ and the pseudorapidity $\eta$ as defined in fig.~\ref{geometry} and explained in section \ref{sectionAmtoKin}.

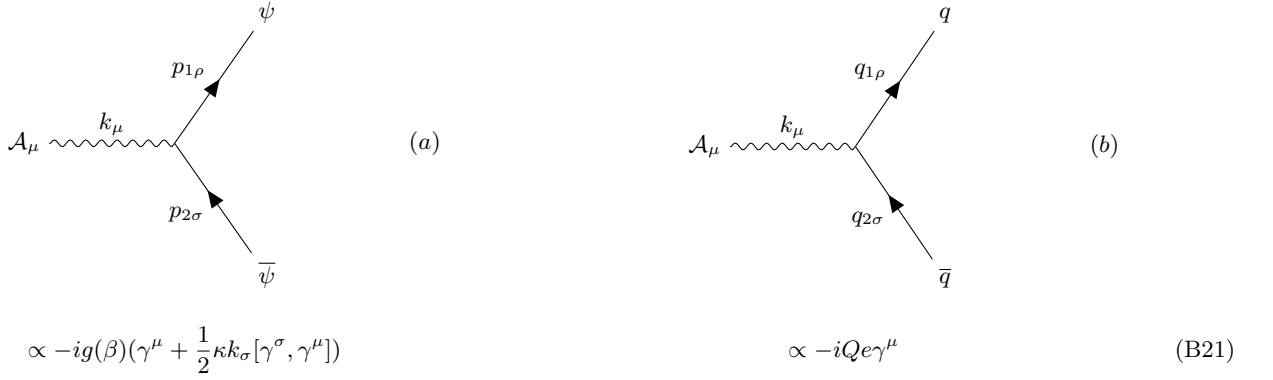
\begin{figure}[ht!]
\justify
\begin{minipage}[t]{0.5\textwidth}
\centering
\begin{tikzpicture}
  \begin{feynman}
  	\vertex (a1) {\(\mathcal{A}_{\mu}\)};
	\vertex[right=2cm of a1] (a2);
	\vertex[above=1.75cm of a2] (a3);
	\vertex[right=1cm of a3] (a4) {\(\psi\)};
	\vertex[below=1.75cm of a2] (a5);
	\vertex[right=1cm of a5] (a6) {\(\overline{\psi}\)};
	\vertex[right=3cm of a2] (b1) {\((a)\)};

	\diagram* {
       		(a6) -- [fermion, edge label=\(p_{2\sigma}\)] (a2) -- [fermion, edge label=\(p_{1\rho}\)] (a4),
		(a1) -- [boson, edge label=\(k_{\mu}\)] (a2),
      };
\end{feynman}
\end{tikzpicture}
\end{minipage}
\begin{minipage}[t]{0.5\textwidth}
\centering
\begin{tikzpicture}
  \begin{feynman}
  	\vertex (a1) {\(\mathcal{A}_{\mu}\)};
	\vertex[right=2cm of a1] (a2);
	\vertex[above=1.75cm of a2] (a3);
	\vertex[right=1cm of a3] (a4) {\(q\)};
	\vertex[below=1.75cm of a2] (a5);
	\vertex[right=1cm of a5] (a6) {\(\overline{q}\)};
	\vertex[right=3cm of a2] (b1) {\((b)\)};
	
	\diagram* {
       		(a6) -- [fermion, edge label=\(q_{2\sigma}\)] (a2) -- [fermion, edge label=\(q_{1\rho}\)] (a4),
		(a1) -- [boson, edge label=\(k_{\mu}\)] (a2),
      };
\end{feynman}
\end{tikzpicture}
\end{minipage}
\begin{equation} \label{vertexSpinor}
\propto-ig(\beta) (\gamma^{\mu}+\frac{1}{2}\kappa k_{\sigma}[\gamma^{\sigma},\gamma^{\mu}]) \qquad \qquad \qquad \qquad \qquad \qquad \qquad \qquad \qquad \propto -iQe \gamma^{\mu} \qquad \qquad \qquad \qquad
\end{equation}
\caption{(a) Spin-\half fermionic monopoles interact with photons at the three-point vertex, where $k_{\sigma}$ is the photon momentum and $g(\beta)$ is the magnetic coupling. (b) The quarks couple to the photon as expected form Dirac QED for spinors with fractional charges, $Qe$, where $Q$ is the charge fraction relevant to the quarks involved, $Q=\frac{1}{3},\frac{2}{3}$. The corresponding Feynman rule is given in \eqref{vertexSpinor}. }\label{vetexSpinora}
\end{figure}

 Moving on now to monopole pair production by DY, the matrix amplitude is derived from the Feynman rule \eqref{vertexSpinor} shown in fig.~\ref{vetexSpinora}.
 \begin{equation} \label{Matampspinor}
\mathcal{M}_{DY}=u_{\beta }(-iQe\gamma^{\mu})\overline{v}_{\beta' }(\frac{-ig_{\mu\nu}}{k^{2}})a_{\alpha}(-ig(\beta))(\gamma^{\nu}+\frac{1}{2}\kappa k_{\sigma}[\gamma^{\sigma},\gamma^{\nu}]) \overline{b}_{\alpha'}
\end{equation}

 The squared matrix amplitude is given by:
 \begin{equation}
 \begin{split}\label{squmatamp12Dy}
|\overline{\mathcal{M}}|^2_{DY}=&\sum_{\beta\beta'}\sum_{\alpha\alpha'}3\frac{5}{9}\frac{1}{4}\frac{1}{3}\Big[u_{\beta}(e\gamma^{\mu})\overline{v}_{\beta' }(\frac{g_{\mu\nu}}{k^{2}})a_{\alpha}(g(\beta))(\gamma^{\nu}+\frac{1}{2}\kappa k_{\sigma}[\gamma^{\sigma},\gamma^{\nu}]) \overline{b}_{\alpha'}\Big]\\
&\Big[ b_{\alpha'}(\gamma^{\nu'}+\frac{1}{2}\kappa k_{\sigma'}[\gamma^{\sigma'},\gamma^{\nu'}])(g(\beta))\overline{a}_{\alpha}(\frac{g_{\mu'\nu'}}{k^{2}})v_{\beta' }(e\gamma^{\mu'}) \overline{u}_{\beta}\Big]\\
	=&\frac{5e^2g(\beta)^2}{36} \Tr\Big[u_{\beta}(\gamma^{\mu})\overline{v}_{\beta' }v_{\beta' }(\gamma^{\mu'}) \overline{u}_{\beta }\Big] (\frac{g_{\mu\nu}}{k^{2}})\\
	&(\frac{g_{\mu'\nu'}}{k^{2}}) \Tr\Big[a_{\alpha}(\gamma^{\nu}+\frac{1}{2}\kappa k_{\sigma}[\gamma^{\sigma},\gamma^{\nu}]) \overline{b}_{\alpha'} b_{\alpha'}(\gamma^{\nu'}+\frac{1}{2}\kappa k_{\sigma'}[\gamma^{\sigma'},\gamma^{\nu'}])\overline{a}_{\alpha}\Big],
\end{split}
 \end{equation}
where on the right-hand-side in the first line, the various symmetry factors are indicated explicitly for the convenience of the reader. 
Specifically, the factors $\frac{1}{4}$ and $\frac{1}{3}$ come from the averaging over spinor states and colour states, respectively. The factor of $3$ accounts for the flavour multiplicity and the factor $\frac{5}{9}$ is the sum over squared quark charges $Q= \frac{2}{3}, \frac{1}{3}$. The sums over spinor states become traces which are evaluated by \Math using the spinor sum rules and the properties of $\gamma$ matrices \eqref{spinsum} in appendix~\ref{app:basic}. The quark trace is evaluated explicitly in \eqref{evaltrace2} of appendix~\ref{app:basic}.  
 
 As in the preceding cases, the differential cross section distributions are computed using \Math as defined in section~\ref{sectionAmtoKin}. As $\beta_p=|\vec{p}| / E_p$ represents the monopole boost, $\beta_q= |\vec{q}| / E_q$ represents the quark boost, and ${E_p}={E_q}$ in the centre-of-mass frame defined in fig.~\ref{geometry}, $|\vec{p}_1| / |\vec{q}_1| = \beta_p / \beta_q$, and hence
\begin{align}
	\frac{\dd\sigma^{S=\frac{1}{2}}_{q\overline{q}\rightarrow M\overline{M}}}{\dd\Omega} &=\frac{1}{64\pi^{2} \sqq}\frac{\beta_p}{\beta_q}|\overline{\mathcal{M}}|^{2}_{DY},
\end{align}
where $\beta_q\rightarrow1$ for quarks of negligible mass (compared to the heavy monopoles), in which case the index can be dropped on $\beta_p\rightarrow\beta$. The kinematic distributions are once again expressed in terms of the scattering angle $\theta$ and the pseudorapidity $\eta$ for various values of $\kappa$. 

\subsection{Spin-1 monopole cross section}

Spin-1 monopole pair production via PF proceeds in three ways: a $t$-channel process, a $u$-channel process and a 4-point interaction, as depicted in fig.~\ref{FeynGraphsFig}. The related Feynman rules are given in fig.~\ref{FeynRulesFig}, where the boson polarisation vectors are shown for the gauge field, $\epsilon(q_1)_{\lambda},\epsilon(q_2)_{\lambda'}$, and the monopole field, $\Upsilon(p_1)_{\kappa}, \Upsilon(p_2)_{\kappa'}$. The monopoles are assumed to have mass $M$ and momentum 4-vectors $p_{1\mu}$ and $p_{2\mu}$, where (on mass-shell): $p_{1,2}^2=M^2$. The photon  momentum 4-vectors are $q_{1\mu}$ and $q_{2\mu}$, $q_{1,2}^2=0$. From \eqref{spin1feyngraphs}, the matrix amplitude for each process is derived below: 
\begin{subequations} \label{MatrixAmp}
\begin{align}
\begin{split}
	\mathcal{M}_{t} = \epsilon_{\lambda'}^{\nu}\Upsilon_{\kappa'}^{\sigma}&(-ig(\beta))(-g_{\sigma\nu}(p_2+\kappa q_2)_{\pi}-g_{\nu\pi}(p_2-(\kappa+1)q_2)_{\sigma}-g_{\sigma\pi}(q_2-2p_2)_{\nu})\frac{(-ig^{\pi\delta}+i\frac{k^{\pi}k^{\delta}}{M^{2}})}{k^{2}-M^{2}}
	\\ &(-ig(\beta))(-g_{\rho\mu}(p_{1}+\kappa q_1)_{\delta}-g_{\mu\delta}(p_1-(\kappa+1)q_1)_{\rho}-g_{\rho\delta}(q_1-2p_1)_{\mu})\epsilon_{\lambda}^{\mu}\Upsilon_{\kappa}^{\rho}
\end{split}
\\
\begin{split}
		\mathcal{M}_{u} = \epsilon_{\lambda'}^{\mu}\Upsilon_{\kappa'}^{\sigma}&(-ig(\beta))(-g_{\sigma\mu}(p_2+\kappa q_1)_{\pi}-g_{\mu\pi}(p_2-(\kappa+1)q_1)_{\sigma}-g_{\sigma\pi}(q_1-2p_2)_{\mu})\frac{(-ig^{\pi\delta}+i\frac{\tilde{k}^{\pi}\tilde{k}^{\delta}}{M^{2}})}{\tilde{k}^{2}-M^{2}}
	\\ &(-ig(\beta))(-g_{\rho\nu}(p_{1}+\kappa q_2)_{\delta}-g_{\nu\delta}(p_1-(\kappa+1)q_2)_{\rho}-g_{\rho\delta}(q_2-2p_1)_{\nu})\epsilon_{\lambda}^{\nu}\Upsilon_{\kappa}^{\rho}
\end{split}	
\\
\begin{split}
	\mathcal{M}_{4} = \epsilon_{\lambda'}^{\nu}\epsilon_{\lambda}^{\mu}\Upsilon& _{\kappa'}^{\sigma}\Upsilon_{\kappa}^{\rho}\Big[-2ig^{2}(\beta)(g_{\mu\nu}g_{\sigma\rho})+ig^{2}(\beta)(g_{\mu\sigma}g_{\nu\rho}+g_{\mu\rho}g_{\nu\sigma})\Big].
\end{split}	
\\ \nonumber	
\end{align}
\end{subequations}

\begin{figure}[ht!]
\justify
\begin{minipage}[t]{0.5\textwidth}
\centering
\begin{tikzpicture}
\begin{feynman}
  	\vertex (a1){\(\mathcal{A}_{\mu}, \epsilon_{\lambda}\)};
	\vertex[below=2cm of a1] (a2);
	\vertex[below=1.3cm of a2] (a3);
	\vertex[left=1.3cm of a3] (a4) {\(W_{\nu}, \Upsilon_{\kappa}\)};
	\vertex[right=1.3cm of a3] (a5) {\(W_{\rho}, \Upsilon_{\kappa'}\)};
	\diagram* {
       		(a4) -- [charged boson, edge label=\(p_{1\nu}\)] (a2) -- [boson, edge label=\(q_{\mu}\)] (a1);
		(a2) -- [charged boson, edge label=\(p_{2\rho}\)] (a5)
      };
\end{feynman}
\end{tikzpicture}
\begin{equation} \notag (a) \end{equation}
\end{minipage}
\begin{minipage}[t]{0.5\textwidth}
\centering
\begin{tikzpicture}
\begin{feynman}
  	\vertex (a1);
	\vertex[above=1.3cm of a1] (a2);
	\vertex[left=1.3cm of a2] (a3) {\(\mathcal{A}_{\mu},  \epsilon^{\mu}_{\lambda}\)};
	\vertex[right=1.3cm of a2] (a4) {\(\mathcal{A}_{\nu}, \epsilon^{*\nu}_{\lambda'}\)};
	\vertex[below=1.3cm of a1] (b2);
	\vertex[left=1.3cm of b2] (b3) {\(W_{\sigma}, \Upsilon^{\rho}_{\kappa}\)};
	\vertex[right=1.3cm of b2] (b4) {\(W_{\rho}, \Upsilon^{*\sigma}_{\kappa'}\)};
	\diagram* {
       		(a3) -- [boson, edge label=\(q_{1\mu}\)] (a1) -- [boson, edge label=\(q_{2\nu}\)] (a4);
		(b3) -- [charged boson, edge label=\(p_{1\sigma}\)] (a1) -- [charged boson, edge label=\(p_{2\rho}\)] (b4)
      };
\end{feynman}
\end{tikzpicture}
\begin{equation} \notag (b) \end{equation}
\end{minipage}
\begin{align}\label{spin1feyngraphs}
	\begin{split}
	\propto -ig(\beta)&(-g^{\nu\mu}(-\kappa p_2+\kappa p_1+p_1)^{\rho}
	\\&-g^{\mu\rho}(p_2+\kappa p_2- \kappa p_1)^{\nu}+g^{\rho\nu}(p_1+p_2)^{\mu})
	\end{split}
	\quad \quad \quad \propto -2ig(\beta)^{2}(g^{\mu\nu}g^{\sigma\rho})+ig(\beta)^{2}(g^{\mu\sigma}g^{\nu\rho}+g^{\mu\rho}g^{\nu\sigma}) 
\end{align}
\caption{Feynman rules for the three- and four-point couplings of the spin-1 field and gauge field in the Lee-Yang model. Wavy lines with arrows indicate the vector monopole field, while wavy lines without arrows represent the photon. The corresponding rules are given in \eqref{spin1feyngraphs}.}
\label{FeynRulesFig}
\end{figure}

Following the discussion in section \ref{sectionAmtoKin}, the probability of production is proportional to the square of the matrix amplitude, $|\overline{\mathcal{M}}|_{PH}^{2}$, averaged over the initial photon polarisations $\lambda^{(')}$,  and summed over the final state monopole polarisations $\kappa^{(')}$, where the matrix amplitude $\mathcal{M}_{PH}$ is equal to the sum of the amplitudes for each contributing process $\mathcal{M}_{PH}=\mathcal{M}_{t}+\mathcal{M}_{u}+\mathcal{M}_{4}$. One has, therefore:
\begin{equation}\label{TotalSquaredMatrixAmp}
	|\overline{\mathcal{M}}|^{2}_{PH}=\frac{1}{4}\sum_{\lambda\lambda'}\sum_{\kappa\kappa'}\Big[\mathcal{M}_{t}\mathcal{M}_{t}^{*}+\mathcal{M}_{u}\mathcal{M}_{u}^{*}+\mathcal{M}_{4}\mathcal{M}_{4}^{*}+2\mathcal{M}_{t}\mathcal{M}_{u}^{*}+2\mathcal{M}_{t}\mathcal{M}_{4}^{*}+2\mathcal{M}_{4}\mathcal{M}_{u}^{*}\Big].
\end{equation}
Equation~\eqref{TotalSquaredMatrixAmp} can be re-written by factoring out the polarisations and defining the factorised matrix element as the product of vertices $\Gamma_{\alpha\beta\delta}$ and photon propagators $\Delta_{\gamma}^{\alpha\beta}$:
\begin{align}\label{TotalSquaredMatrixAmpFact}
\notag	|\overline{\mathcal{M}}|^{2}_{PH}&=\frac{1}{4}\sum_{\lambda\lambda'}\epsilon^{*\nu}_{\lambda'}\epsilon^{\mu}_{\lambda}\epsilon^{\nu'}_{\lambda'}\epsilon^{*\mu'}_{\lambda}\sum_{\kappa\kappa'}\Upsilon^{*\sigma}_{\kappa'}\Upsilon^{\rho}_{\kappa}\Upsilon^{\sigma'}_{\kappa'}\Upsilon^{*\rho'}_{\kappa}\Big[\mathcal{M}_{\mu\nu\rho\sigma}\mathcal{M}_{\mu'\nu'\rho'\sigma'}^{*}\Big]~,\\
\mathcal{M}_{\mu\nu\rho\sigma}&=\Gamma_{\nu\sigma\pi}\Delta_{\gamma}^{\pi\delta}\Gamma_{\delta\mu\rho} +\Gamma_{\mu\sigma\pi}\Delta_{\gamma}^{\pi\delta}\Gamma_{\delta\nu\rho}+\Gamma_{\mu\nu\sigma\rho}~.
\end{align}
We assume that the monopoles have mass $M$, polarisation vectors $\Upsilon(p_1)_{\kappa}, \Upsilon(p_2)_{\kappa'}$ and are characterised by momentum 4-vectors $p_{1\mu}$ and $p_{2\mu}$, where (on mass-shell): $p_{1,2}^2=M^2$. The photons with polarisation vectors $\epsilon(q_1)_{\lambda},\epsilon(q_2)_{\lambda'}$ have momentum 4-vectors $q_{1\mu}$ and $q_{2\mu}$, $q_{1,2}^2=0$. The polarisation sums are evaluated by \Math following the standard sum rules (listed in appendix~\ref{app:basic}, see eqs.~\eqref{sumphot}), \eqref{evaltrace1}) such that
\begin{equation}
|\overline{\mathcal{M}}|^{2}_{PH}=\frac{1}{4}(-g^{\rho\rho'}+\frac{p_1^{\rho}p_1^{\rho'}}{M^2})(-g^{\sigma\sigma'}+\frac{p_2^{\sigma}p_2^{\sigma'}}{M^2})(-g^{\mu\mu'})(-g^{\nu\nu'})\Big[\mathcal{M}_{\mu\nu\rho\sigma}\mathcal{M}_{\mu'\nu'\rho'\sigma'}^{*}\Big]~,
\end{equation}
where $g^{\mu\nu}$ is the Minkowski space-time metric tensor.
The expression for $|\overline{\mathcal{M}}|^{2}_{PH}$ is reduced by \Math and inserted into the definition of the differential cross section in eq.~\eqref{Xsecdeff},
\begin{equation}
	\frac{\dd\sigma^{S=1}_{\gamma\gamma\rightarrow\overline{M}M}}{\dd\Omega} =\frac{1}{64\pi^{2} \sgg}\beta|\overline{\mathcal{M}}|^{2}_{PH},
\end{equation}
where $\beta=|\vec{p}| / E_p$ is the monopole boost,  and for which $|\vec{q}|=E_q$ is the photon energy, such that $E_p=E_q$ in this setup. 

Following a similar procedure to the treatment above and using the set of Feynman rules given in \eqref{FeyRule1DYeq}, kinematic distributions for monopole pair production by DY are discussed next. The relevant Feynman rules are drawn in fig.~\ref{FeyRule1DY}, where fig.~\ref{FeyRule1DY}(a) is the vertex for the $W_{\sigma}W_{\rho}^{\dagger}\mathcal{A}_{\nu}$ interaction and fig.~\ref{FeyRule1DY}(b) shows the $\psi\overline{\psi}\mathcal{A}_{\nu}$ vertex which comes directly from the ordinary QED with quarks as Dirac spinors. The quarks each have a mass $m$ considered small compared to the monopole mass, $M$, and are characterised by momentum 4-vectors $q_{1\mu}$ and $q_{2\mu}$, where on mass shell one has $q_{1,2}^2=m^2$. Similarly, the mesons have mass $M$ each and are characterised by momentum 4-vectors $p_{1\mu}$ and $p_{2\mu}$, where on mass-shell  one has $p_{1,2}^2=M^2$.

The boson polarisation vectors are indicated for the monopoles by $\Upsilon(p_1)_{\kappa}^{\rho}, \Upsilon(p_2)_{\kappa'}^{\sigma}$. The matrix amplitude is derived from \eqref{FeyRule1DYeq}:
\begin{align}
 	\mathcal{M}_{DY} & = \Upsilon_{\kappa}^{\rho}\Upsilon_{\kappa'}^{*\sigma} u_{\alpha}(-iQe\gamma_{\mu})\overline{v}_{\beta}(\frac{-ig^{\mu\nu}}{k^{2}})(-ig(\beta))\left(-g_{\sigma\nu}(-\kappa p_2+\kappa p_1+p_1)_{\rho}-g_{\nu\rho}(p_2+\kappa p_2- \kappa p_1)_{\sigma}+g_{\rho\sigma}(p_1+p_2)_{\nu}\right),  \label{MatAmpSpin1}
\end{align}
where  $u_{\alpha}, v_{\beta}, \overline{u}_{\alpha}, \overline{v}_{\beta}$ are the fermionic spinors representing the quarks, $g^{\mu\nu}$ is the Minkowski metric tenor and $\Upsilon_{\kappa}$ is the monopole polarisation vector. 

\begin{figure}[ht!]
\justify
\begin{minipage}[t]{0.5\textwidth}
\centering
\begin{tikzpicture}
\begin{feynman}
  	\vertex (a1){\(\mathcal{A}_{\nu}\)};
	\vertex[right=2cm of a1] (a2);
	\vertex[below=1.75cm of a2] (a3);
	\vertex[right=1cm of a3] (a4) {\(W_{\rho}^{\dagger},  \Upsilon^{*\sigma}_{\kappa'}\)};
	\vertex[above=1.75cm of a2] (a5);
	\vertex[right=1cm of a5] (a6) {\(W_{\sigma}, \Upsilon_{\kappa}^{\rho}\)};
	\vertex[right=3cm of a2] (b1) {\((a)\)};
	\diagram* {
       		(a4) -- [charged boson, edge label=\(p_{2\sigma}\)] (a2) -- [boson, edge label=\(k_{\nu}\)] (a1);
		(a2) -- [charged boson, edge label=\(p_{1\rho}\)] (a6)
      };
\end{feynman}
\end{tikzpicture}
\end{minipage}
\begin{minipage}[t]{0.5\textwidth}
\centering
\begin{tikzpicture}
  \begin{feynman}
  	\vertex (a1) {\(\mathcal{A}_{\mu}\)};
	\vertex[right=2cm of a1] (a2);
	\vertex[above=1.75cm of a2] (a3);
	\vertex[right=1cm of a3] (a4) {\(\psi\)};
	\vertex[below=1.75cm of a2] (a5);
	\vertex[right=1cm of a5] (a6) {\(\overline{\psi}\)};
	\vertex[right=3cm of a2] (b1) {\((b)\)};
	
	\diagram* {
       		(a6) -- [fermion, edge label=\(q_{1\mu}\)] (a2) -- [fermion, edge label=\(q_{2\mu}\)] (a4),
		(a1) -- [boson, edge label=\(k_{\mu}\)] (a2),
      };
\end{feynman}
\end{tikzpicture}
\end{minipage}
\begin{align}\label{FeyRule1DYeq}
\begin{split}
	\propto -ig(\beta)(-&g^{\sigma\mu}(-\kappa p_2+\kappa p_1+p_1)^{\rho}-\\&g^{\mu\rho}(p_2+\kappa p_2- \kappa p_1)^{\sigma}+g^{\rho\sigma}(p_1+p_2)^{\mu}) 
\end{split}
	\Quad\Quad \Quad \Quad \Quad\Quad\Quad \propto -iQe \gamma^{\mu}\Quad\Quad\Quad\Quad\Quad\Quad\Quad\Quad
\end{align}
\caption{(a) Spin-1 bosonic monopoles interacting with gauge field at the three-point vertex, where $g(\beta)$ is the magnetic coupling. (b) The quarks couple to the photon as expected form Dirac QED for spinors with fractional charges, $Qe$, where $Q$ is the charge fraction relevant to the quarks involved, $Q=\frac{1}{3},\frac{2}{3}$.  The corresponding Feynman rules are given in \eqref{FeyRule1DYeq}. }\label{FeyRule1DY}
\end{figure}

The averaged matrix amplitude squared is then written as
\begin{equation}
\begin{split}
 	|\overline{\mathcal{M}^{2}}|_{DY} & = \frac{1}{3}\, \frac{1}{4}\, \frac{5}{9}\, \frac{3 \, e^2g^2(\beta)}{k^4}\, 
	\sum_\kappa \, \sum_{\kappa'} \, 
	\Big[\Upsilon_{\kappa}^{\rho}\Upsilon_{\kappa'}^{\sigma*}\Upsilon_{\kappa}^{*\rho'}\Upsilon_{\kappa'}^{\sigma'}\Big]\, \Tr\Big[(u_{\alpha}\gamma_{\mu}\overline{v}_{\beta})(v_{\beta}\gamma_{\mu'}\overline{u}_{\alpha})\Big]\\
& (g^{\mu\nu})\left(-g_{\sigma\nu}(-\kappa p_2+\kappa p_1+p_1)_{\rho}-g_{\nu\rho}(p_2+\kappa p_2- \kappa p_1)_{\sigma}+g_{\rho\sigma}(p_1+p_2)_{\nu}\right)\\
& (g^{\mu'\nu'})\left(-g_{\sigma'\nu'}(-\kappa p_2+\kappa p_1+p_1)_{\rho'}-g_{\nu'\rho'}(p_2+\kappa p_2- \kappa p_1)_{\sigma'}+g_{\rho'\sigma'}(p_1+p_2)_{\nu'}\right).\label{squaredMatAmp1}
\end{split}
\end{equation}
In the above expression, the factors $\frac{1}{3}$ and $\frac{1}{4}$ are attributed to the averaging over colour and quark spins states, respectively. Indeed, taking the example of a red up quark, if it meets an anti-up quark, it has a one in three chance of this anti-quark having the colour anti-red, so the cross section is reduced by a factor $\frac{1}{3}$. In the same way, each sum over quark spins contributes a factor of a half to the cross section as well. The factor of 3 accounts for flavour multiplicity and the factor $\frac{5}{9}$ comes from the summation over quark charge fractions. 

The various traces are evaluated in appendix~\ref{app:basic} (see eqs.~\eqref{evaltrace1}), \eqref{spinsum} and \eqref{evaltrace2}). After substitution into the squared matrix amplitude \eqref{squaredMatAmp1}, the relevant distributions are evaluated in the centre-of-mass frame as explained in section \ref{sectionAmtoKin}, using \eqref{Xsecdeff}, 
\begin{equation}
	\frac{\dd\sigma^{S=1}_{q\overline{q}\rightarrow M\overline{M}}}{\dd\Omega} =\frac{1}{64\pi^{2} \sqq}\frac{\beta_p}{\beta_q}|\overline{\mathcal{M}}|^{2}, \nonumber
\end{equation}
where the substitution $|\vec{p}_1| / |\vec{q}_1| = \beta_p / \beta_q$ was made because $\beta_p=|\vec{p}| / E_p$ represents the monopole boost, $\beta_q= |\vec{q}| / E_q$ represents the quark boost, and ${E_p}={E_q}$ in the centre-of-mass frame defined in fig.~\ref{geometry}.


\end{document}